\documentclass[a4paper,11pt]{article}
\usepackage[top=2.5cm, bottom=2.5cm, left=2cm, right=2cm]{geometry}
\usepackage{settings_custom}
\usepackage{authblk}
\usepackage[sort&compress,numbers]{natbib}
\allowdisplaybreaks

\begin{document}

\title{Fitting an ellipsoid to random points: predictions using the replica method}
\date{\today}

\author[1]{Antoine Maillard\thanks{Email: \texttt{antoine.maillard@math.ethz.ch}.}}
\author[2]{Dmitriy Kunisky\thanks{Email: \texttt{dmitriy.kunisky@yale.edu}. Partially supported by ONR Award N00014-20-1-2335 and a Simons Investigator Award to Daniel Spielman.}}

\affil[1]{Department of Mathematics, ETH Zurich}
\affil[2]{Department of Computer Science, Yale University}

\maketitle

\begin{abstract}
    We consider the problem of fitting a centered ellipsoid to $n$ standard Gaussian random vectors in $\bbR^d$, as $n, d \to \infty$ with 
    $n/d^2 \to \alpha > 0$.
    It has been conjectured that this problem is, with high probability, satisfiable (SAT; that is, there exists an ellipsoid passing through all $n$ points) for $\alpha < 1/4$, and unsatisfiable (UNSAT) for $\alpha > 1/4$.
    In this work we give a precise analytical argument, based on the non-rigorous replica method of statistical physics, 
    that indeed predicts a SAT/UNSAT transition at $\alpha = 1/4$, as well as the shape of a typical fitting ellipsoid in the SAT phase (i.e., the lengths of its principal axes).
    Besides the replica method, our main tool is the dilute limit of extensive-rank ``HCIZ integrals'' of random matrix theory.
    We further study different explicit algorithmic constructions of the matrix characterizing the ellipsoid.
    In particular, we show that a procedure based on minimizing its nuclear norm 
    yields a solution in the whole SAT phase. Finally, we characterize the SAT/UNSAT transition for ellipsoid fitting of a large class of rotationally-invariant random vectors.
    Our work suggests mathematically rigorous ways to analyze fitting ellipsoids to random vectors, which is the topic of a companion work.
\end{abstract}

\section{Introduction}\label{sec:intro}
\noindent
We consider the following constraint satisfaction problem: given $n$ random \change{standard Gaussian} points in dimension $d$, when is there an ellipsoid (centered at $0$) whose boundary goes through all the points?
More formally, we want to solve the problem:
\begin{equation}\label{eq:P}
   (\rm{P}) \, : \, \begin{dcases} 
    \{\bx_\mu^\T \bS \bx_\mu = d\}_{\mu=1}^n, \\ 
    \bS \succeq 0,
   \end{dcases}  
\end{equation}
where $\bS \in \bbR^{d \times d}$ is a symmetric matrix,
and $(\bx_1, \cdots, \bx_n)$ is the set of points to fit.
Geometrically speaking, the eigenvectors of $\bS$ give the directions of the principal axes of the ellipsoid, while 
its eigenvalues $(\lambda_i)_{i=1}^d$ are related to the lengths $(r_i)_{i=1}^d$ of its principal (semi-)axes by
$r_i = \sqrt{d} \lambda^{-1/2}$.
In what follows, we will refer to the rescaled quantities $r'_i = r_i / \sqrt{d}$ as the lengths of the ellipsoid axes, 
effectively rescaling distances so that the sphere with $\bS = \Id_d$ has all axes of length $1$.

\myskip
This question was first raised in the context of statistical estimation
\cite{saunderson2011subspace,saunderson2012diagonal,saunderson2013diagonal}, and has received significant attention 
very recently \cite{potechin2023near,kane2023nearly,hsieh2023ellipsoid,tulsiani2023ellipsoid,bandeira2023fitting}. 
In the original series of works on this problem \cite{saunderson2011subspace,saunderson2012diagonal,saunderson2013diagonal},
it was conjectured that this constraint satisfaction problem had a sharp satisfiability/unsatisfiability (or SAT/UNSAT) 
transition: if $n, d \to \infty$ with $n/d^2 \to \alpha > 0$, then $(\rm P)$ typically admits solutions 
if $\alpha < 1/4$, and no solution if $\alpha > 1/4$.
The conjecture arose from numerical simulations, as well as the remark that if the linear measurements were Gaussian (i.e., if the rank-one matrices $\bx_\mu \bx_\mu^\T$ were to be replaced by standard Gaussian matrices), 
then this modified problem would provably have a SAT/UNSAT transition at $\alpha = 1/4$,
corresponding to the statistical dimension of the cone of positive semidefinite matrices \cite{gordon1988milman,amelunxen2014living,potechin2023near}.
See below for further details about this argument.

\myskip
In the present work, we leverage the non-rigorous replica method of statistical physics to
analytically characterize the SAT/UNSAT transition.
The replica method unveils a universality property of a quantity known as the free entropy, which allows us to precisely match the problem of ellipsoid 
fitting to a variant of the ``Gaussian measurement'' model described above.
To the best of our knowledge, our work is the first analytical derivation of the $\alpha = 1/4$ satisfiability transition in the original ellipsoid fitting problem.
Our calculation leverages 
a ``dilute'' limit of objects known in the random matrix theory literature as extensive-rank HCIZ\footnote{Standing for Harish-Chandra-Itzykson-Zuber \cite{harish1957differential,itzykson1980planar}.} integrals \cite{matytsin1994large,guionnet2002large,bun2014instanton}. Using these, we provide analytical formulas for $(i)$ the size of the solution space to $(\rm P)$ at any given $\alpha < 1/4$, 
$(ii)$ the typical spectral density of a uniformly-sampled solution $\bS$ (i.e., the geometrical shape of a typical ellipsoid fit), and $(iii)$ the minimal length $\ell^\star(\alpha)$ of the longest principal axis of any ellipsoid fit\footnote{The purpose of the latter results is to illustrate how fitting ellipsoids, when they exist, become more eccentric as the number of points to interpolate increases.}.
In a companion work \cite{maillard2023fitting}, the first author and A.\ Bandeira use the mathematical insights on universality presented here to rigorously prove that a slightly modified version of $(\rm P)$ indeed undergoes a SAT/UNSAT transition at $\alpha = 1/4$.

\myskip 
Most recent progress on the ellipsoid fitting conjecture has relied on the analysis of explicit solutions $\bS^\star$ to the first 
set of constraints in eq.~\eqref{eq:P} (these constraints are linear in $\bS$), and arguing that $\bS^\star \succeq 0$ with high probability if $\alpha = n/d^2$ is small enough \cite{potechin2023near,kane2023nearly,hsieh2023ellipsoid,tulsiani2023ellipsoid,bandeira2023fitting}. 
Using the replica framework, we are able to analytically predict the range of $\alpha$ for which a wide class of these methods succeed, and we characterize the typical spectral density of $\bS^\star$ for such methods. 
It follows from our analysis that a minimal nuclear-norm estimator (i.e., $\bS^\star = \argmin \|\bS\|_{\NN}$, in the notation defined below, over all $\bS$ satisfying the linear constraints in eq.~\eqref{eq:P}) 
is typically positive semidefinite in the whole SAT phase $\alpha = n/d^2 < 1/4$.
Our analyses are supported by finite-$d$ simulations for both for the SAT/UNSAT transition and the explicit constructions just mentioned,
showing very good agreement with our analytical predictions.

\myskip 
Finally, we consider the more general problem of fitting ellipsoids to non-Gaussian but rotationally-invariant i.i.d.\ vectors $\{\bx_\mu\}_{\mu=1}^n$, with an arbitrarily-fluctuating norm. 
We show, again using our replica framework, that the SAT/UNSAT transition point $\alpha_c$ depends solely on the parameter $\tau \coloneqq d^{-1} \, \Var[\|\bx_1\|^2]$, 
and we compute the corresponding critical $\alpha_c(\tau)$.

\myskip 
\textbf{Notations --} 
\change{We denote $[n] \coloneqq \{1, \cdots, n\}$ the set of integers from $1$ to $n$.}
$\lesssim$ denotes inequality up to a global constant (not depending on $d$).
We denote by $\mcS_d$ the set of $d \times d$ real symmetric matrices.
For a function $V : \bbR \to \bbR$, and $\bS \in \mcS_d$ with eigenvalues $(\lambda_i)_{i=1}^d$, we define $V(\bS)$ as the matrix with the same eigenvectors as $\bS$ 
and eigenvalues $(V(\lambda_i))_{i=1}^d$. 
We write $|\bS| = V(\bS)$ for $V(x) = |x|$.
For $\gamma \in [1, \infty]$ and $\bS \in \mcS_d$ we denote by $\|\bS\|_{S_\gamma} \coloneqq (\sum_i |\lambda_i|^\gamma)^{1/\gamma}$ the Schatten-$\gamma$ norm.
We denote by $\|\bS\|_{\op} \coloneqq \|\bS\|_{S_\infty}$ the operator norm, and $\|\bS\|_\NN \coloneqq \|\bS\|_{S_1} = \Tr[|\bS|]$ the nuclear norm.
We say that a random matrix $\bY \in \mcS_d$ is generated from the \emph{Gaussian Orthogonal Ensemble} $\GOE(d)$ if
\begin{equation*}
    Y_{ij} \iid \mcN(0, (1+\delta_{ij})/d) \textrm{ for } i \leq j.
\end{equation*}
For any $B \subseteq \bbR$ we denote by $\mcM_1^+(B)$ the set of real probability distributions on $B$.
Finally, we define $\rho_{\sci} \in \mcM_1^+(\bbR)$ as the standard semicircular density, i.e.,
\begin{equation*}
  \rho_\sci(x) \coloneqq \frac{1}{2\pi} \sqrt{4-x^2} \, \indi\{|x| \leq 2\}.
\end{equation*}

\subsection{General remarks}

\noindent
Before discussing the existing literature on ellipsoid fitting and presenting our main results, 
we make a couple of basic remarks.

\myskip 
\textbf{Structure of the solution space --} Notice that $(\rm P)$ is a convex program (it is an example of a semidefinite program). 
In particular, the set of solutions to $(\rm P)$ is convex, a remark that will greatly simplify its characterization via the replica method.

\myskip 
\textbf{The case $\alpha > 1/2$ --}
Since $(\rm P)$ almost surely contains a set of $n$ linearly independent linear constraints\footnote{The linear independence can be shown easily; see, e.g., Lemma~B.1 of \cite{potechin2023near}.}
on $\bS$, $(\rm P)$ will be unsatisfiable if $n > \dim(\mcS_d) = d(d+1)/2$, that is, if $\alpha > 1/2$.
Despite the simplicity of this argument, the condition $\alpha > 1/2$ is still the best-known rigorous bound on the UNSAT phase.

\myskip
\textbf{The trace of solutions --}
Assume that $\bS$ is a solution to $\bx_\mu^\T \bS \bx_\mu = d$ for all $\mu \in [n]$ (without necessarily having $\bS \succeq 0$).
Then we must have 
\begin{align*}
    \left|\frac{1}{d}\Tr[\bS] - 1\right| &= \frac{1}{d}\left| \Tr\left[\bS \left(\frac{1}{n} \sum_{\mu=1}^n \bx_\mu \bx_\mu^\T - \Id_d\right)\right]\right|, \\ 
    &\leq\left(\frac{1}{d} \Tr |\bS|\right)\cdot \left\| \frac{1}{n} \sum_{\mu=1}^n \bx_\mu \bx_\mu^\T - \Id_d\right\|_\op.
\end{align*}
Moreover, it is well-known that the empirical covariance matrix of $n$ $d$-dimensional Gaussian vectors concentrate around identity
as $n/d \to \infty$, and more precisely we have with high probability \cite{vershynin2018high}:
\begin{equation*}
    \left\| \frac{1}{n} \sum_{\mu=1}^n \bx_\mu \bx_\mu^\T - \Id_d\right\|_\op \lesssim \sqrt{\frac{d}{n}}.
\end{equation*}
Therefore, for any such $\bS$ we have
\begin{equation*}
    \left|\frac{1}{d}\Tr[\bS] - 1\right| \lesssim \sqrt{\frac{d}{n}} \cdot \left(\frac{1}{d} \Tr |\bS|\right).
\end{equation*}
In particular, if $\bS \succeq 0$, or if $\Tr[|\bS|] = \mcO(d)$, this implies:
\begin{equation}\label{eq:trace_close_1}
    \left|\frac{1}{d}\Tr[\bS] - 1\right| \lesssim \sqrt{\frac{d}{n}} \lesssim \frac{1}{\sqrt{d}},
\end{equation}
where the last inequality holds in the regime $n = \Theta(d^2)$ we will consider.
Thus, subject even just to the linear constraints of $(\rm P)$, the trace of any satisfying $\bS$ must be strongly concentrated.
\footnote{Notice that in the Gaussian case $\Delta \sim \mcN(0, 2)$, so we recover $\bz \sim \mcN(0, 2\bQ)$.}

\change{
\myskip
\textbf{Almost exact ellipsoid fitting --}
In the current manuscript we consider the problem of \emph{exactly} fitting an ellipsoid to random points.
We notice that allowing an error (even rather small) in the fit can drastically change the behavior of the set of solutions. 
Indeed, by elementary concentration arguments
one can show (see Appendix~\ref{subsec_app:almost_exact_fit} for details) that, with high probability as $d,n \to \infty$ (for some constants $C_1, C_2 > 0$):
\begin{equation}\label{eq:almost_exact_fit}
   \frac{C_1}{d} \leq \frac{1}{n} \sum_{\mu=1}^n \left(\frac{\|\bx_\mu\|^2}{d} - 1\right)^2 \leq \frac{C_2}{d}.
\end{equation}
We can interpret eq.~\eqref{eq:almost_exact_fit} as showing that the sphere (with $\bS = \Id_d$) is an ellipsoid fit to $n$ Gaussian points in dimension $d$ with high probability, 
up to a mean squared error of $\Theta(1/d)$: this almost exact ellipsoid fitting problem thus always admits this trivial solution.
On the other hand, it is believed that if we ask that ellipsoids have a fitting error at most $\smallO(1/d)$, then the problem exhibits a sharp 
satisfiability transition for $n \simeq d^2/4$, as in exact ellipsoid fitting.
The rigorous analysis of the companion work to this manuscript~\cite{maillard2023fitting} tackles such an ``almost exact'' ellipsoid fitting problem
(where the fitting has to be much smaller than the one of $\bS = \Id_d$), and we refer to this work for more details on this point.
}

\subsection{Related literature}\label{subsec:literature}

\subsubsection{Ellipsoid fitting}
The ellipsoid fitting problem is a seemingly elementary question of random geometry: for instance, if $(\bx_\mu)_{\mu=1}^n$ possess a (centered) ellipsoid fit, 
then the points $(\pm \bx_\mu)_{\mu=1}^n$ all lie on the boundary of their convex hull.
Perhaps more surprisingly, ellipsoid fitting also bears interesting connections 
to statistical learning and theoretical computer science.

\myskip 
To the best of our knowledge, the first such motivation to consider ellipsoid fitting of random vectors arose from the study of a procedure known as Minimum Trace Factor Analysis (MTFA) \cite{saunderson2011subspace,saunderson2012diagonal,saunderson2013diagonal}, 
which aims to decompose an observed data matrix $\bM$ as $\bM = \bL + \bD$, with $\bL \succeq 0$ and low-rank, and $\bD$ diagonal.
More precisely, it was proven that the probability of finding an ellipsoid fit is directly related to the probability of success of MTFA in a random setup.
\change{
Furthermore, it was also shown~\cite{saunderson2012diagonal,potechin2023near} that 
ellipsoid fitting is linked to an average-case discrepancy problem, more precisely to the performance of a semidefinite relaxation 
of the non-convex optimization problem $\min_{\bepsilon \in \{\pm 1\}^d} \|\bA \bepsilon\|_\infty$ when $\bA$ is a real symmetric $d \times d$ random matrix, 
see Appendix~A of \cite{potechin2023near}.}
Another connection can be made to the problem of overcomplete independent component analysis \cite{podosinnikova2019overcomplete}.
For a more exhaustive and detailed account of the numerous connections of ellipsoid fitting to problems in theoretical computer science 
and statistics we refer the interested reader to the very complete exposition of \cite{potechin2023near}.

\begin{figure}[!t]
  \centering
\begin{tikzpicture}
  % Define the parameters

  % Draw the axis (large horizontal arrow)
  \draw[->,thick,line width = 1.5pt] (-2.5,0) -- (13,0) node[below=5pt,font=\large] {$n$};

  % Define the positions and labels for the ticks
  \pgfmathsetmacro\tickPositionA{-2.}
  \pgfmathsetmacro\tickPositionE{-0.25}
  \pgfmathsetmacro\tickPositionF{1.75}
  \pgfmathsetmacro\tickPositionB{4}
  \pgfmathsetmacro\tickPositionC{8}
  \pgfmathsetmacro\tickPositionD{11}
  \def\tickLabels{{{"$d^{6/5-\eps}$","\phantom{}\cite{saunderson2012diagonal}"},{"$d^{3/2-\eps}$","\phantom{}\cite{ghosh2020sum}"},{"$d^2/\plog(d)$","\phantom{}\cite{potechin2023near,kane2023nearly}"},{"$d^2/C$","\phantom{}\cite{hsieh2023ellipsoid,tulsiani2023ellipsoid,bandeira2023fitting}"}, {"$d^2/4$", "\phantom{}\cite{maillard2023fitting}"}, {"$d^2/2$",""}}}

  % Draw the ticks and labels
  \foreach \pos [count=\i] in {\tickPositionA, \tickPositionE, \tickPositionF, \tickPositionB, \tickPositionC, \tickPositionD} {
    \draw[line width=1.5pt] (\pos,0.1) -- (\pos,-0.1) node[below] {\pgfmathparse{\tickLabels[\i-1][0]}\pgfmathresult};
    \node at (\pos,-0.7) [below, anchor=north] {\pgfmathparse{\tickLabels[\i-1][1]}\pgfmathresult};
  };

    % Add colored bands
  \fill[green, opacity=0.4] (-2.5,0.2) rectangle (\tickPositionB, -0.1);
  \fill[yellow, opacity=0.4] (\tickPositionB,0.2) rectangle (\tickPositionC, -0.1);
  \fill[orange, opacity=0.4] (\tickPositionC,0.2) rectangle (\tickPositionD, -0.1);
  \fill[red, opacity=0.4] (\tickPositionD,0.2) rectangle (12.7, -0.1);

    % Text labels above colored regions
  \node[green!75!black, anchor=south, font = \small] at ({(-2 + \tickPositionB)/2},0.65) {SAT};
  \node[green!75!black, anchor=south, font = \small] at ({(-2 + \tickPositionB)/2},0.25) {(rigorous)};
  \node[yellow!80!black, anchor=south, font = \small] at ({\tickPositionB+(\tickPositionC-\tickPositionB)/2},0.65) {SAT};
  \node[yellow!80!black, anchor=south, font = \small] at ({\tickPositionB+(\tickPositionC-\tickPositionB)/2},0.25) {(conjecture)};
  \node[orange!100!black, anchor=south,font = \small] at ({\tickPositionC+(\tickPositionD-\tickPositionC)/2},0.65) {UNSAT};
  \node[orange!100!black, anchor=south,font = \small] at ({\tickPositionC+(\tickPositionD-\tickPositionC)/2},0.25) {(conjecture)};
  \node[red!100!black, anchor=south,font = \small] at ({\tickPositionD+(13-\tickPositionD)/2},0.65) {UNSAT};
  \node[red!100!black, anchor=south,font = \small] at ({\tickPositionD+(13-\tickPositionD)/2},0.25) {(rigorous)};

  \end{tikzpicture}
\caption{
\label{fig:summary}    
A summary of the current state of the ellipsoid fitting conjecture. 
In red, we show regions which are rigorously known to be in the UNSAT phase, and in orange regions which are conjectured to be.
Similarly, we show in green regions rigorously known to be in the SAT phase, and in yellow regions which are conjectured to be so.
The companion work to this manuscript \cite{maillard2023fitting} closes these gaps by proving rigorously that a slightly modified problem has a transition at $n \simeq d^2/4$.
}
\end{figure}
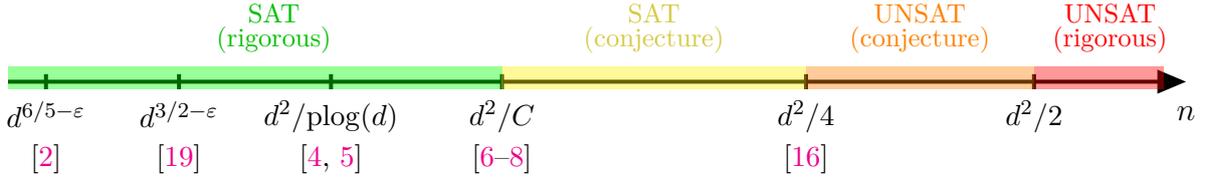

\myskip 
\textbf{Rigorous bounds --}
The series of work that introduced the ellipsoid fitting of random vectors \cite{saunderson2011subspace,saunderson2012diagonal,saunderson2013diagonal} 
provided a first bound: it was proven that $(\rm P)$ is feasible (with high probability) as long as $n = \mcO(d^{6/5 - \eps})$, 
for any $\eps > 0$. This bound was subsequently improved to $n = \mcO(d^{3/2-\eps})$ \cite{ghosh2020sum}, then 
to $n = \mcO(d^2 / \plog(d))$ (where $\plog(d)$ is a polynomial in $\log d$) \cite{potechin2023near,kane2023nearly}, 
and very recently to $n \leq d^2 / C$, for a (large) constant $C > 0$ \cite{hsieh2023ellipsoid,tulsiani2023ellipsoid,bandeira2023fitting}.
As we mentioned above, all these results relied on an explicit candidate, i.e.\ an explicit solution $\bS^\star$ to the linear constraints in eq.~\eqref{eq:P}: 
the task reduces then to prove that $\bS^\star \succeq 0$ (typically) if $n/d^2$ is small enough.
Finally, in a companion work to this manuscript \cite{maillard2023fitting}, it is proven that a slightly relaxed version of $(\rm P)$ 
undergoes a sharp SAT/UNSAT transition exactly at $n \simeq d^2/4$.
We summarize in Fig.~\ref{fig:summary} the current knowledge on the ellipsoid fitting problem.

\myskip
\textbf{Gordon's theorem and a Gaussian equivalent problem --}
As noticed in \cite{potechin2023near}, the satisfiability of the problem $({\rm P})$ becomes a classical problem of random geometry 
if one replaces the rank-one matrices $\bx_\mu \bx_\mu^\T$ by $\GOE(d)$ matrices $(\bG_\mu)_{\mu=1}^n$, that is: 
\begin{equation}\label{eq:P_Gaussian}
   (\rm{P}_\Gauss) \, : \, \begin{dcases} 
    \{\Tr[\bG_\mu \bS] = d\}_{\mu=1}^n, \\ 
    \bS \succeq 0,
   \end{dcases}  
\end{equation}
Indeed, $(\rm P_\Gauss)$ is equivalent to finding a point in the intersection 
of a randomly-oriented affine subspace of $\mcS_d$ (defined by the linear constraints) with the convex cone $\mcS_d^+$ of positive semidefinite matrices.
Such problems are precisely the focus of Gordon's celebrated ``escaping through a mesh'' theorem \cite{gordon1988milman}, 
and are known to exhibit a SAT/UNSAT transition as $n$ crosses the ``statistical dimension'' of the cone, which
approaches $d^2/4$ as $d \to \infty$ in the case of $\mcS_d^+$ \cite{amelunxen2014living}.
However, in the ellipsoid fitting problem the structure of the rank-one matrices $\bx_\mu \bx_\mu^\T$ implies that this random affine subspace is not uniformly oriented, preventing us from using these results. 
Nevertheless, we will show by leveraging the non-rigorous replica method that a quantity known as the free entropy of the space of solutions can be mapped to its counterpart in a variation of the problem 
$(\rm P_\Gauss)$ above, and we will make explicit which features of the distribution of the vectors $\{\bx_\mu\}_{\mu=1}^n$ are crucial for this ``universality'' to hold. 
This remark is at the heart of the rigorous analysis of \cite{maillard2023fitting}, see also the discussion in Section~\ref{sec:conclusion}.

\subsubsection{Statistical physics approaches}

The statistical physics of disordered systems, which originated in the 1970's from the study of models known as \emph{spin glasses} \cite{mezard1987spin}, has since 
seen numerous applications throughout statistical learning, information theory, and theoretical computer science. 
Many of these applications are thoroughly reviewed in the recent book \cite{charbonneau2023spin}.
Without aiming to be exhaustive, examples include error-correcting codes \cite{sourlas1989spin,nishimori2001statistical,montanari2007modern,mezard2009information},
random constraint satisfaction problems \cite{mezard2009information,krzakala2007landscape,mezard2009constraint}, 
a wide class of inference tasks \cite{mezard2009information,zdeborova2016statistical}, optimization of high-dimensional random functions \cite{fyodorov2004complexity,maillard2020landscape}, 
and learning in neural networks \cite{aubin2019committee,gabrie2020mean,mannelli2020marvels}.
While often based on non-rigorous techniques (such as the replica method), the statistical physics approaches are widely believed to be exact.
Recently, many of these predictions have been established mathematically, with consequences in all the aforementioned fields, 
and this line of work is still very active \cite{guerra2002thermodynamic,talagrand2006parisi,panchenko2013parisi,auffinger2013random,subag2017complexity,barbier2019optimal}.

\myskip 
\textbf{Statistical physics and semidefinite programming --}
As a combination of linear constraints with a global matrix positivity constraint,
ellipsoid fitting belongs to the general class of semidefinite programs (SDPs). 
Such problems have been analyzed with tools of statistical physics in previous works \cite{montanari2016semidefinite,javanmard2016phase};
however, these studies rely on an application of the celebrated Grothendieck inequality of functional analysis \cite{khot2012grothendieck}, 
which essentially implies that one can restrict the rank of the matrix $\bS$ to be at most $r = \mcO_d(1)$ as $d \to \infty$ (the limit $r \to \infty$ is then taken \emph{after} $d \to \infty$), 
a setting particularly well-suited to a statistical physics approach.
In our current setting, such a Grothendieck inequality is not available; indeed, we will see that close to the SAT/UNSAT transition solutions typically have rank approximately $d/2$, 
so that we do not expect finite-rank solutions to exist.
Breaking with prior work, in our replica computation we will not restrict the rank of solutions: while this will create additional difficulties, it will allow us to characterize the solution space for all values of $\alpha$.

\subsection{Summary of main results}\label{subsec:summary_main_results}

We now summarize the main results of our derivations using the non-rigorous replica method. 
We present claims for our main findings, while details of the derivation as well as more extensive analyses (both analytical and numerical) are presented in Sections~\ref{sec:replica} and \ref{sec:dilute_hciz}.

\subsubsection{The SAT/UNSAT transition and solutions in the SAT phase}\label{subsubsec:main_results_sat_unsat}

Our first main result concerns the solution space of the ellipsoid fitting problem.
\begin{claim}[Solution space of ellipsoid fitting]\label{claim:solution_space}
  Let $n,d \to \infty$, with $n/d^2 \to \alpha > 0$. We define $\Gamma = \Gamma(\bx_1, \cdots, \bx_n)$ as the (convex) set of solutions to $(\rm P)$: 
  \begin{equation*}
      \Gamma \coloneqq \{\bS \in \mcS_d \, : \, \bS \succeq 0 \text{ and } \bx_\mu^\T \bS \bx_\mu = d \text{ for all } \mu \in [n]) \}.
  \end{equation*}
  \change{We consider $\bx_1, \cdots, \bx_n \iid \mcN(0, \Id_d)$, and} denote $\Phi_d \coloneqq (1/d^2) \EE \log |\Gamma|$, with $|\Gamma|$ the volume of $\Gamma$.
  Then:
  \begin{itemize}
    \item If $\alpha < 1/4$, then $\Phi_d \to \Phi(\alpha) > -\infty$ as $d \to \infty$, and $\Phi(\alpha)$ is given by eq.~\eqref{eq:phi_RS_general}, 
    taking $\beta = 1$ and $V(x) = +\infty \cdot \indi\{x < 0\}$.
    \item $\Phi(\alpha) \to -\infty$ as $\alpha \uparrow 1/4$.
  \end{itemize}
  For $\alpha < 1/4$, let $\bS \sim \Unif(\Gamma)$, and $\mu_\bS$ its empirical spectral distribution. Then: 
  \begin{itemize}
    \item $\mu_\bS$ converges weakly \change{(and in probability over $\{\bx_1, \cdots, \bx_n\}$ and the law of $\bS$)} 
    as $d \to \infty$ to $\mu = \mu[\alpha]$ given by the solution to eq.~\eqref{eq:se_general} (taking again $\beta = 1$ and $V(x) = +\infty \cdot \indi\{x < 0\}$).\footnote{That is, for any continuous test function of bounded support $f: \mathbb{R} \to \mathbb{R}$, $\int f d\mu_{\bS} \to \int f d\mu[\alpha]$ in probability.}
    \item As $\alpha \uparrow 1/4$, $\mu[\alpha]$ approaches $\mu_c$, given by:
    \begin{equation*}
    \mu_c(x) \coloneqq \frac{1}{2} \delta(x) + \frac{4\sqrt{9\pi^2 - 4 x^2}}{9 \pi^3} \indi\left\{0 \leq x \leq \frac{3\pi}{2}\right\}.
    \end{equation*}
  \end{itemize}
\end{claim}
\begin{figure}[!t]
     \centering
     \begin{subfigure}[t]{0.53\textwidth}
         \centering
         \includegraphics[width=\textwidth]{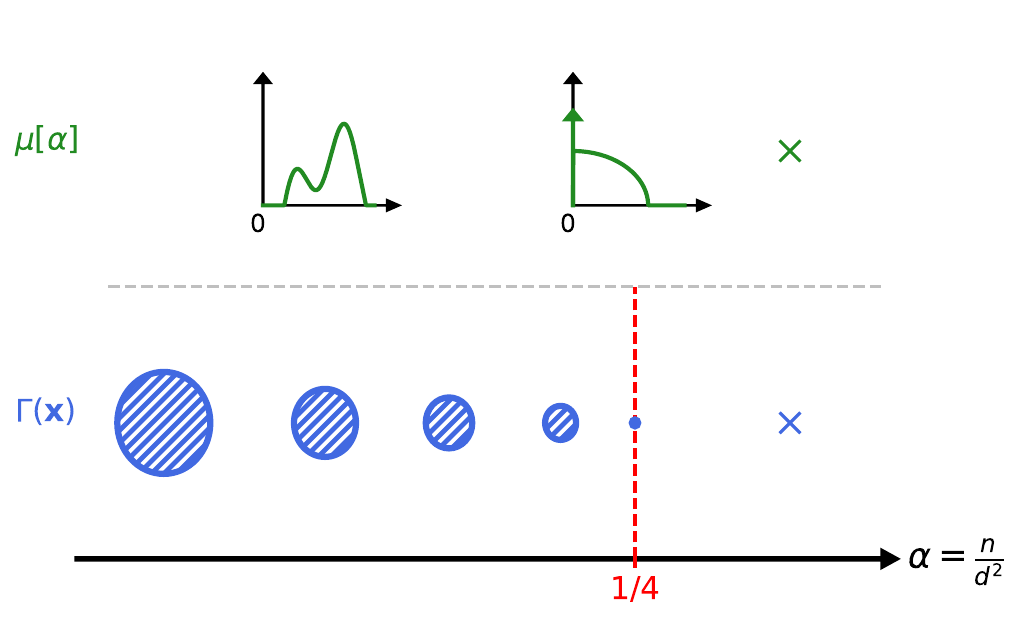}
         \caption{\label{subfig:sketch_sat_unsat}Sketch of the evolution of the solution set and typical spectral density of solutions as a function of $\alpha$. 
         For $\alpha \uparrow 1/4$ the solution space shrinks to a point, and the support of the spectral density touches $0$.
         }
     \end{subfigure}
     \hfill
     \begin{subfigure}[t]{0.45\textwidth}
         \centering
         \includegraphics[width=\textwidth]{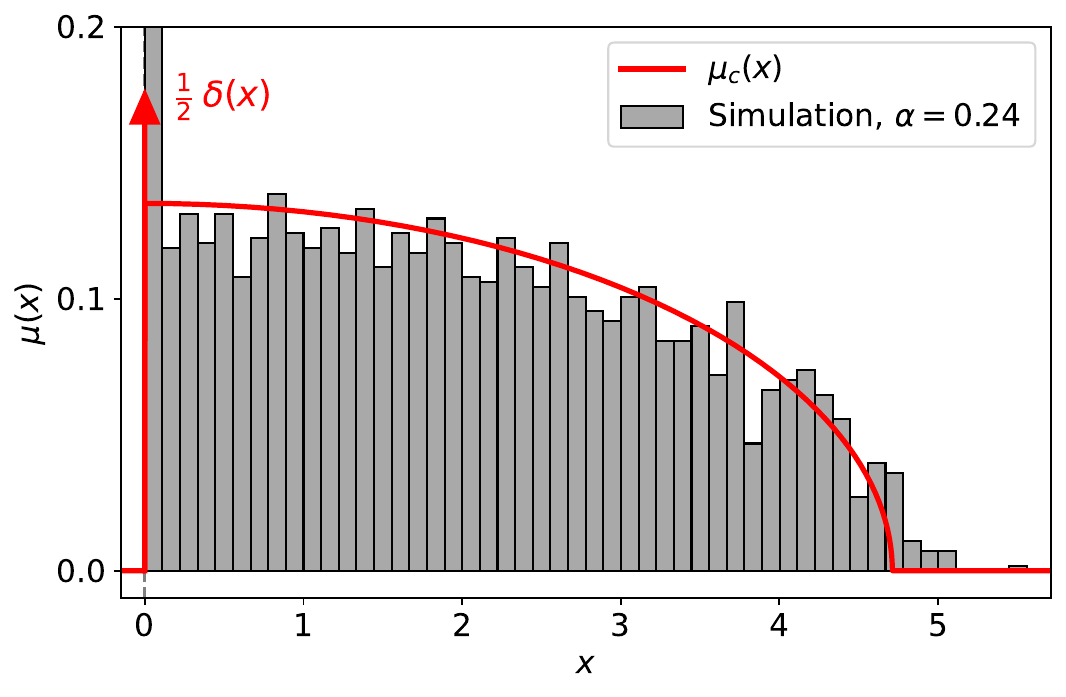}
         \caption{The typical spectral density of solutions for $\alpha \uparrow 1/4$. We compare $\mu_c(x)$ to finite-size simulations for $\alpha = 0.24$
         and $d = 100$, combining $50$ realizations. There is a fraction 
         $\simeq 0.52$ of near-zero eigenvalues ($|\lambda| < 10^{-3}$).}
     \end{subfigure}
  \caption{
  Evolution of the solution set and typical spectral density of solutions (left), and the typical spectral density near the transition (right).
  For any $\alpha < 1/4$, the spectral density and the volume of the solution space are characterized by the ``replica'' equations~\eqref{eq:phi_RS_general},\eqref{eq:se_general}. 
  \label{fig:main_results_solution_space}}
\end{figure}
\noindent
Claim~\ref{claim:solution_space} characterizes the size of the (convex) solution space of ellipsoid fitting in the SAT phase, as well as the spectral density of a typical solution.
In Fig.~\ref{fig:main_results_solution_space} we illustrate these predictions, and compare our prediction for $\mu_c(x)$ to finite-size simulations for $\alpha \simeq 0.25$, showing a very good agreement.
We note that the spectral density $\mu_c$ at the transition point is a surprisingly simple one, its absolutely continuous part having the shape of a quarter-circle of radius $3\pi / 2$.
This density also has an atom of weight 1/2 at zero; geometrically, as $\alpha \uparrow 1/4$, this means that fitting ellipsoids have $d/2$ ``degenerate'' principal axes, i.e., whose lengths diverge to infinity, so that the fitting ellipsoids are in fact ``elliptical cylinders.''

\myskip
\textbf{Remark I --} While we do not prove it here, the statistical physics analysis of $\Phi$ is based on the idea that $(1/d^2) \log |\Gamma|$ concentrates around $\Phi$ as $d \to \infty$ (``self-averages'' in the physics language), 
a claim which has been proved in several other models using standard concentration inequalities \cite{talagrand2010mean,boucheron2013concentration}.
The results of Claim~\ref{claim:solution_space} should thus be understood as characterizing the \emph{typical} size of the solution space.

\myskip
\textbf{Remark II --} We provide in eq.~\eqref{eq:se_general} a set of asymptotic equations (i.e., that do not depend on $n, d$) satisfied by the limiting spectral density $\mu[\alpha]$. These equations involve a complicated functional $I_\HCIZ$ related to the spherical HCIZ integrals of random matrix theory \cite{guionnet2022rare}. While this functional has a mathematically established expression (see Appendix~\ref{subsec_app:hciz}), it is very cumbersome, and we rather leverage a perturbative expansion of $I_\HCIZ$ to analyze the solution space as $\alpha \uparrow 1/4$; see Section~\ref{subsec:sat_unsat}.

\begin{figure}
    \centering
    \includegraphics[width=0.8\textwidth]{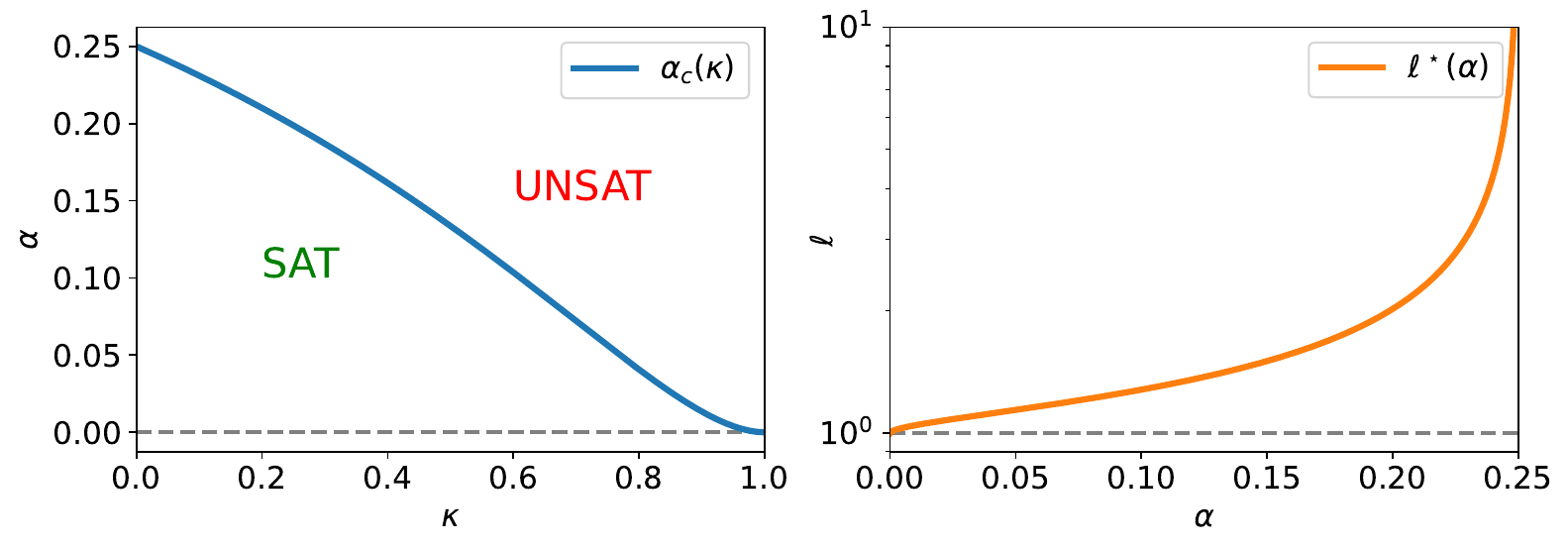}
    \caption{\label{fig:alphac_kappa} The evolution of $\alpha_c(\kappa)$ (left) and of \change{$\ell^\star(\alpha) = [\kappa^\star(\alpha)]^{-1/2}$}, the minimal value 
    of the longest principal axis of any ellipsoid fit (right).}
\end{figure}

\myskip
\textbf{Minimum length of the longest principal axis --}
Claim~\ref{claim:solution_space} can be generalized 
to the case in which the positivity constraint is replaced by $\bS \succeq \kappa \Id_d$, for some $\kappa \geq 0$.
We show there is a corresponding SAT/UNSAT threshold $\alpha_c(\kappa) \leq 1/4$ for this problem, characterized by another simple set of scalar equations (see eq.~\eqref{eq:alphac_lstar_kappa}).
Conversely, if $\alpha < 1/4$, any ellipsoid fit must satisfy $\lambda_{\min}(\bS) \leq \kappa^\star(\alpha) \coloneqq \alpha_c^{-1}(\alpha)$, \change{with $\alpha_c^{-1}$ the inverse function of $\alpha_c$, i.e.\ $\alpha_c[\kappa^\star(\alpha)] = \alpha$}.
Geometrically speaking, this means all ellipsoid fits have a principal axis of length at least $\ell^\star(\alpha) \coloneqq [\kappa^\star(\alpha)]^{-1/2}$ (and that this minimal value is reached).
We show in Fig.~\ref{fig:alphac_kappa} the value of both functions $\alpha_c(\kappa)$ and $\ell^\star(\alpha)$.
Note that $\alpha_c(\kappa) \to 0$ as $\kappa \to 1$: indeed, eq.~\eqref{eq:trace_close_1} is impossible to satisfy if 
$\bS \succeq \kappa \Id_d$ with $\kappa > 1$.

\subsubsection{The thresholds of different explicit approaches}\label{subsubsec:main_results_explicit_approaches}
We also characterize the performance of explicit constructions of the type
\begin{equation}\label{eq:P_V}
   (\rm{P}_V) \, : \,
    \min_{\bS \, : \, \{\bx_\mu^\T \bS \bx_\mu = d\}_{\mu=1}^n} \Tr[V(\bS)],
\end{equation}
for a convex ``potential'' function $V$. While our analysis holds for any $V(x)$, we have in mind two classes of functions: 
\begin{itemize}
  \item $V(x) = |x|^\gamma$ for some $\gamma \geq 1$. This amounts to considering the solution to the linear constraints with minimal Schatten-$\gamma$ norm.
  In particular, if $\gamma$ is close to $1$, this approach should favor solutions with lower rank (i.e., a sparser vector of eigenvalues, or a greater number of degenerate principal axes), which the above analysis suggests will help find solutions when $\alpha$ is close to $1/4$.
  \item $V(x) = |x-1|^\gamma$ for some $\gamma \geq 1$. The motivation to consider such $V(x)$ arises as $\bS = \Id_d$ satisfies the linear constraints in eq.~\eqref{eq:P} up to an error $\mcO(\sqrt{d})$ by standard concentration arguments.
  It thus seems sensible to consider $\Id_d$ as a ``starting point'' and to try to perturb it to find a fitting ellipsoid.
\end{itemize}

\myskip
Our main result for these explicit constructions is as follows.
We note that, even when $1/4 < \alpha < 1/2$ and $(\rm P)$ is typically unsatisfiable, the problem $(\rm{P}_V)$ is still feasible, and below we characterize the spectral density of the resulting $\bS$ (which is no longer positive semidefinite) even in this case.

\begin{claim}[Performance of explicit constructions]\label{claim:explicit_constructions}
  Let $\Gamma(V)$ be the set of minimizers of $(\rm P_V)$ in eq.~\eqref{eq:P_V}, 
  and $\hbS$ drawn uniformly from $\Gamma(V)$. Then, the empirical spectral distribution of $\hbS$
  converges weakly in probability to a deterministic distribution $\nu(x)$, which is given as the solution to the closed replica equations 
  derived in Section~\ref{subsec:large_beta}. 
  Specifically, for the potentials considered above, we have:
  \begin{itemize}
    \item[$(i)$] If $V(x) = |x|$ and $\alpha < 1/4$, $\nu = \mu[\alpha]$ given in Claim~\ref{claim:solution_space}. In particular, $\hbS$ is typically positive semidefinite.
    \item[$(ii)$] If $V(x) = |x|$ and $\alpha > 1/4$, $\nu(x)$ is given by eq.~\eqref{eq:nu_NN_unsat}.
    \item[$(iii)$] If $V(x) = |x|^\gamma$ for $\gamma \in (1, \infty)$, $\nu(x)$ is given by eq.~\eqref{eq:min_Sgamma}.
    \item[$(iv)$] If $V(x) = |x-1|^\gamma$ for $\gamma \in (1, \infty)$, $\nu(x)$ is given by eq.~\eqref{eq:min_Sgamma_2}.
  \end{itemize}
  Moreover, in cases $(ii),(iii),(iv)$, there is a unique minimizer $\hbS$ of $(\rm{P}_V)$ as $d \to \infty$, and thus the statement means that this particular matrix has asymptotic spectral density $\nu(x)$.
\end{claim}

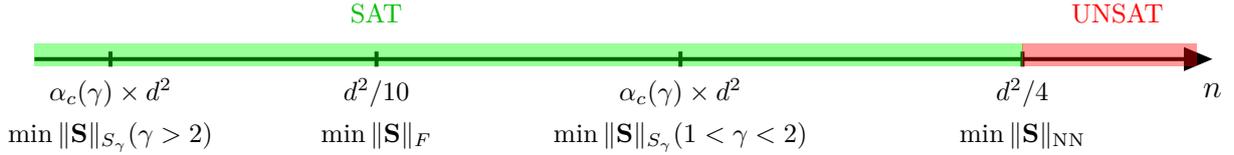
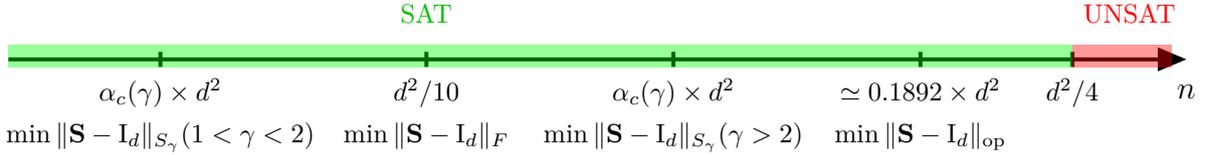
\begin{figure}[!t]
\begin{subfigure}[t]{\textwidth}
  \centering
\begin{tikzpicture}
  % Define the parameters

  % Draw the axis (large horizontal arrow)
  \draw[->,thick,line width = 1.5pt] (-4,0) -- (11.5,0) node[below=5pt,font=\large] {$n$};

  % Define the positions and labels for the ticks
  \pgfmathsetmacro\tickPositionA{0.5}
  \pgfmathsetmacro\tickPositionB{4.5}
  \pgfmathsetmacro\tickPositionC{9}
  \pgfmathsetmacro\tickPositionD{-3}
  \def\tickLabels{{{"$d^2/10$","$\min \|\bS\|_F$"},{"$\alpha_c(\gamma) \times d^2$","$\min \|\bS\|_{S_\gamma} (1 < \gamma < 2)$"}, {"$d^2/4$", "$\min \|\bS\|_\NN$"}, {"$\alpha_c(\gamma) \times d^2$","$\min \|\bS\|_{S_\gamma} (\gamma > 2)$"}}}

  % Draw the ticks and labels
  \foreach \pos [count=\i] in {\tickPositionA, \tickPositionB, \tickPositionC, \tickPositionD} {
    \draw[line width=1.5pt] (\pos,0.1) -- (\pos,-0.1) node[below, font = \small] {\pgfmathparse{\tickLabels[\i-1][0]}\pgfmathresult};
    \node at (\pos,-0.7) [below, anchor=north, font = \small] {\pgfmathparse{\tickLabels[\i-1][1]}\pgfmathresult};
  };

    % Add colored bands
  \fill[green, opacity=0.4] (-4,0.2) rectangle (\tickPositionC, -0.1);
  \fill[red, opacity=0.4] (\tickPositionC,0.2) rectangle (11.3, -0.1);

    % Text labels above colored regions
  \node[green!75!black, anchor=south, font = \small] at (\tickPositionA,0.35) {SAT};
  \node[red!100!black, anchor=south,font = \small] at ({\tickPositionC+(11.5-\tickPositionC)/2},0.35) {UNSAT};

  \end{tikzpicture}
  \caption{\label{subfig:threshold_min_S_gamma}Thresholds for different methods of the type $\hbS = \argmin \|\bS\|_{S_\gamma}$, for $\bS$ a solution to the linear constraints 
  in eq.~\eqref{eq:P}.}
\end{subfigure} 
\par\vspace{0.5cm} % force a bit of vertical whitespace
\begin{subfigure}[t]{\textwidth}
  \centering
\begin{tikzpicture}
  % Define the parameters

  % Draw the axis (large horizontal arrow)
  \draw[->,thick,line width = 1.5pt] (-4,0) -- (11.5,0) node[below=5pt,font=\large] {$n$};

  % Define the positions and labels for the ticks
  \pgfmathsetmacro\tickPositionA{1.5}
  \pgfmathsetmacro\tickPositionD{4.75}
  \pgfmathsetmacro\tickPositionE{-2}
  \pgfmathsetmacro\tickPositionB{8}
  \pgfmathsetmacro\tickPositionC{10}
  \def\tickLabels{{{"$d^2/10$","$\min \|\bS - \Id_d\|_F$"},{"$\simeq 0.1892 \times d^2$","$\min \|\bS - \Id_d\|_\op$"}, {"$d^2/4$", ""}, {"$\alpha_c(\gamma) \times d^2$", "$\min \|\bS - \Id_d\|_{S_\gamma} (\gamma > 2)$"}, {"$\alpha_c(\gamma) \times d^2$", "$\min \|\bS - \Id_d\|_{S_\gamma} (1 < \gamma < 2)$"}}}

  % Draw the ticks and labels
  \foreach \pos [count=\i] in {\tickPositionA, \tickPositionB, \tickPositionC, \tickPositionD, \tickPositionE} {
    \draw[line width=1.5pt] (\pos,0.1) -- (\pos,-0.1) node[below, font = \small] {\pgfmathparse{\tickLabels[\i-1][0]}\pgfmathresult};
    \node at (\pos,-0.7) [below, anchor=north, font = \small] {\pgfmathparse{\tickLabels[\i-1][1]}\pgfmathresult};
  };

    % Add colored bands
  \fill[green, opacity=0.4] (-4,0.2) rectangle (\tickPositionC, -0.1);
  \fill[red, opacity=0.4] (\tickPositionC,0.2) rectangle (11.3, -0.1);

    % Text labels above colored regions
  \node[green!75!black, anchor=south, font = \small] at (\tickPositionA,0.35) {SAT};
  \node[red!100!black, anchor=south,font = \small] at ({\tickPositionC+(11.5-\tickPositionC)/2},0.35) {UNSAT};

  \end{tikzpicture}
  \caption{\label{subfig:threshold_min_S_gamma_shifted}Thresholds for different methods of the type $\hbS = \argmin \|\bS-\Id_d\|_{S_\gamma}$, for $\bS$ a solution to the linear constraints 
  in eq.~\eqref{eq:P}. }
\end{subfigure} 
\par\vspace{0.5cm} % force a bit of vertical whitespace
\caption{ 
  \label{fig:threshold_explicit_constructions}
  The thresholds for different explicit constructions of a solution to the linear constraints.
  For any given method with threshold $\alpha_c$, the solution ceases to be positive semidefinite for $n> \alpha_c d^2$. 
  Notice that the minimal nuclear norm approach succeeds in the whole SAT phase.
  For each method, we analytically derive $\alpha_c$, and we furthermore predict the asymptotic spectral density of solutions, see Claim~\ref{claim:explicit_constructions}.
}
\end{figure}

\myskip
For any $V$ as described above, we characterize the threshold $\alpha_c(V)$ as the smallest $\alpha$ such that $\nu(x)$ has negative numbers in its support.
We summarize in Fig.~\ref{fig:threshold_explicit_constructions} how these thresholds evolve when $V(x) = |x|^\gamma$ (Fig.~\ref{subfig:threshold_min_S_gamma})
and $V(x) = |x-1|^\gamma$ (Fig.~\ref{subfig:threshold_min_S_gamma_shifted}).
Their detailed characterization (and numerical evaluations) are performed in Sections~\ref{subsec:nn} and \ref{subsec:close_identity}.

\myskip 
\textbf{Minimal nuclear norm --} A somewhat surprising prediction of our results is that the minimal nuclear norm solution (i.e., eq.~\eqref{eq:P_V} with $V(x) = |x|$) 
is a positive semidefinite matrix with high probability in the whole SAT phase $\alpha < 1/4$.
In other words, minimizing the nuclear norm appears to solve the ellipsoid fitting problem in the entire regime where a solution typically exists at all.
Although this minimizer is likely hard to characterize mathematically, it could provide an interesting approach to rigorously showing the existence of solutions 
in the conjectured SAT phase.

\myskip 
\textbf{The least-squares solution --}
Using numerical simulations, \cite{potechin2023near} predicts that the ``least-squares'' approach -- i.e.\ $\hbS_\LS = \argmin \|\bS\|_F$ -- ceases to be positive semidefinite 
for $\alpha > \alpha_c \simeq 1/17$; see Figs.~1 and 9 in \cite{potechin2023near}.
As the data presented is rather noisy, we believe that this prediction of $\alpha_c$ is influenced by finite-size effects, and that the actual threshold for $d \to \infty$ is $\alpha_c = 1/10$, as illustrated in Fig.~\ref{fig:threshold_explicit_constructions}. 
We derive this prediction analytically in Section~\ref{subsec:nn}, and compare it to numerical simulations in Fig.~\ref{fig_app:ls_comparison_numerics} in the appendix, strengthening our findings.
Interestingly, $1/10$ is also the threshold numerically observed for another ``identity perturbation'' explicit construction \cite{potechin2023near}.

\subsubsection{Beyond Gaussian vectors: general norm fluctuations}\label{subsubsec:main_results_rot_inv}

Finally, we consider the ellipsoid fitting problem when $(\bx_\mu)_{\mu=1}^n$ are taken i.i.d.\ from a general rotationally-invariant distribution with a fluctuating norm. 
We present our results for the characterization of the SAT/UNSAT transition; the characterization of the performance of explicit algorithms (summarized in Claim~\ref{claim:explicit_constructions})
can be generalized to this setting as well.
We take the following i.i.d.\ model for $(\bx_\mu)_{\mu=1}^n$: 
\change{
\begin{equation}\label{eq:model_rot_inv}
    \bx_\mu = \sqrt{\chi_\mu} \bomega_\mu,
\end{equation}
}
in which $\chi_\mu, \bomega_\mu$ are independent, and $\bomega_\mu \sim \Unif[\mcS^{d-1}(\sqrt{d})]$, in particular $\|\bomega_\mu\|^2 = d$.
Without loss of generality (as the ellipsoid fitting property is invariant under a global rescaling of all the vectors) 
we take $\EE[\chi_\mu] = 1$, so that $\EE \| \bx_\mu\|^2 = d$, and the vectors are isotropic: $\EE[\bx_\mu \bx_\mu^\T] = \Id_d$.
We consider a regime in which the typical fluctuations of $\|\bx_\mu\|^2$ around $d$ are in the scale $\Theta(\sqrt{d})$, and we define 
\begin{equation}\label{eq:def_tau}
    \tau \coloneqq \lim_{d \to \infty} \left\{d \, \EE \left[\left(\frac{\|\bx\|^2}{d} - 1\right)^2\right]\right\} = \lim_{d \to \infty} \{d \, \EE [(\chi - 1)^2]\}.
\end{equation}
Notice that for $\bx_\mu \sim \mcN(0, \Id_d)$ we have $\tau_{\mathrm{Gaussian}} = 2$. 
We can now state our main result on the satisfiability of ellipsoid fitting in this model:

\begin{claim}[SAT/UNSAT transition for general rotationally-invariant distributions]\label{claim:rot_inv} For the model of eq.~\eqref{eq:model_rot_inv}, the SAT/UNSAT transition $\alpha_c$ of ellipsoid fitting depends solely on $\tau$.
  \change{
Moreover, $\alpha_c(\tau)$ is given by eq.~\eqref{eq:alphac_tau_2}, where $X(\tau)$ is the solution to eq.~\eqref{eq:alphac_tau_1}, given by:
\begin{subnumcases}{\label{eq:alphac_tau}}
\label{eq:alphac_tau_2}
  \alpha_c(\tau) = -\frac{1}{2}\left(1 - \frac{\tau}{2}\right) \left(\int_{-X(\tau)}^2 \rho_\sci(v) [X(\tau)+v] \rd v\right)^2 + \frac{1}{2} \int_{-X(\tau)}^2 \rho_\sci(v) [X(\tau) + v]^2 \rd v, & \\
\label{eq:alphac_tau_1}
  X(\tau) = \left(1 - \frac{\tau}{2}\right)  \int_{-X(\tau)}^2 \rho_\sci(v) [X(\tau)+v] \rd v. &
\end{subnumcases}
  }
\end{claim}

\begin{figure}
     \centering
     \begin{subfigure}[t]{0.49\textwidth}
    \centering
    \includegraphics[width=\textwidth]{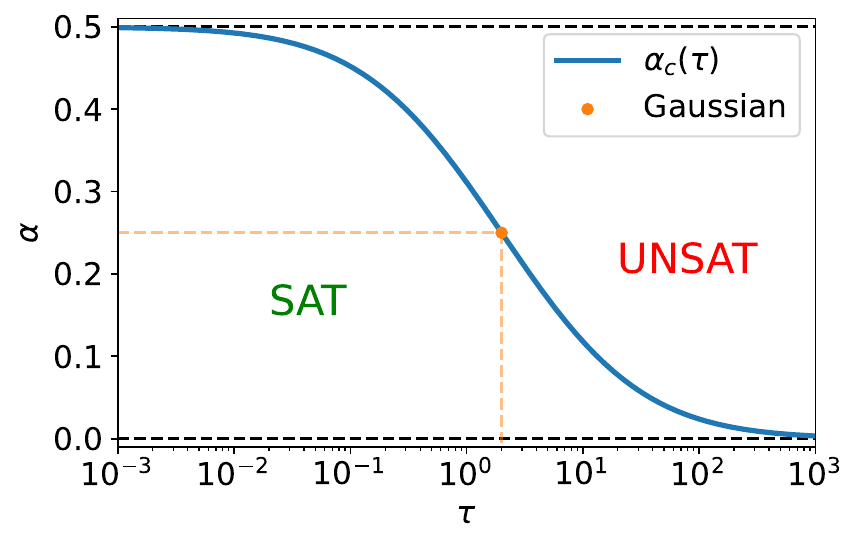}
    \caption{The evolution of $\alpha_c(\tau)$, as obtained by solving eq.~\eqref{eq:alphac_tau}.}
    \label{subfig:alphac_tau}
     \end{subfigure}
     \hfill
     \begin{subfigure}[t]{0.49\textwidth}
    \includegraphics[width=\textwidth]{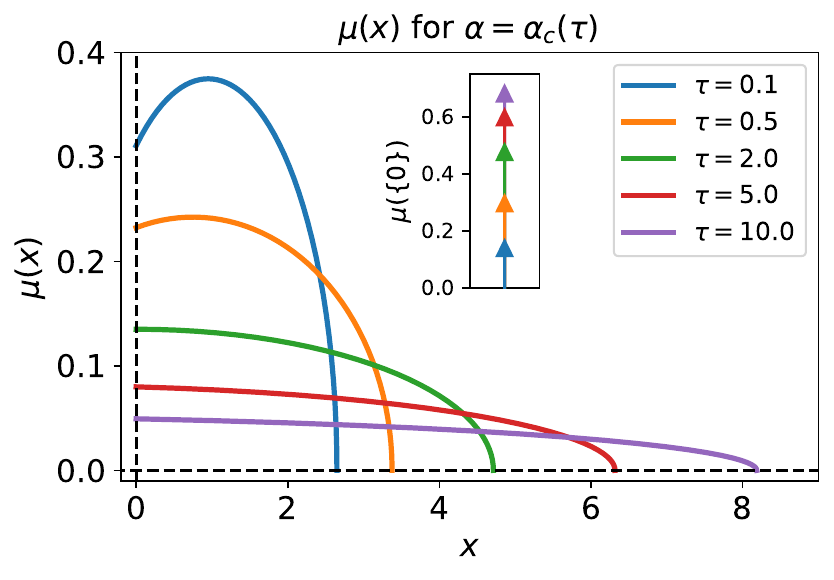}
    \caption{The spectral density of solutions at the SAT/UNSAT threshold $\alpha = \alpha_c(\tau)$. 
    In inset, we show the weight of the $0$ eigenvalue.}
    \label{subfig:mu_alphac_tau}
     \end{subfigure}
    \caption{The SAT/UNSAT transition for rotationally-invariant vectors with fluctuating norm.}
    \label{fig:transition_fluctuating_norm}
\end{figure}

\myskip
\textbf{Discussion --}
In Fig.~\ref{subfig:alphac_tau} we show the function $\alpha_c(\tau)$ obtained by a numerical solver of eq.~\eqref{eq:alphac_tau}.
In Fig.~\ref{subfig:mu_alphac_tau} we also plot the spectral density of typical solutions 
in the SAT phase near the SAT/UNSAT threshold. For $\tau = 2$ we recover that $\mu(x)$ is given by $\mu_c$ in Claim~\ref{claim:solution_space}.
Besides recovering \change{$\alpha_c(2) = 1/4$}, we also observe that $\alpha_c(\tau \to \infty) = 0$ and $\alpha_c(\tau \to 0) = 1/2$\footnote{
  \change{
  Indeed one can see from eq.~\eqref{eq:alphac_tau_1} that $X(0) = 2$ and $\lim_{\tau \to \infty} X(\tau) = -2$.
  }
}. 
The first limit shows that, as one expects, when the norm fluctuations are larger ($\tau \to \infty$) then ellipsoid fitting is harder, and it becomes impossible to fit an ellipsoid to $n = \Theta(d^2)$ vectors when the norm fluctuations are strictly larger than $\sqrt{d}$.
On the other hand, when $\tau \to 0$, it becomes easier and easier to find an ellipsoid fit. Notice that for any $\tau > 0$ finding a solution to the linear constraints 
is impossible for $\alpha > 1/2$ by linear independence of these constraints, which explains why $\alpha_c(\tau) \to 1/2$ as $\tau \to 0$.
However, if the norm of $\bx_\mu$ has exactly zero fluctuations (\change{e.g.\ if $\bx_\mu \sim \Unif(\{\pm 1\}^d)$, or if $\bx_\mu \sim \Unif(\mcS^{d-1}(\sqrt{d}))$}), then the sphere \change{with $\bS = \Id_d$} is always an ellipsoid fit for any value of $n$, 
\change{and the ellipsoid fitting problem is trivial for such vectors}.
It is therefore likely
that there is a transition in $\alpha_c$ from $1/2$ to $\infty$ if one consider a regime in which $\tau \to 0$ as $d \to \infty$ (i.e., if the fluctuations of the norm are $\smallO(\sqrt{d})$). 
We leave the investigation of this regime for future work. 

\myskip 
\change{
  While the fluctuations of the norm of $\bx_\mu$ are crucial to characterize the satisfiability of ellipsoid fitting as we 
  discussed, extending our results to non-rotationally invariant vectors (i.e.\ $\|\bomega_\mu\|^2 = d$ in eq.~\eqref{eq:model_rot_inv}, 
  but not uniformly sampled) is a challenging open problem. While some delocalization of the distribution of $\bomega_\mu$ is certainly necessary (as can be seen by the extreme case in which
  $\bomega_\mu$ is a fixed vector, in which case no ellipsoid fit exists), we leave to future work a precise characterization of more general distributions for which Claim~\ref{claim:rot_inv} is expected to hold.
}

\subsection{Organization of the paper}\label{subsec:organization_paper}

In Section~\ref{sec:replica} we carry out a general calculation, based on the non-rigorous replica method of statistical physics,
to compute the asymptotic entropy (i.e., the logarithm of the volume) of solutions to eq.~\eqref{eq:P}, as well as the entropy of minimizers of the 
convex program $(\rm{P}_V)$ of eq.~\eqref{eq:P_V}. 
Leveraging a dilute limit of objects known in random matrix theory as spherical HCIZ integrals, we 
derive in Section~\ref{sec:dilute_hciz} the SAT/UNSAT transition of $(\rm P)$ as well as the typical spectral density of solutions (see Section~\ref{subsec:sat_unsat}), as summarized in Claim~\ref{claim:solution_space}. 
We discuss in Section~\ref{subsec:minimal_length_max_axis} the generalization of the problem to the constraint $\bS \succeq \kappa \Id_d$, 
deriving the minimal length of the longest principal axis of any ellipsoid fit, as summarized in Fig.~\ref{fig:alphac_kappa}.
We then characterize the performance of $(\rm P_V)$ for different choices of the potential function $V$; see Sections~\ref{subsec:nn}-\ref{subsec:close_identity} and Claim~\ref{claim:explicit_constructions}.
Finally, we generalize our calculation in Section~\ref{subsec:rot_inv_fluctuating_norm} to non-Gaussian vectors satisfying rotational invariance with an arbitrarily fluctuating norm, obtaining in particular Claim~\ref{claim:rot_inv}.
We conclude in Section~\ref{sec:conclusion} by discussing some mathematically rigorous approaches that may stem from our analysis.

\myskip 
\textbf{Reproducibility -- }
The finite-size numerical experiments used to produce the different figures in this paper were performed using the general-purpose convex optimization solver \texttt{Mosek} through the \texttt{cvxpy} interface in the \texttt{Python} language.
The code is available in a public \href{https://github.com/AnMaillard/fitting_ellipsoid_replicas}{GitHub repository} \cite{github_repo}.

\section{Replica-symmetric analysis of the entropy of solutions}\label{sec:replica}
\noindent
\subsection{The free entropy}\label{subsec:free_entropy}
Let us first rewrite the problem slightly.
We notice that the equation $\bx_\mu^\T \bS \bx_\mu = d$ can be rewritten as 
$\Tr[\bW_\mu \bS] = (d - \Tr[\bS])/\sqrt{d}$, with $\bW_\mu \coloneqq (\bx_\mu \bx_\mu^\T - \Id_d)/\sqrt{d}$. 
One can check easily that for $\bx_\mu \iid \mcN(0, \Id_d)$, $\bW = \bW_\mu$ has for its first two moments
$\EE[\bW] = 0$ and $\EE[W_{ij} W_{kl}] = \delta_{ik}\delta_{jl}(1+\delta_{ij})/d$, 
for $i \leq j$ and $k \leq l$, which we note coincide with the first two moments of the $\GOE(d)$ distribution.

\myskip
Therefore, given $n$ random matrices $\bW = (\bW_1, \cdots, \bW_n)$ drawn i.i.d., 
we aim to characterize the set of matrices $\bS$ that satisfy all equations $\Tr[\bW_\mu \bS] = (d - \Tr[\bS])/\sqrt{d}$.
For any choice of a convex potential $V(x)$ and any $\beta \geq 0$ we define the \emph{partition function}, and the associated Gibbs measure:
\begin{align}\label{eq:partition_function}
    \mcZ(\beta, \bW) &\coloneqq \int_{\mcS_d} \rd \bS \, e^{- \beta d \Tr[V(\bS)]} \, \prod_{\mu=1}^n \delta\left(\Tr[\bW_\mu \bS] - \frac{d - \Tr \, \bS}{\sqrt{d}}\right), \\
    \label{eq:def_Gibbs}
    \bbP_{\beta, \bW}(\bS) &\coloneqq \frac{1}{\mcZ(\beta, \bW)} e^{- \beta d \Tr[V(\bS)]} \, \prod_{\mu=1}^n \delta\left(\Tr[\bW_\mu \bS] - \frac{d - \Tr \, \bS}{\sqrt{d}}\right),
\end{align}
for $\delta(\cdot)$ the Dirac delta function \change{ and $\rd \bS \coloneqq \prod_{i \leq j} \rd S_{ij}$ the Lebesgue measure on $\mcS_d$.}
\change{As we will see, for an appropriate choice of $V$ and $\beta$, we will interpret $\mcZ(\beta, \bW)$ as the \emph{volume} of the set $\Gamma$ of ellipsoid fits defined in Claim~\ref{claim:solution_space}. 
Characterizing the asymptotics of this volume will be a direct way to probe whether the ellipsoid fitting problem has any solutions. }

\myskip
Following the statistical physics terminology, we call
\begin{equation}\label{eq:def_free_entropy}
  \Phi(\alpha,\beta) \coloneqq \lim_{d \to \infty} \frac{1}{d^2} \EE \log \mcZ(\beta,\bW)
\end{equation}
the \emph{free entropy} of the system.
Our main goal will be the computation of $\Phi(\alpha, \beta)$ in eq.~\eqref{eq:def_free_entropy}. 
We will carry out our analysis for a generic $V(x)$ and $\beta$, having in mind three distinct setups: 
    \begin{enumerate}[label=\textbf{(\roman*)},ref=(\roman*)]
    \item\label{sdp_pot} $V(x) = +\infty \cdot \indi\{x < 0\}$, and $\beta = 1$.
    Then the term $\exp\{-\beta d \Tr[V(\bS)]\}$ will restrict the support of the Gibbs measure in eq.~\eqref{eq:def_Gibbs} to positive definite matrices.
    In this case the free entropy $\Phi(\alpha)$ is exactly the entropy of the set of solutions $\Gamma$ in Claim~\ref{claim:solution_space}, 
    while the Gibbs measure of eq.~\eqref{eq:def_Gibbs} is the uniform measure on $\Gamma$. 
    In particular, we will characterize the SAT/UNSAT point $\alpha_c$ as the value of $\alpha$ in which the convex solution space shrinks and disappears, 
    i.e.\ $\Phi(\alpha) \to - \infty$ as $\alpha \uparrow \alpha_c$.
    \item\label{nn_pot} $V(x) = |x|^\gamma$ for some $\gamma \geq 1$, and $\beta \to \infty$.
    In this limit, the Gibbs measure of eq.~\eqref{eq:def_Gibbs} will approach the uniform measure on the minimizers of the convex program $\min_{\bS \in \mcC(\bx)} \| \bS\|_{S_\gamma}$, 
    with $\mcC(\bx) \coloneqq \{\bS \, : \, \bx_\mu^\T \bS \bx_\mu = d, \, \forall \mu \in [n]\}$.
    The case $\gamma = 1$ corresponds to minimal nuclear norm solutions, while
    $\gamma = 2$ is the least-squares estimator of \cite{saunderson2013diagonal,potechin2023near}.
    \item\label{ls_pot} $V(x) = |x-1|^\gamma$ for some $\gamma \geq 1$, and $\beta \to \infty$. 
    In this limit, the Gibbs measure of eq.~\eqref{eq:def_Gibbs} will concentrate on the uniform measure on the minimizers of
    $\min_{\bS \in \mcC(\bx)} \| \bS - \Id_d\|_{S_\gamma}$.
\end{enumerate}
\change{Thus, while the choice \ref{sdp_pot} will lead us to characterize the asymptotic volume of the set of ellipsoid fits, 
the generality of our analysis of $\mcZ(\beta, \bW)$ will enable us to tackle settings~\ref{nn_pot} and \ref{ls_pot} under the same formalism. 
This will allow deducing properties 
of the minimizers of convex objectives that can be used as potential ellipsoid fits, see Section~\ref{subsubsec:main_results_explicit_approaches}.}

\myskip 
\textbf{Additional notations --}
We use the probabilistic notation $\EE$ for the average under $\bW = (\bW_1, \cdots, \bW_n)$, 
while we use brackets $\langle \cdot \rangle$ for the average under the (random) Gibbs measure of eq.~\eqref{eq:def_Gibbs}.

\myskip
\textbf{Rewriting of the free entropy --}
Let us first perform a useful rewriting of the partition function.
As the trace of $\bS$ plays a special role in what follows,
we first make use of the following ``polar'' change of variables, for any function $f$: 
\begin{equation}\label{eq:polar_matrices}
    \int_{\mcS_d} \rd \bS f(\bS) = \int_{\mcS_d} \rd \bR \, \delta\left(\frac{1}{d} \Tr[\bR] - 1\right) \int_\bbR \rd m \, |m|^{\frac{d(d+1)}{2} - 1} \, f(m\bR).
\end{equation}
Using eq.~\eqref{eq:polar_matrices} in eq.~\eqref{eq:partition_function} we get:
\begin{equation*}
    \mcZ(\beta, \bW) =\!\! \int  \! \rd \bR \, \delta\left(\frac{1}{d} \Tr[\bR] - 1\right) \int \rd m |m|^{\frac{d(d+1)}{2}-n-1} \, e^{-\beta d \Tr[V(m\bR)]} \prod_{\mu=1}^n \delta\left(\Tr[\bW_\mu \bR] - \frac{(1-m)\sqrt{d}}{m}\right).
\end{equation*}
We change variables to $m = \Tr[\bS]/d = 1 + u /\sqrt{d}$:
\begin{align}
    \label{eq:Z_W_1}
    \nonumber
    \mcZ(\beta, \bW) &= \int \rd \bR \, \delta\left(\frac{1}{d} \Tr[\bR] - 1\right) \, \int \frac{\rd u}{\sqrt{d}} \, e^{\left(\frac{d(d+1)}{2}-n-1\right) \log \left|1 + \frac{u}{\sqrt{d}}\right| -\beta d \Tr[V((1+u/\sqrt{d})\bR)]} \\ 
    & \times \change{\prod_{\mu=1}^n} \delta\left(\Tr[\bW_\mu \bR] + \frac{u}{1 + \frac{u}{\sqrt{d}}}\right).
\end{align}
By eq.~\eqref{eq:trace_close_1}, we know that if $\bS$ is a solution to the linear constraints with $\bS \succeq 0$, or $\Tr |\bS| \lesssim d$ as $d \to \infty$, 
then $u$ remains of order $\mcO(1)$ as $n, d \to \infty$. We assume that this holds in all the cases we consider: in~\ref{sdp_pot} we have $\bS \succeq 0$, while 
in~\ref{nn_pot} and \ref{ls_pot}, the potential $V$ prevents the spectrum of $\bS$ from growing to infinity.
Similarly, we assume that $\Tr[\bR^2]/d = \mcO(1)$ as $d \to \infty$ (and more generally we will see that the spectrum of $\bR$ remains bounded), \change{One sees then} 
that many terms involving $u$ in eq.~\eqref{eq:Z_W_1} are of order $\exp\{\smallO(d^2)\}$, and can thus be dropped when looking at the free entropy:
\begin{equation}\label{eq:simplification_logZ}
    \Phi(\alpha, \beta) = \lim_{d \to \infty} \frac{1}{d^2} \, \EE \log \int \rd \bR \, \int \rd u \,\delta\left(\frac{1}{d} \Tr[\bR] - 1\right) \, e^{-\beta d \Tr[V(\bR)]} \prod_{\mu=1}^n\delta\left(\Tr[\bW_\mu \bR] + u\right).
\end{equation}
Similarly, one can deduce of eq.~\eqref{eq:simplification_logZ} the corresponding simplifications of the partition function and Gibbs measure of eqs.~\eqref{eq:partition_function} and \eqref{eq:def_Gibbs}, at leading order.

\subsection{Replicated free entropy}\label{subsec:replicated_fentropy}

\noindent
\textbf{A word on the replica method --}
The replica method is based on the \emph{replica trick}, a heuristic use of the following formula, for a random variable $X > 0$:
\begin{equation}\label{eq:replica_trick}
    \EE \log X = \lim_{r \to 0} \frac{\EE X^r - 1}{r} = \frac{\partial}{\partial r} [\log \EE X^r]_{r = 0}.
\end{equation}
The replica method is based on several heuristics, and leverages the fact that it is often possible to compute the moments 
$\EE[\mcZ(\beta, \bW)^r]$ for \emph{integer $r$}. It proceeds as follows: 
\begin{itemize}[leftmargin=25pt]
    \item[$(i)$] Assume that by eq.~\eqref{eq:replica_trick} we have $\Phi(\alpha,\beta) = \partial_r [\Phi(\alpha,\beta;r)]_{r = 0}$, with  
    \begin{equation}\label{eq:def_Phir}
        \Phi(\alpha,\beta;r) \coloneqq \lim_{d \to \infty} \frac{1}{d^2} \log \EE [\mcZ(\beta, \bW)^r].
    \end{equation}
    \item[$(ii)$] Compute $\Phi(\alpha,\beta;r)$ for \change{\emph{integer $r \in \bbN$}}.
    \item[$(iii)$] Use these values to analytically expand $\{\Phi(\alpha,\beta;r)\}_{r \in \bbN}$ to all $r \geq 0$.
    \item[$(iv)$] Compute $\Phi(\alpha,\beta) = \partial_r [\Phi(\alpha,\beta;r)]_{r = 0}$ from the analytic continuation above.
\end{itemize}
The replica trick
is obviously non-rigorous given the inversion of limits $r \downarrow 0$ and $d \to \infty$,
as well as the analytic continuation to arbitrary $r > 0$ of the $r$-th moment. 
However, it is widely believed to be exact and has achieved tremendous success, e.g.\ in the study of spin glasses and in high-dimensional statistics, 
and its predictions have been rigorously established in a wide range of settings \cite{mezard1987spin,charbonneau2023spin}.

\myskip
We now compute the replicated free entropy $\Phi(\alpha,\beta;r)$ of eq.~\eqref{eq:def_Phir}, for integer $r \geq 0$. 
We start from the simplified expression of eq.~\eqref{eq:simplification_logZ} 
(we drop the $d \to \infty$ limit, which is implicit in the subsequent equations): 
\begin{equation*}
   e^{d^2 \Phi(\alpha, \beta ; r)} = \int \prod_{a=1}^r \rd \bR^a \, \rd u^a \,\delta\left(\frac{1}{d} \Tr[\bR^a] - 1\right) \, e^{-\beta d \sum_{a=1}^r\Tr[V(\bR^a)]} \, \EE \,\prod_{\mu=1}^n \prod_{a=1}^r \delta\left(\Tr[\bW_\mu \bR^a] + u^a\right).
\end{equation*}
The average over $\{\bW_\mu\}_{\mu=1}^n$ reduces to: 
\begin{equation}\label{eq:average_constraint}
    \EE \,\prod_{\mu=1}^n \prod_{a=1}^r \delta\left(\Tr[\bW_\mu \bR^a] + u^a\right) =
    \left[\EE_{\bW} \prod_{a=1}^r \delta\left(\Tr[\bW \bR^a] + u^a\right)\right]^n.
\end{equation}
We introduce the so-called \emph{overlap} matrix:
\begin{equation}\label{eq:def_overlap}
    Q_{ab} \coloneqq \frac{1}{d} \Tr[\bR^a \bR^b].
\end{equation}
Let us now argue that, if $\bQ \in \mcS_r$ is fixed, then as $d \to \infty$ the variables $z^a \coloneqq \Tr[\bW \bR^a]$ are approximately jointly Gaussian, 
with zero mean and covariance $\EE[z^a z^b] = 2 Q_{ab}$, when $\bW \deq (\bx \bx^\T - \Id_d)/\sqrt{d}$ and $\bx \sim \mcN(0, \Id_d)$.
Indeed, by rotation invariance of the standard Gaussian distribution:
\begin{equation*}
    z^a = \Tr[\bW \bR^a] \deq \frac{1}{\sqrt{d}} \sum_{i=1}^d \lambda_i(\bR^a) (x_i^2-1).
\end{equation*}
Applying the central limit theorem (recall that \change{$r \in \bbN$ is fixed}), we see that as $d \to \infty$, we have $(z^a)_{a=1}^r \sim \mcN(0, 2\bQ)$.
The term in eq.~\eqref{eq:average_constraint} thus becomes at leading order as $d \to \infty$:
\begin{equation}\label{eq:average_constraint_after_clt}
    \EE_{\bW} \prod_{a=1}^r \delta\left(\Tr[\bW \bR^a] + u^a\right)
    \simeq \frac{1}{(4\pi)^{r/2} \sqrt{\det \bQ}} \exp\left\{-\frac{1}{4} \bu^\T \bQ^{-1} \bu\right\}.
\end{equation}
Integrating over the different possible values of $\bQ$, we get
that, up to $\smallO_d(1)$ terms:
\begin{align}\label{eq:phi_beta_r_after_average}
    \Phi(\alpha,\beta;r) &= \frac{1}{d^2} \log \int \prod_{a \leq b} \rd Q_{ab} \prod_{a=1}^r \rd \bR^a \, \rd u^a  \, \delta\left(\frac{1}{d} \Tr[\bR^a] - 1\right) \, 
    \, e^{-\beta d \sum_{a=1}^r \Tr[V(\bR^a)]} \\ 
    \nonumber
    & \times \prod_{a \leq b} \delta(d \Tr[\bR^a \bR^b] - d^2 Q_{ab}) \times
    \exp\left\{- \frac{nr}{2} \log 4 \pi - \frac{n}{2} \log \det \bQ - \frac{n}{4} \bu^\T \bQ^{-1} \bu\right\}.
\end{align}
The integral over $\{u^a\}_{a=1}^r$ can now be carried out explicitly, and gives a negligible term at leading order in the replicated free entropy\footnote{
More precisely, if one was to include the terms that were neglected in eq.~\eqref{eq:simplification_logZ}, 
 one sees that the integral over $\{u^a\}_{a=1}$ is of the type $\int_{\bbR^r} \rd \bu \exp\{-\frac{n}{2} \bu^\T \bQ^{-1} \bu + \smallO(d^2) F(\bu)\}$.
By Laplace's method one can show this integral is of the type $\exp\{\smallO(d^2)\}$, and thus negligible at leading order in the free entropy.
}. One obtains:
\begin{align*}
    \Phi(\alpha,\beta;r) &= \frac{1}{d^2} \log \int \prod_{a \leq b} \rd Q_{ab} \prod_{a=1}^r \rd \bR^a  \, \delta\left(\frac{1}{d} \Tr[\bR^a] - 1\right) 
    \, e^{-\beta d \sum_{a=1}^r \Tr[V(\bR^a)]} \\ 
    & \times \prod_{a \leq b} \delta(d \Tr[\bR^a \bR^b] - d^2 Q_{ab}) \times
    \exp\left\{\frac{n r}{2} \log 4\pi - \frac{n}{2} \log \det \bQ\right\}.
\end{align*}
We can now perform Laplace's method on the so-called \emph{order parameters} $\{Q_{ab}\}$, as classical in the replica method \cite{mezard1987spin}. 
We also introduce Lagrange multipliers $\hQ_{ab}$ to enforce the constraints of eq.~\eqref{eq:def_overlap}, 
and obtain (recall $n = \alpha d^2$), as $d \to \infty$: 
\begin{equation}\label{eq:Phi_r_after_saddle}
    \Phi(\alpha,\beta;r) = \extr_{\bQ \in \mcS_r^+, \hbQ \in \mcS_r} \Big[\frac{\alpha r}{2} \log 4\pi - \frac{\alpha}{2} \log \det \bQ + \frac{1}{4} \Tr[\bQ \hbQ] + J(\hbQ) \Big],
\end{equation}
where we wrote the variational problem as an extremum condition over $(\bQ, \hbQ)$,
with 
\begin{equation*}
    J(\hbQ) \coloneqq \lim_{d \to \infty}\frac{1}{d^2} \log \int \prod_{a=1}^r \rd \bR^a \, \delta\left(\frac{1}{d}\Tr[\bR^a] - 1\right) \, e^{-\beta d \sum_a\Tr[V(\bR^a)]  - \frac{d}{4} \sum_{a,b} \hQ_{ab} \Tr[\bR^a \bR^b]}.
\end{equation*}

\myskip
\textbf{The replica-symmetric ansatz --} 
We can now introduce the \emph{replica-symmetric} ansatz in the variational 
problem of eq.~\eqref{eq:Phi_r_after_saddle}. This amounts to assume that the variables attaining the extremum satisfy
$Q_{aa} = Q, Q_{ab} = q, \hQ_{aa} = \hQ, \hQ_{ab} = -\hq$ for all $a \neq b$.
Replica symmetry is a classical assumption in replica theory: while it fails to hold in many disordered systems, 
here one can justify its use.
Indeed, since the potential $V$ is assumed to be convex, the Gibbs measure of eq.~\eqref{eq:partition_function} is log-concave.
In particular, replica symmetry is satisfied,  since the solution space is composed of a single cluster of solutions \cite{mezard1987spin}.
We will discuss this point further in the conclusion, as this remark might allow for an easier rigorous treatment of the replica predictions in our model.

\myskip
Assuming the replica-symmetric ansatz yields (again, the limit $d \to \infty$ is implicit):
\begin{equation*}
    J(\hbQ) = \frac{1}{d^2} \log \int \prod_{a=1}^r \rd \bR^a \, \delta\left(\frac{1}{d}\Tr[\bR^a] - 1\right) \, e^{ -\beta d \sum_a \Tr[V(\bR^a)]- \frac{d (\hQ + \hq)}{4} \sum_{a} \Tr[(\bR^a)^2] + \frac{d \hq}{4} \Tr[(\sum_{a} \bR^a)^2]}.
\end{equation*}
We now make use of the identity:
\begin{equation*}
    \EE_{\bY \sim \mathrm{GOE}(d)} \left[\exp\left\{\frac{d}{2} \Tr[\bM \bY]\right\}\right] = \exp\left\{\frac{d}{4} \Tr[\bM^2]\right\}.
\end{equation*}
Plugging it in the definition of $J(\hbQ)$, we reach the following expression, which is analytic in $r$:
\begin{equation}\label{eq:J_hQ_analytic}
    J(\hbQ) =  \frac{1}{d^2} \log \EE_{\bY} \Bigg\{\Bigg(\int \rd \bR \, \delta\left(\frac{1}{d} \Tr[\bR] - 1 \right) e^{-\beta d \Tr[V(\bR)]} \, e^{ - \frac{d (\hQ + \hq)}{4} \Tr[\bR^2] + \frac{d \sqrt{\hq}}{2} \Tr[\bR \bY]}\Bigg)^{r}\Bigg\}.
\end{equation}
Using eq.~\eqref{eq:J_hQ_analytic} in eq.~\eqref{eq:Phi_r_after_saddle}, along with $\det \bQ = (Q-q)^{r-1} (Q+ (r-1) q)$,
we finally reach:
\begin{align}\label{eq:phi_r_final}
    \nonumber
    \Phi(\alpha,\beta;r) &= \extr_{Q, q,\hQ, \hq} \Bigg[\frac{\alpha r}{2} \log 4 \pi- \frac{\alpha (r-1)}{2} \log(Q-q) - \frac{\alpha}{2} \log(Q+(r-1)q)+ \frac{r}{4} Q \hQ - \frac{r(r-1)}{4} q \hq \\ 
    & +\frac{1}{d^2} \log \EE_{\bY} \Bigg\{\Bigg(\int \rd \bR \, \delta\left(\frac{1}{d} \Tr[\bR] - 1 \right) e^{-\beta d \Tr[V(\bR)]} \, e^{ - \frac{d (\hQ + \hq)}{4} \Tr[\bR^2] + \frac{d \sqrt{\hq}}{2} \Tr[\bR \bY]}\Bigg)^{r}\Bigg\} \Bigg].
\end{align}

\subsection{\texorpdfstring{The $r \to 0$ limit and extensive-rank HCIZ integrals}{}}\label{subsec:r_0_HCIZ}

\noindent
We now perform the last step of the replica method:
taking the derivative with respect to $r$, and then the $r \to 0$ limit yields 
\begin{align}
    \label{eq:phi_RS}
    \Phi(\alpha, \beta) &= \extr_{Q, q, \hQ, \hq} \Bigg[\frac{\alpha}{2} \log 4 \pi - \frac{\alpha}{2} \log(Q-q) - \frac{\alpha q}{2 (Q-q)} 
    + \frac{Q \hQ}{4} + \frac{q \hq}{4} + \lim_{d \to \infty} \frac{1}{d^2} \EE_{\bY} \log I(\bY) \Bigg], \\ 
    \nonumber
    I(\bY) &\coloneqq \int \rd \bR \, \delta\left(\frac{1}{d} \Tr[\bR] - 1 \right) e^{-\beta d \Tr[V(\bR)]} \, e^{ - \frac{d (\hQ + \hq)}{4} \Tr[\bR^2] + \frac{d \sqrt{\hq}}{2} \Tr[\bR \bY]}.
\end{align}
Recall that $\bY \sim \GOE(d)$.
Physically, the parameters $q$ and $Q$ correspond, when $\bR$ is sampled under the effective Gibbs measure associated to $I(\bY)$ in 
eq.~\eqref{eq:phi_RS}, to 
\begin{equation}\label{eq:interpretation_Q_q}
    \begin{dcases} 
        Q &= \frac{1}{d} \EE \langle \Tr [\bR^2] \rangle_\eff, \\
        q &= \frac{1}{d} \EE \Tr [\langle \bR \rangle_\eff^2],
    \end{dcases}    
\end{equation}
where we have defined the effective Gibbs measure: 
\begin{equation}\label{eq:def_effective_gibbs}
    \change{
    \langle \cdot \rangle_\eff \coloneqq \frac{\int \rd \bR \, (\cdot) \, \delta(d^{-1}\Tr[\bR] - 1) \, e^{-\beta d \Tr[V(\bR)] - \frac{d (\hQ + \hq)}{4} \Tr[\bR^2] + \frac{d \sqrt{\hq}}{2} \Tr[\bR \bY]}}{\int \rd \bR \, \delta(d^{-1}\Tr[\bR] - 1) \, e^{- \beta d \Tr[V(\bR)] - \frac{d (\hQ + \hq)}{4} \Tr[\bR^2] + \frac{d \sqrt{\hq}}{2} \Tr[\bR \bY]}}.
    }
\end{equation}
Eq.~\eqref{eq:interpretation_Q_q} can be computed from the stationary equations on $\hq, \hQ$: we detail this computation in Appendix~\ref{subsec_app:interpretation_q_Q}.
In particular, 
\begin{equation*}
    Q - q = \frac{1}{d} \EE \left\{\Tr \left[\langle \bR^2 \rangle_\eff - \langle \bR \rangle_\eff^2\right]\right\} = \frac{1}{d} \EE \sum_{i,j} [\langle R_{ij}^2 \rangle_\eff - \langle R_{ij} \rangle_\eff^2] \geq 0 .
\end{equation*}
Therefore, we have $0 \leq q \leq Q$.

\myskip 
\textbf{Computing $I(\bY)$ --}
We make a change of variables in $I(\bY)$ into eigenvectors and eigenvalues \cite{anderson2010introduction} (see also \cite{livan2018introduction} for a physics point of view and an introduction to the Coulomb gas approach to 
eigenvalues of random matrices), so that up to a multiplicative constant 
that only depends on $n, d$:
\begin{equation*}
     I(\bY) \propto \int_{\bbR^d} \rd \bLambda \, \delta\left(\frac{1}{d} \sum_{i=1}^d \lambda_i - 1 \right) \prod_{i < j} |\lambda_i - \lambda_j| \, e^{- \beta d \sum_{i=1}^d V(\lambda_i) - \frac{d (\hQ + \hq)}{4} \sum_{i=1}^d \lambda_i^2} \int_{\mcO(d)} \mcD \bO \, e^{\frac{d \sqrt{\hq}}{2} \Tr[\bO \bLambda \bO^\T \bY]},
\end{equation*}
where $\mcD \bO$ is the Haar measure on the orthogonal group $\mcO(d)$, and $\bLambda = (\lambda_i)_{i=1}^d$.
Since all the terms inside the integral only depend on $\bLambda$ via its empirical distribution $d^{-1} \sum_i \delta_{\lambda_i}$, we furthermore change variables in the integration 
to this empirical distribution: in the large-$d$ limit the Jacobian of this change only introduces a term in the exponential scale $\exp(\Theta(d))$ -- related to the entropy of the empirical distribution -- which we therefore neglect%
\footnote{See e.g.\ \cite{livan2018introduction} for a physics description of this point, or \cite{arous1997large} for a rigorous proof in a similar setting.}.
Performing Laplace's method in the space of probability distributions, we finally reach 
that (up to an additive constant that only depends on $\alpha$):
\begin{equation}
    \label{eq:E_logI}
    \frac{1}{d^2} \EE_{\bY} \log I(\bY) = \!\!\!\! \sup_{\substack{\mu \in \mcM_1^+(\bbR) \\ \int \mu(\rd x) x = 1}} \Big\{\frac{1}{2} \Sigma(\mu) - \int \mu(\rd x) \Big[\beta V(x) + \frac{\hQ + \hq}{4} x^2 \Big] + \frac{1}{2} I_\HCIZ(\sqrt{\hq}, \mu, \rho_\sci)\Big\},
\end{equation}
with $\Sigma(\mu) \coloneqq \int \mu(\rd x) \mu(\rd y) \, \log|x-y|$ Voiculescu's non-commutative entropy \cite{voiculescu1993analogues}.
Moreover, we defined the \emph{Harish-Chandra-Itzykson-Zuber} (HCIZ) integral \cite{harish1957differential,itzykson1980planar}:
\begin{equation}\label{eq:def_hciz}
    I_\HCIZ(\theta, \bR, \bY) \coloneqq \frac{2}{d^2} \log \int \mcD \bO \exp\Big\{\frac{\theta d}{2} \Tr[\bO \bR \bO^\intercal \bY]\Big\}.
\end{equation} 
The large-$d$ behavior of $I_\HCIZ(\theta, \bR, \bY)$ has been studied extensively, both in the case in which one of the matrices has rank $\smallO(d)$ \cite{guionnet2005fourier,collins2007new}, 
or when the two matrices have extensive rank, as in our setting \cite{matytsin1994large,guionnet2002large}.
In this latter setup, it is known that the large-$d$ limit of $I_\HCIZ(\theta, \bR, \bY)$ only depends on the limit spectral density of 
$\bR$ and $\bY$, justifying our notation $I_\HCIZ(\theta, \mu, \nu)$ in eq.~\eqref{eq:E_logI}.
Analytical formulas for $I_\HCIZ(\theta, \mu, \nu)$ are known \cite{matytsin1994large,guionnet2002large}, and involve 
the solution to a complex partial differential equation known as Burgers' equation, which we recall for completeness in 
Appendix~\ref{subsec_app:hciz}.
Unfortunately, this exact expression is very tedious to analyze (both analytically and numerically), and further simplifications of $I_\HCIZ(\theta, \mu, \nu)$ are only available in some restricted cases \cite{maillard2022perturbative,troiani2022optimal,pourkamali2023matrix}. 

\myskip
\change{
\textbf{Concentration of the empirical spectral distribution --}
We briefly comment on how our computation also gives access to the typical spectral density of 
ellipsoid fits (or more generally of samples under the Gibbs measure of eq.~\eqref{eq:def_Gibbs}).
An important insight of the replica method is that, once averaged over their respective disorders, the effective Gibbs measure of eq.~\eqref{eq:def_effective_gibbs} approximates the original measure of eq.~\eqref{eq:def_Gibbs}.
Moreover, as a by-product of the use of Laplace's method in eq.~\eqref{eq:E_logI}, one can see that the density $\mu \in \mcM_1^+(\bbR)$ maximizing the variational principle of 
eq.~\eqref{eq:E_logI} is the typical asymptotic spectral density of a sample $\bR \sim \EE_{\bY} \langle \cdot \rangle_\eff$ (recall the definition of the effective Gibbs measure in eq.~\eqref{eq:def_effective_gibbs}).
Combining these two facts imply that $\mu$ is also the typical asymptotic spectral density of a sample $\bS$ under the original Gibbs measure of 
eq.~\eqref{eq:def_Gibbs}, a fact which we use in Claims~\ref{claim:solution_space} and \ref{claim:explicit_constructions}.
We give in Appendix~\ref{subsec_app:concentration_spectrum} a more detailed justification of this conclusion, detailing the points sketched above.
}

\subsection{Final result}\label{subsec:final_result_replica}

\noindent
Let us summarize our formula for the free entropy, using eqs.~\eqref{eq:phi_RS} and \eqref{eq:E_logI}, and introducing $t \coloneqq q/Q \in [0,1]$. 
We omit additive constants that only depend on $\alpha$.
\begin{align}
    \label{eq:phi_RS_general}
    \nonumber
    \Phi(\alpha, \beta) = \extr_{Q, t, \hQ, \hq} \sup_{\substack{\mu \in \mcM_1(\bbR),\\ \int \mu(\rd x) x = 1}} \Bigg[&
    -\frac{\alpha}{2} \log Q - \frac{\alpha}{2} \log (1 - t) - \frac{\alpha t}{2 (1-t)} + 
    \frac{Q(\hQ + t \hq)}{4} + \frac{1}{2} \Sigma(\mu) \\ 
    &
    - \int \mu(\rd x) \left[\beta V(x) + \frac{\hQ + \hq}{4} x^2 \right]
    +\frac{1}{2} I_\HCIZ(\sqrt{\hq}, \mu, \rho_{\sci})
    \Bigg].
\end{align}
Introducing Lagrange multipliers $\lambda_0, \lambda_1$ the so-called replica-symmetric equations 
read then: 
\begin{subnumcases}{\label{eq:se_general}}
    \label{eq:se_general_1}
    \frac{Q \hq}{2} = \frac{\alpha t }{(1-t)^2}, & \\ 
    \label{eq:se_general_2}
    (1-t)  Q = \frac{1}{\sqrt{\hq}} \partial_\theta I_\HCIZ(\theta = \sqrt{\hq}, \mu, \rho_{\sci}), &\\
    \label{eq:se_general_3}
    \hQ + t \hq = \frac{2 \alpha}{Q}, &\\
    \label{eq:se_general_4}
    \int \mu(x) \rd x = 1, & \\ 
    \label{eq:se_general_5}
    \int \mu(x) \, x  \, \rd x = 1, & \\
    \label{eq:se_general_6}
    \int \mu(x) \, x^2  \, \rd x = Q, & \\
    \label{eq:se_general_7}
    \int \mu(y) \log | x-y |\, \rd y - \beta V(x) - \frac{\hQ + \hq}{4} x^2 - \lambda_1 x - \lambda_0 + \frac{1}{2} \frac{\partial}{\partial \mu(x)}  I_\HCIZ(\sqrt{\hq}, \mu, \rho_{\sci}) = 0.&
\end{subnumcases}
Eq.~\eqref{eq:se_general_7} is valid for all $x \in \supp(\mu)$. In particular if $V(x) = +\infty$ then $x \notin \supp(\mu)$.
Moreover, because of the application of Laplace's method in $I(\bY)$, cf.\ eq.~\eqref{eq:E_logI}, $\mu(x)$ is the limiting spectral density 
of a sample $\bS$ under the Gibbs measure of eq.~\eqref{eq:def_Gibbs} (the concentration of this spectral density being also a consequence of our analysis).

\section{The dilute limit of HCIZ integrals and consequences}\label{sec:dilute_hciz}
\noindent
We start from the general result of eq.~\eqref{eq:se_general}.
As we mentioned above solving these equations in general is hard, as the function $I_\HCIZ(\sqrt{\hq},\mu,\sigma_\sci)$ has an explicit 
but very tedious expression involving the solution of a transport problem between the two distributions $\mu$ and $\rho_{\sci}$, see Appendix~\ref{subsec_app:hciz}.

\myskip
However, recall that when studying the SAT/UNSAT transition (i.e.\ \ref{sdp_pot}), the set of solutions shrinks and disappears as we approach the SAT/UNSAT transition $\alpha \uparrow \alpha_c$.
Similarly, when studying the minimizers under a potential $V(x)$ (\ref{nn_pot} and \ref{ls_pot}), we will take the limit $\beta \to \infty$: 
in this limit, if there is a unique solution to the linear constraints that minimizes the potential energy, 
we also expect the Gibbs measure to concentrate its mass around it.
\change{Thus, in all these cases (namely~\ref{sdp_pot}, and in both \ref{nn_pot} and \ref{ls_pot} if there is a unique minimal-energy solution)}
we have $t \to 1$, as $t = q/Q$ represents the ``angular width'' of the asymptotic support of the Gibbs measure, 
cf.\ eq.~\eqref{eq:interpretation_Q_q}.
Through eq.~\eqref{eq:se_general_1} this will imply that $\hq \to \infty$ if the variance of $\mu$ remains finite.
This will allow us to leverage so-called ``dilute'' $\theta \to \infty$ expansions of the HCIZ integral \cite{bun2014instanton} in order to analytically solve eq.~\eqref{eq:se_general}
either close to the SAT/UNSAT transition point or in the $\beta \to \infty$ limit.

\subsection{The dilute limit of the HCIZ integral}\label{subsec:dilute_hciz}

\noindent
The dilute limit of the HCIZ integral reads, 
for any $\rho_A, \rho_B$ (assuming $\EE_{A}[X] = 0$ or $\EE_B[X] = 0$) \cite{bun2014instanton} : 
\begin{align}\label{eq:dilute_expansion_hciz}
    \nonumber
    I_\HCIZ(\theta, \rho_A, \rho_B) &= \theta \int_0^1 X_A(p) X_B(p) \rd p - \frac{1}{2} \log \theta - \frac{1}{2}[\Sigma(\rho_A) + \Sigma(\rho_B)] - \frac{3}{4} \\ 
    &- \frac{\pi^2}{6 \theta} \int_0^1 \rho_A[X_A(p)] \rho_B[X_B(p)]\rd p +  \mcO_{\theta\to \infty}(\theta^{-2}).
\end{align}
Here, $X_A(p) = F_A^{-1}(p)$, with $F_A(x) = \int_{-\infty}^x \rho_A(u) \rd u$ the cumulative distribution function of $\rho_A$.

\myskip 
\textbf{The first order of eq.~\eqref{eq:dilute_expansion_hciz} --}
In what follows, we will mainly need the first order of the dilute expansion.
Its form can easily be understood:
as $\theta \to \infty$, the integral in eq.~\eqref{eq:def_hciz} is dominated by the matrix $\bO$ that 
maximizes $\Tr[\bA \bO \bB \bO^\intercal]$: this matrix is such that $\bA$ and $\bO \bB \bO^\intercal$ are diagonal in the same basis, both of them having ordered eigenvalues. 
This precisely corresponds to the leading order term in eq.~\eqref{eq:dilute_expansion_hciz}.

\subsection{The SAT/UNSAT transition of ellipsoid fitting}\label{subsec:sat_unsat}

\noindent
We assume here that $\beta = 1$ and $V(x) = + \infty \times \indi\{x < 0\}$, i.e.\ we are in case~\ref{sdp_pot}, and we will show the statements of Claim~\ref{claim:solution_space} 
from the general replica results of eq.~\eqref{eq:phi_RS_general}.
Notice that the statement of Claim~\ref{claim:solution_space} when $\alpha < \alpha_c$ are consequences of the general replica analysis as discussed in Section~\ref{subsec:final_result_replica}, 
and we therefore focus on the claims related to the SAT/UNSAT transition, in particular showing that $\alpha_c = 1/4$. 
As we argued above, we assume that $t \to 1$ as $\alpha \uparrow \alpha_c$. 
Moreover, we assume that the spectral density of solutions remains of order $1$ as $\alpha \uparrow \alpha_c$ (this hypothesis is justified both from numerical simulations, 
and because we must still satisfy the constraint $\int \mu(x) \, x \, \rd x = 1$).
By eq.~\eqref{eq:se_general_6}, $Q$ then also has a finite limit as $\alpha \uparrow \alpha_c$.

\myskip
From eq.~\eqref{eq:se_general_1}, one sees that $\hq \sim \hq_0 (1-t)^{-2}$ as $\alpha \uparrow \alpha_c$, 
with $\hq_0 = 2\alpha / Q$.
Using it in eq.~\eqref{eq:se_general_2} along with the expansion of eq.~\eqref{eq:dilute_expansion_hciz} (which we can use since $\hq \to \infty$) 
we have 
\begin{equation}\label{eq:se_2_transition}
    \sqrt{2 \alpha\int \mu(x) x^2 \rd x} = \int_0^1 X_\mu(p) X_{\sci}(p) \rd p.
\end{equation}
Furthermore, \change{combining $\hq \sim \hq_0(1-t)^{-2}$ with} eq.~\eqref{eq:se_general_3}, we have $\hQ + \hq \sim R_0 (1-t)^{-1}$, where
\begin{equation}\label{eq:R0_hq0_sat_unsat}
R_0 = \hq_0 = \frac{2\alpha}{Q}.
\end{equation}

\myskip
We finally focus on eqs.~\eqref{eq:se_general_4},\eqref{eq:se_general_5} and \eqref{eq:se_general_7}.

\noindent
\textbf{Singularities in $\mu(x)$ --}
Given the ``hard-wall'' potential that influences $\mu$, we assume that $\mu(x)$ can only possibly develop singular delta peaks 
in $x = 0$: indeed, there is a hard constraint that $\supp(\mu) \subseteq \bbR_+$, while there is a term  $(1-t)^{-1} R_0 x^2/4$ as $t \to 1$ in eq.~\eqref{eq:se_general_7}, which pushes eigenvalues close to $0$. 
Therefore, eigenvalues may ``accumulate'' near zero, creating a delta peak as $t \to 1$.

\myskip
In Appendix~\ref{subsec_app:nc_entropy_near_sat_unsat} we justify that, although such a delta peak makes the first term in eq.~\eqref{eq:se_general_7} singular, 
it only diverges as $\mcO(\log(1-t))$.
Using the expansion of eq.~\eqref{eq:dilute_expansion_hciz}, the dominant order of eq.~\eqref{eq:se_general_7} as $t \to 1$ is thus of order $(1-t)^{-1}$, and we get that for all 
$x \in \supp(\mu)$ (rescaling also the Lagrange multipliers \change{$\lambda_1 \to (1-t)^{-1} \lambda_1$ and $\lambda_0 \to (1-t)^{-1} \lambda_0$}):
\begin{equation}\label{eq:se_7_expansion}
    \frac{\hq_0}{4} x^2 + \lambda_1 x + \lambda_0 = \frac{\sqrt{\hq_0}}{2} \int_0^1 \frac{\partial X_{\mu}(p)}{\partial \mu(x)} X_{\sci}(p) \rd p.
\end{equation}
Recall that $X_\mu(p) = F_\mu^{-1}(p)$, with $F_\mu$ the cumulative distribution function of $\mu$.
Differentiating the relation 
\begin{equation*}
    p = \int_{-\infty}^{X_\mu(p)} \mu(u) \rd u
\end{equation*}
with respect to $\mu(x)$ we get, for $x \in \supp \mu$: 
\begin{equation*}
    \frac{\partial X_\mu(p)}{\partial \mu(x)} = - \frac{\indi[x \leq X_\mu(p)]}{\mu[X_\mu(p)]}.
\end{equation*}
Plugging it back into eq.~\eqref{eq:se_7_expansion} and differentiating with respect to $x$ we get:
\begin{equation*}
    \frac{\hq_0}{2} x + \lambda_1 = -\frac{\sqrt{\hq_0}}{2}\frac{\partial}{\partial x} \int_0^1 \frac{\indi[x \leq X_\mu(p)]}{\mu[X_\mu(p)]}X_{\sci}(p) \rd p.
\end{equation*}
We can then change variables to $p = F_\mu(u)$, and get:
\begin{equation*}
    \sqrt{\hq_0} x + \frac{2\lambda_1}{\sqrt{\hq_0}} = - \frac{\partial}{\partial x}\int_{x}^{\max \supp \mu} X_{\sci}[F_\mu(u)] \rd u = X_{\sci}[F_\mu(x)].
\end{equation*}
Equivalently, for all $x \in \supp(\mu)$:
\begin{equation}\label{eq:F_mu_transition}
   F_\mu(x) = F_\sci\left(\sqrt{\hq_0} x + \frac{2\lambda_1}{\sqrt{\hq_0}}\right) = F_{\sci}\left(\frac{x-m}{\sigma}\right),
\end{equation}
with $\sigma \coloneqq \hq_0^{-1/2}$ and $m \coloneqq - 2 \lambda_1 / \hq_0$.
Eq.~\eqref{eq:F_mu_transition} means that $\mu(x)$ is a semicircular density with mean $m$ and variance $\sigma^2$, 
truncated (if necessary) so that its support is included in $\bbR_+$.
The truncated mass is put as a delta peak in $0$, with mass $F_\mu(0) = F_\sci(-m/\sigma)$.

\myskip 
Let us now leverage eqs.~\eqref{eq:se_general_5}, \eqref{eq:se_2_transition}, and \eqref{eq:R0_hq0_sat_unsat}, to deduce the values of $\sigma, m$, and subsequently $\alpha_c$.
We rewrite these three equations as:
\begin{equation}\label{eq:constraints_spectrum_transition}
    \begin{dcases}
        \int \mu(x) \, x \, \rd x &= 1, \\
        \int \mu(x) \, x^2 \, \rd x &= 2 \alpha \sigma^2, \\
        2 \alpha \int \mu(x) \, x^2 \, \rd x &= \left(\int_0^1 X_\mu(p) X_\sci(p) \rd p\right)^2.
    \end{dcases}
\end{equation}
Using eq.~\eqref{eq:F_mu_transition} and changing variables $x = X_\mu(p)$ (using that $X_\mu(p) = 0$ for $p \leq F_\mu(0)$):
\begin{align*}
    \int_0^1 X_\mu(p) X_\sci(p) \rd p &= \int X_\sci[F_\mu(x)] \, x \, \mu(x) \, \rd x, \\ 
    &= \frac{1}{\sigma} \int x^2 \mu(x) \rd x - \frac{m}{\sigma}, \\ 
    &= 2 \alpha \sigma - \frac{m}{\sigma},
\end{align*}
using the first two equations of eq.~\eqref{eq:constraints_spectrum_transition}. 
Comparing with the last one, we get that $m = 0$. 
This implies that $\mu(\{0\}) = F_\mu(0) = 1/2$: close to the transition, solutions are typically half-rank!
The first equation of eq.~\eqref{eq:constraints_spectrum_transition} gives then:
\begin{equation*}
    \sigma = \left[\int_{0}^2 x \rho_\sci(x) \rd x\right]^{-1} = \frac{3 \pi}{4}.
\end{equation*}
Moreover, the second equation of eq.~\eqref{eq:constraints_spectrum_transition} implies $\sigma^2 / 2 = 2 \alpha \sigma^2$,
i.e.\ that the SAT/UNSAT transition point $\alpha = \alpha_c$ is
\begin{equation}\label{eq:ellipse_threshold}
    \alpha_c = \frac{1}{4},
\end{equation}
validating the original ``ellipse'' conjecture which motivated this calculation. 
Moreover, the asymptotic spectral density of solutions $\mu(x) = \mu_c(x)$ near the SAT/UNSAT threshold is 
given by
\begin{equation}\label{eq:mu_limit}
    \mu_c(x) = \frac{1}{2} \delta(x) + \frac{4\sqrt{9\pi^2 - 4 x^2}}{9 \pi^3} \indi\left\{0 \leq x \leq \frac{3\pi}{2}\right\}.
\end{equation}
This ends our derivation of Claim~\ref{claim:solution_space}. In Fig.~\ref{fig:main_results_solution_space} we illustrate these predictions, 
and compare them to finite-$d$ numerical simulations.
We note that one may use the next orders of the dilute expansion of eq.~\eqref{eq:dilute_expansion_hciz} to compute the perturbative expansion 
of $\mu(x)$ in powers of $(1-4\alpha)$, beyond the limit $\mu_c(x)$. We leave such an investigation for future work.

\subsection{Minimal length of the longest principal axis}\label{subsec:minimal_length_max_axis}

\noindent
We now consider the constraint $\bS \succeq \kappa \Id_d$, for some $\kappa \geq 0$.
In a geometric point of view, this constrains the principal axes of the ellipsoid to have length at most $\kappa^{-1/2}$.
Note that this is still a convex constraint on $\bS$, and one obtains a SAT/UNSAT transition
$\alpha_c(\kappa)$ characterized as well by eqs.~\eqref{eq:F_mu_transition},\eqref{eq:constraints_spectrum_transition}, 
with the density being now truncated to $\{x \geq \kappa\}$ rather than $\{x \geq 0\}$.
This yields the following set of simple scalar equations (their derivation from eqs.~\eqref{eq:F_mu_transition},\eqref{eq:constraints_spectrum_transition} is detailed in Appendix~\ref{subsec_app:kappa}): 
\begin{equation}\label{eq:alphac_lstar_kappa}
    \begin{dcases}
    m &= -\frac{\kappa}{1 - \kappa}, \\
    1 &= m + (\kappa - m) G_0\left[\frac{\kappa-m}{\sigma}\right] - \sigma G_1\left[\frac{\kappa-m}{\sigma}\right], \\
    \alpha_c(\kappa) &= \frac{m^2 + \sigma^2}{2\sigma^2} + \frac{\kappa^2 - m^2}{2\sigma^2} G_0\left[\frac{\kappa-m}{\sigma}\right] - \frac{m}{\sigma} G_1\left[\frac{\kappa-m}{\sigma}\right] - \frac{1}{2} G_2\left[\frac{\kappa-m}{\sigma}\right],
    \end{dcases}
\end{equation}
where we defined the functions (for $k \geq 0$):
\begin{equation*}
    G_k(r) \coloneqq \int_{-2}^r \rho_\sci(u) \, u^k \, \rd u,
\end{equation*}
Given $\kappa$, one solves the first and second equations of eq.~\eqref{eq:alphac_lstar_kappa} to obtain $(m, \sigma)$, 
yielding then $\alpha_c(\kappa)$ by the last equation. This procedure leads to Fig.~\ref{fig:alphac_kappa}.

\subsection{The limit \texorpdfstring{$\beta \to \infty$}{} for a general potential}\label{subsec:large_beta}

\noindent
In this section, we take the zero-temperature limit (i.e.\ $\beta \to \infty$) in the free entropy of eq.~\eqref{eq:phi_RS_general}, assuming a 
generic convex potential $V(x)$.
We denote the intensive ``energy'' of a matrix $\bS$ as $e(\bS) \coloneqq d^{-1} \Tr V(\bS)$, 
and we define the \emph{ground state energy} as
\begin{equation*}
    e_\GS(V) \coloneqq \lim_{d \to \infty} \EE\left[\min_{\bS \in \mcC(\bx)} e(\bS)\right],
\end{equation*}
with $\mcC(\bx) \coloneqq \{\bS \in \mcS_d \, : \, \Tr[\bW_\mu \bS] = (d-\Tr[\bS])/\sqrt{d}, \, \, \forall \mu \in [n]\}$.
As $\beta \to \infty$ the Gibbs measure concentrates its mass on the set of matrices $\bS$ 
such that $e(\bS) \simeq e_\GS(V)$.
Because $V$ is convex, there are two possibilities for the behavior of the Gibbs measure at low temperature:
\begin{enumerate}[label=\textbf{(\alph*)},ref=(\alph*)]
    \item\label{zero_entropy_gs} As $\beta \to \infty$, the support of the Gibbs measure shrinks around the global minimizer of $e(\bS)$ with $\bS \in \mcC(\bx)$, which is unique (in the sense 
    that the set of matrices with the same intensive energy has \change{zero entropy}, i.e.\ $s(\beta) \coloneqq \Phi(\alpha, \beta) + \beta e(\beta) \to - \infty$, with $e(\beta)$ the typical energy of a sample under the Gibbs measure at temperature $\beta$).
    Then we have $t = q/Q \to 1$ as $\beta \to \infty$, and in this limit $\mu(x)$ is the asymptotic spectral density of this global minimizer.
    \item\label{pos_entropy_gs} The set of minimizers of $e(\bS)$ with $\bS \in \mcC(\bx)$ has a finite entropy: 
    in this case, the Gibbs measure tends as $\beta \to \infty$ to the uniform measure on the set of minimizers (the ground states).
    This implies that $t$ remains bounded \change{away} from $1$ as $\beta \to \infty$, and $\mu(x)$ is the typical spectral density 
    of a matrix uniformly sampled in this set.
\end{enumerate}
We analyze the limit of the replica-symmetric equations as $\beta \to \infty$ in these two cases. In Sections~\ref{subsec:nn} and \ref{subsec:close_identity} 
we will detail which of these cases apply for the potentials considered in Claim~\ref{claim:explicit_constructions}, and perform a detailed derivation of its statements.

\myskip
\subsubsection{Case~\ref{zero_entropy_gs}}
As $t \to 1$ as $\beta \to \infty$,
we assume a scaling $1 - t \sim h \beta^{-\eta_t}$ for some $\eta_t > 0$.
We also assume that $\mu(x)$ has a finite limit as $\beta \to \infty$, and thus $Q$ as well.
Combining all equations in eq.~\eqref{eq:se_general} one finds easily that if we denote 
$\hq \sim \hq_0 \beta^{\eta_\hq}$, we have $\eta_{t} = \eta_{\hq}/2 = 1$. 
Moreover, we have at leading order, using the dilute expansion of eq.~\eqref{eq:dilute_expansion_hciz} and proceeding similarly as in Section~\ref{subsec:sat_unsat}: 
\begin{subnumcases}{\label{eq:se_beta_inf}}
    \label{eq:se_beta_inf_1}
    \frac{\hq_0}{2} \int \mu(x) \, x^2 \, \rd x = \frac{\alpha}{h^2}, & \\ 
    \label{eq:se_beta_inf_2}
    h \sqrt{\hq_0} \int \mu(x) \, x^2 \, \rd x = \int_0^1 X_\mu(p) X_{\sci}(p) \rd p, &\\
    \label{eq:se_beta_inf_3}
    \int \mu(x) \rd x = 1, & \\ 
    \label{eq:se_beta_inf_4}
    \int \mu(x) \, x  \, \rd x = 1, & \\
    \label{eq:se_beta_inf_5}
    - V(x) - \frac{h \hq_0}{4} x^2 - \lambda_1 x - \lambda_0 + \frac{\sqrt{\hq_0}}{2} \int_0^1 \frac{\partial X_\mu(p)}{\partial \mu(x)} X_{\sci}(p) \rd p = 0.&
\end{subnumcases}
We also rescaled the Lagrange multipliers with $\beta$, to ensure the two conditions of eq.~\eqref{eq:se_beta_inf_3},\eqref{eq:se_beta_inf_4}.
It is easy to see that by eq.~\eqref{eq:se_beta_inf_5} we have, performing exactly as in the derivation of eq.~\eqref{eq:F_mu_transition}:
\begin{equation*}
    V'(x) + \frac{h \hq_0}{2} x + \lambda_1 = \frac{\sqrt{\hq_0}}{2} X_{\sci}[F_\mu(x)],
\end{equation*}
which can also be written as:
\begin{equation}\label{eq:se_beta_inf_spectrum}
    F_\mu(x) = F_{\sci}\left[\frac{2}{\sqrt{\hq_0}} V'(x) + h \sqrt{\hq_0}  \,x + \frac{2 \lambda_1}{\sqrt{\hq_0}}\right].
\end{equation}
$\mu$ is thus a non-linear transformation of a semicircular law, the non-linearity depending on the potential $V$. 
Notice that in eq.~\eqref{eq:se_beta_inf_spectrum} there might be $\delta$ peaks in $\mu$ at discontinuity points of $V'$, and we will see such an example in what follows, in the case of the minimal nuclear norm solution.

\myskip 
\subsubsection{Case~\ref{pos_entropy_gs}}
    In this case the entropy $s(\beta) = \Phi(\alpha, \beta) + \beta e(\beta)$ remains bounded away from $-\infty$ as $\beta \to \infty$. 
    Since $e(\beta) = \int \mu(x) V(x) \rd x$, we have in particular from eq.~\eqref{eq:phi_RS_general} that $\Sigma(\mu)$ remains bounded as $\beta \to \infty$: the limit of $\mu(x)$ as $\beta \to \infty$ has a continuous support, with 
    no delta peaks.
    Moreover, $t, Q, \hq, \hQ$ all have finite limits as $\beta \to \infty$ by eq.~\eqref{eq:se_general}, and $t$ remains bounded away from $1$.
    Finally, $\mu$ must maximize the free entropy, and thus satisfy eq.~\eqref{eq:se_general_7}, 
    for all $x \in \supp(\mu)$: 
    \begin{equation*}
        \int \mu(y) \log | x-y |\, \rd y - \beta V(x) - \frac{\hQ + \hq}{4} x^2 - \lambda_1 x - \lambda_0 + \frac{1}{2} \frac{\partial}{\partial \mu(x)}  I_\HCIZ(\sqrt{\hq}, \mu, \sigma_{\sci}) = 0.
    \end{equation*}
    Since $\mu(x)$, $\hq$ and $\hQ + \hq$ have finite limits as $\beta \to \infty$, the analysis of the order $\Theta(\beta)$ of this equation implies that all $x \in \supp(\mu)$ must satisfy an equation of the type
    \begin{equation*}
         V(x) = A x + B,
    \end{equation*}
    with $A \coloneqq \lim_{\beta \to \infty} (-\lambda_1/\beta)$ and $B \coloneqq \lim_{\beta \to \infty} (-\lambda_0/\beta)$.
    In other words, \emph{$V(x)$ must be a linear function on the support of $\mu$}.
    Conversely, if this is the case then the term $-\beta \int \mu(x) V(x) \rd x$ is fixed by the constraints $\int \mu(x)\, x \, \rd x = 1$ and $\int \mu(x) \rd x = 1$. 
    The limit $\mu(x)$ is then a solution of the equation, for $x \in \supp(\mu)$:
    \begin{equation}
        \label{eq:mu_pos_entropy_gs}
        \int \mu(y) \log | x-y |\, \rd y - \frac{\hQ + \hq}{4} x^2 - \lambda_1 x - \lambda_0 + \frac{1}{2} \frac{\partial}{\partial \mu(x)}  I_\HCIZ(\sqrt{\hq}, \mu, \sigma_{\sci}) = 0,
    \end{equation}
    alongside with eqs.~\eqref{eq:se_general_1} to \eqref{eq:se_general_6}. 

\subsection{The minimal nuclear norm estimator and generalizations}\label{subsec:nn}

\noindent
We apply in this section the zero-temperature analysis of Section~\ref{subsec:large_beta} to
the case $V(x) = |x|^\gamma$, for $\gamma \geq 1$. This corresponds to solving the convex optimization problem
\begin{equation*}
    \min_{\bS \in \mcC(\bx)} \|\bS\|_{S_\gamma},
\end{equation*}
with $\mcC(\bx) \coloneqq \{\bS \in \mcS_d \, : \, \Tr[\bW_\mu \bS] = (d-\Tr[\bS])/\sqrt{d}, \, \, \forall \mu \in [n]\}$.
We show here the statements of Claim~\ref{claim:explicit_constructions} corresponding to this case, and detail the picture of Fig.~\ref{subfig:threshold_min_S_gamma}.
To keep our notations consistent with Claim~\ref{claim:explicit_constructions}, we denote $\nu(x)$ the asymptotic spectral density of the minimizer, rather than $\mu(x)$.
For reasons that will become clear during the derivation, we separate the cases $\gamma = 1$ (i.e.\ the nuclear norm minimizer) and $\gamma > 1$.

\subsubsection{The case \texorpdfstring{$\gamma = 1$}{}}
We now take $V(x) = |x|$.
We first make a general remark, to clarify which of the two regimes~\ref{zero_entropy_gs} and \ref{pos_entropy_gs} of Section~\ref{subsec:large_beta} 
can occur as a function of $\alpha \in (0,1/2)$.
\begin{itemize}
    \item Assume that $\alpha > 1/4$, so that we are in the UNSAT phase of $(\rm P)$. Then there is no positive semidefinite solution 
    to the linear constraints. In particular, the distribution $\nu(x)$ has a support with both positive and negative eigenvalues.
    Therefore, $V(x)$ can not be linear on the support of $\nu$: 
    we must thus be in the regime~\ref{zero_entropy_gs}, in which the ground states have asymptotically zero entropy.
    \item On the other hand, if $\alpha < 1/4$, there are positive semidefinite solutions to the linear constraints.
    By eq.~\eqref{eq:trace_close_1}, all such solutions have normalized trace asymptotically equal to $1$, and therefore 
    an energy $e(\bS) = d^{-1} \Tr |\bS| \simeq 1$. However, since all solutions to the linear constraints satisfy $d^{-1}\Tr[\bS] \simeq 1$ by eq.~\eqref{eq:trace_close_1},
    they also verify $(1/d)\Tr|\bS| \geq (1/d)|\Tr \bS| \simeq 1$, with equality if and only if $\bS \succeq 0$:
    this implies that the near-ground states of $e(\bS)$ (i.e.\ the minimal nuclear norm solutions) are exactly the positive semidefinite solutions\footnote{
    Note that in finite-size simulations, there will be a unique nuclear norm minimizer. However, as $d \to \infty$, its normalized nuclear norm will approach the one 
    of any positive semidefinite solution.}! 
    Moreover, in this case $V(x)$ is a linear function 
    on the support of all these solutions: we are thus in the regime~\ref{pos_entropy_gs} in which the set of ground states has positive entropy.
    And the typical spectral density of a uniformly-sampled solution is then $\mu[\alpha]$ given in Claim~\ref{claim:solution_space}.
\end{itemize}

\begin{figure}[!t]
     \centering
     \begin{subfigure}[t]{0.53\textwidth}
         \centering
         \includegraphics[width=\textwidth]{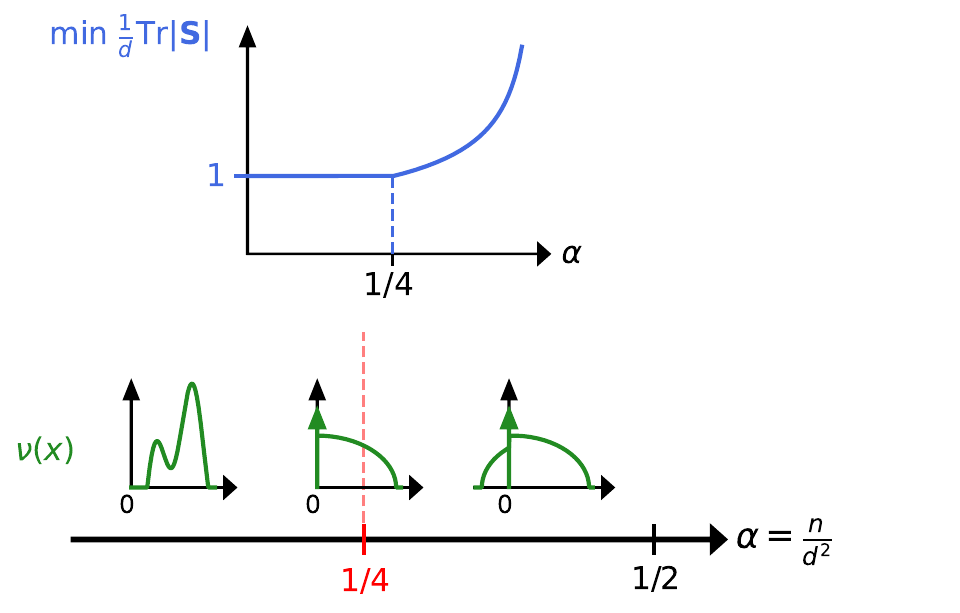}
         \caption{
            \label{subfig:sketch_nn_evolution}
            Sketch of the evolution of the minimal nuclear norm (top), and of the asymptotic spectral density of the minimal nuclear norm solution (bottom).
            For $\alpha < 1/4$, $\nu = \mu[\alpha]$ the typical spectral density of positive semidefinite solutions, see Claim~\ref{claim:solution_space}. 
            For $\alpha > 1/4$, $\nu$ is made of two truncated semicircular laws given by eqs.~\eqref{eq:general_nu_nn},\eqref{eq:se_NN_mds}.
         }
     \end{subfigure}
     \hfill
     \begin{subfigure}[t]{0.45\textwidth}
         \centering
         \includegraphics[width=\textwidth]{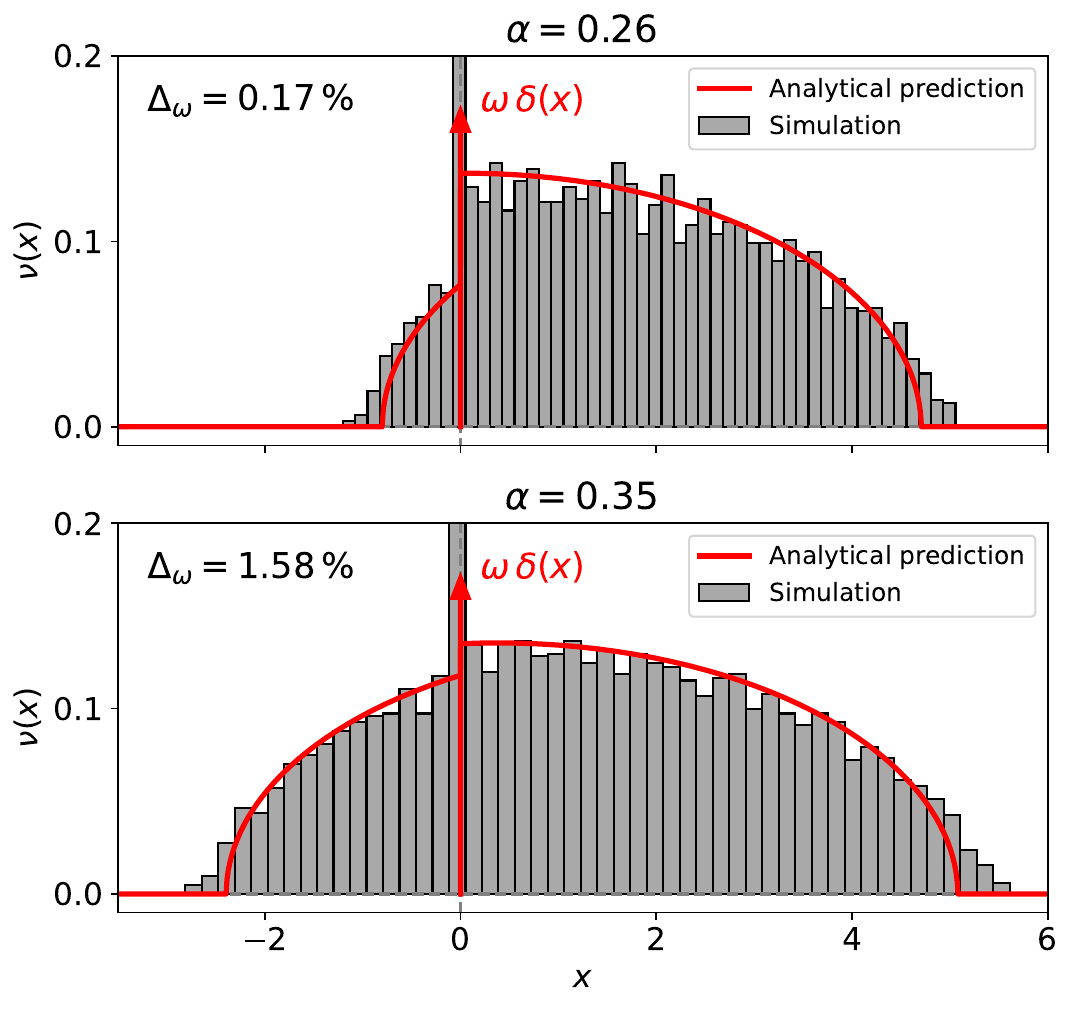}
         \caption{  \label{subfig:nn_alpha_0.26_0.35}
         The typical spectral density of the minimal nuclear norm solution for $\alpha \in \{0.26, 0.35\}$.
         We compare with finite-size simulations 
         using $d = 100$, combining $50$ realizations.
         We also write the relative difference $\Delta_\omega$ between $\omega$ from eq.~\eqref{eq:general_nu_nn} and the fraction of 
         near-zero eigenvalues ($|\lambda| < 10^{-3}$) of the numerical simulation.
         }
     \end{subfigure}
   \caption{\label{fig:nn_solution}
        Results on the minimal nuclear norm solution. We sketch the global evolution of the objective value and the asymptotic spectral density of solutions (left), 
        and compare these predictions quantitatively with finite-$d$ numerical simulations (right).
   }
\end{figure}

\myskip
The discussion above is summarized in Fig.~\ref{subfig:sketch_nn_evolution}.
We now focus on the case $\alpha \in (1/4, 1/2)$.
As we discussed above, we are then in case \ref{zero_entropy_gs} of Section~\ref{subsec:large_beta}.
With $V(x) = |x|$, eq.~\eqref{eq:se_beta_inf_spectrum} becomes: 
\begin{equation}\label{eq:nu_NN_unsat}
    F_\nu(x) = F_{\sci}\left[\frac{2}{\sqrt{\hq_0}} \sign(x) + h \sqrt{\hq_0} x + \frac{2 \lambda_1}{\sqrt{\hq_0}}\right],
\end{equation}
in which $\hq_0, h, \lambda_1$ are given by eqs.~\eqref{eq:se_beta_inf_1}, \eqref{eq:se_beta_inf_2}, \eqref{eq:se_beta_inf_4}: 
\begin{subnumcases}{\label{eq:se_NN_unsat}}
    \label{eq:se_NN_unsat_1}
        \frac{\hq_0}{2} \int \nu(x) \, x^2 \, \rd x = \frac{\alpha}{h^2},& \\ 
    \label{eq:se_NN_unsat_2}
        h \sqrt{\hq_0} \int \nu(x) \, x^2 \, \rd x = \int_0^1 X_\nu(p) X_{\sci}(p) \rd p, &\\
    \label{eq:se_NN_unsat_3}
        \int \nu(x) \, x  \, \rd x = 1. &
\end{subnumcases}
Note that because of the discontinuity of $\sign(x)$, $\nu(x)$ has a delta peak in $x = 0$ (this corresponds to the \change{$\ell_1$} norm promoting a sparse eigenvalue vector).
Denoting $m = - 2 \lambda_1 / (h\hq_0)$, $\delta = 2/(h \hq_0) $ and $\sigma = 1 / (h \sqrt{\hq_0})$, we have 
\begin{equation}\label{eq:general_nu_nn}
    \nu(x) = \omega \delta(x) + \frac{1}{2 \pi \sigma^2} \left[\indi\{x > 0\}\sqrt{4 \sigma^2 - (x - m + \delta)^2} + \indi\{x < 0\}\sqrt{4 \sigma^2 - (x - m - \delta)^2}\right].
\end{equation}
The square roots are understood to be extended to $0$ when their parameter is negative.
Moreover, here $\omega = F_{\sci}((m+\delta)/\sigma) - F_{\sci}((m - \delta)/\sigma)$.
Eq.~\eqref{eq:general_nu_nn} means that the distribution $\nu(x)$ is the combination of two truncated semicircles with different centers and same width. 
One semicircle is truncated to be supported only on negative $x$, while the other is truncated to positive $x$. The possibly missing mass 
to make $\nu$ a probability distribution is put as a $\delta$ peak in $0$. Anticipating on the computation of $(\sigma, m, \delta)$,
one can refer to Fig.~\ref{subfig:nn_alpha_0.26_0.35} for examples of such a distribution.

\myskip
We can then use eq.~\eqref{eq:nu_NN_unsat} to simplify eq.~\eqref{eq:se_NN_unsat}.
In Appendix~\ref{subsec_app:derivation_se_NN_mds}, we show that this leads to the set of three equations 
on the variables $\sigma$ and $x_\pm \coloneqq (m\pm \delta)/\sigma$: 
\begin{equation}\label{eq:se_NN_mds}
    \begin{dcases}
        2 \alpha &= H_2(x_-) + 1 + x_+^2 - H_2(x_+), \\
        (x_+ + x_-) &= \sigma (x_+ - x_-) [H_1(x_-) - (x_+ - H_1(x_+))], \\
        \frac{1}{\sigma} &= H_1(x_-) + x_+ - H_1(x_+),
    \end{dcases}
\end{equation}
in which the functions $(H_k)_{k \in \{1, 2\}}$ have explicit expressions given in eqs.~\eqref{eq_app:H1},\eqref{eq_app:H2} (in appendix).

\myskip 
\textbf{The limit $\alpha \downarrow 1/4$ --}
In the limit $\alpha \downarrow 1/4$, one checks easily that the solution to eq.~\eqref{eq:se_NN_mds} is 
$x_- = 0$, $x_+ \geq 2$, and $\sigma = 3 / 4\pi$. Thus, $\omega = F_\sci(x_+) - F_\sci(x_-) \to 1/2$, and more generally 
$\nu(x) \to \mu_c(x)$ given in eq.~\eqref{eq:mu_limit}. This is coherent with the picture we discussed above: for any $\alpha < 1/4$, 
the minimal nuclear norm solution is positive semidefinite, and thus exactly at the threshold we find that the typical spectrum of this solution coincides 
with the typical spectrum of the positive semidefinite solution.

\myskip 
In Fig.~\ref{fig:nn_solution}, we summarize the behavior of the minimal nuclear norm solution, 
and compare the prediction of eq.~\eqref{eq:general_nu_nn} to finite-$d$ simulations of the minimal nuclear norm estimator, for $\alpha > 1/4$.
We find very good agreement with our predictions.

\subsubsection{The case \texorpdfstring{$\gamma > 1$}{}}
\begin{figure}[!t]
     \centering
     \begin{subfigure}[t]{0.45\textwidth}
         \centering
         \includegraphics[width=\textwidth]{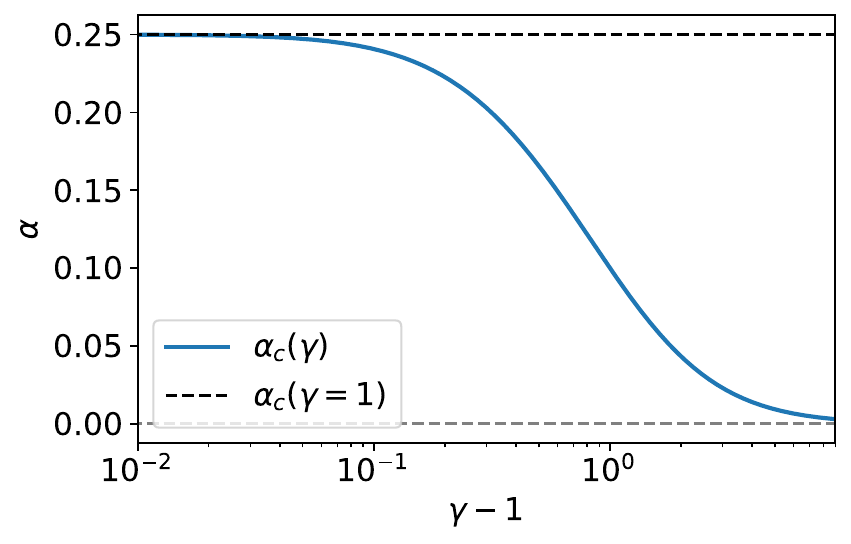}
         \caption{Evolution of $\alpha_c(\gamma)$ as a function of $\gamma$. 
         In dashed black we show the asymptotic $\alpha_c(\gamma = 1) = 1/4$.}
         \label{subfig:alphac_gamma}
     \end{subfigure}
     \hfill
     \begin{subfigure}[t]{0.54\textwidth}
         \centering
         \includegraphics[width=\textwidth]{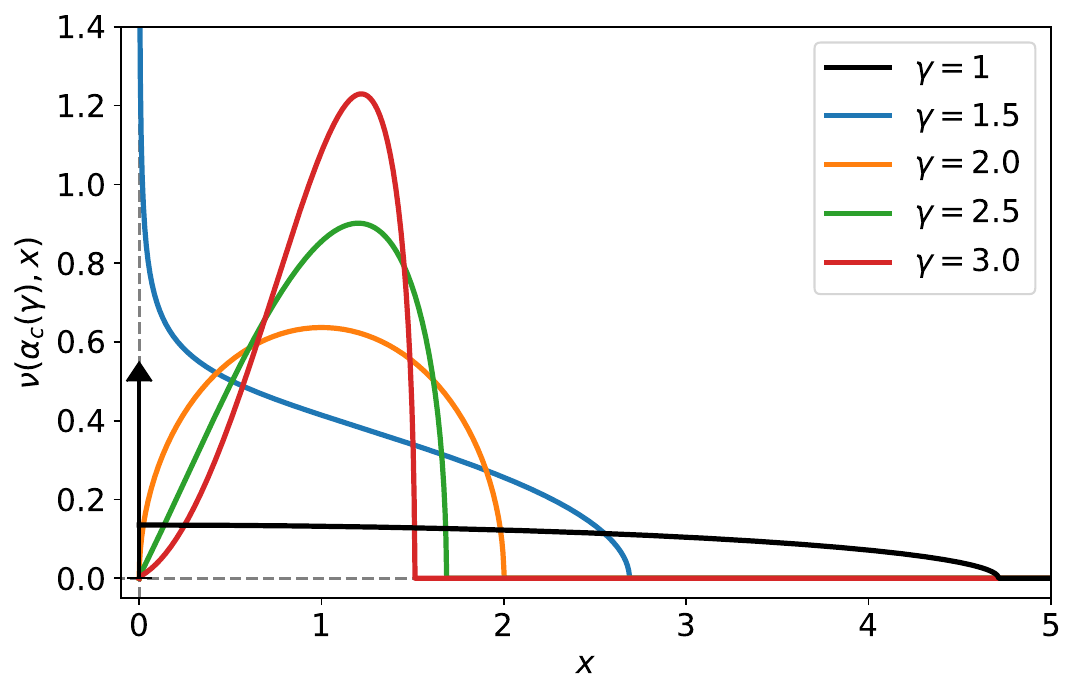}
         \caption{
        The density $\nu(x)$ for $\alpha = \alpha_c(\gamma)$, as a function of $\gamma$. 
        In the limit $\gamma  = 1$ we represent as arrows the delta peak in the density in $0$, with mass $1/2$.
         Note that for $\gamma < 2$ the density diverges in $x = 0$, but remains integrable.}
         \label{subfig:nu_alphac_gamma}
     \end{subfigure}
        \caption{Illustration of our results at the transition point $\alpha = \alpha_c(\gamma)$, 
        for the procedure $\min \|\bS\|_{S_\gamma}$. 
        We recover that $\alpha_c(\gamma = 1) = 1/4$: the minimal nuclear norm estimator is positive semidefinite in the whole SAT phase.}
        \label{fig:alphac_gamma_complete}
\end{figure}

We now consider $V(x) = |x|^\gamma$ with $\gamma > 1$. Since $V(x)$ is not linear on any open interval, we are therefore in the case~\ref{zero_entropy_gs} of Section~\ref{subsec:large_beta}.
Specializing eqs.~\eqref{eq:se_beta_inf} and \eqref{eq:se_beta_inf_spectrum} to this setting (and again changing notations $\mu \to \nu$), we find 
the following set of equations:
\begin{equation}\label{eq:min_Sgamma}
    \begin{dcases}
        \nu(x) &= \left[\frac{2 \gamma (\gamma-1)}{\sqrt{\hq_0}} |x|^{\gamma-2} + h \sqrt{\hq_0}\right] \rho_\sci\left(\frac{2\gamma}{\sqrt{\hq_0}} \frac{|x|^\gamma}{x} + h \sqrt{\hq_0} x + \frac{2\lambda_1}{\sqrt{\hq_0}}\right), \\
        \int \nu(x) \, x \, \rd x &= 1, \\
        \frac{\hq_0}{2}\int \nu(x) \, x^2 \, \rd x &= \frac{\alpha}{h^2} , \\
        \lambda_1 &= - \gamma \int \nu(x) \, |x|^\gamma \, \rd x.
    \end{dcases}
\end{equation}
One can solve eq.~\eqref{eq:min_Sgamma} numerically to obtain the three parameters $(h, \hq_0, \lambda_1)$ as a function of $\alpha$ and $\gamma$. 
Moreover, the infimum $\lambda_{\min}$ of the support of $\nu$ is the solution to 
\begin{equation*}
    \frac{2\gamma}{\sqrt{\hq_0}} \frac{|\lambda_{\min}|^\gamma}{\lambda_{\min}} + h\sqrt{\hq_0} \, \lambda_{\min} + \frac{2\lambda_1}{\sqrt{\hq_0}} = - 2.
\end{equation*}
The support of $\nu$ ceases to be a subset of $\bbR_+$ when $\lambda_{\min}$ crosses $0$, i.e.\ for $\alpha = \alpha_c(\gamma)$ we have:
\begin{equation}\label{eq:condition_threshold_Sgamma}
    \lambda_1 = - \sqrt{\hq_0}.
\end{equation}
Solving (numerically) eq.~\eqref{eq:condition_threshold_Sgamma} on top of eq.~\eqref{eq:min_Sgamma} yields the predicted threshold $\alpha_c(\gamma)$ 
such that $\nu$ is not positively supported for $\alpha > \alpha_c(\gamma)$. 
We show the behavior of the threshold, as well as the spectral density $\nu(x)$ for $\alpha = \alpha_c(\gamma)$, in Fig.~\ref{fig:alphac_gamma_complete}.

\myskip 
\textbf{Example: $\gamma = 2$ --} This corresponds to a ``least-squares'' solution $\min_{\bS \in \mcC(\bx)} \|\bS\|_F$ \cite{saunderson2013diagonal,potechin2023near}.
By eq.~\eqref{eq:min_Sgamma}, $\nu$ is a semicircular density, with mean $1 = - 2 \lambda_1 / (4 + h \hq_0)$ and variance $\sigma^2 = \hq_0 / (4 + h \hq_0)^2$.
Solving then the last three equations of eq.~\eqref{eq:min_Sgamma} easily yields, for $\alpha < 1/2$:
\begin{equation}\label{eq:complete_sol_LS}
    \sigma = \sqrt{\frac{2\alpha}{1-2\alpha}}.
\end{equation}
$\nu(x)$ is thus a semicircular law with mean $1$ and variance $2\alpha / (1 - 2\alpha)$, see Fig~\ref{subfig:nu_alphac_gamma}.
The critical value $\alpha_c(\gamma = 2)$ at which the solution ceases to be positive semidefinite can then
be computed as the solution to the equation $2 \sigma = 1$, yielding $\alpha_c(\gamma = 2) = 1/10$.
We show in Fig.~\ref{fig_app:ls_comparison_numerics} (in appendix) how the predicted spectrum compares to analytical simulations, for various values of $\alpha$. We find a very good agreement, in particular with 
our prediction $\alpha_c(\gamma = 2) = 1/10$.

\subsection{Estimators of the type \texorpdfstring{$\min \|\bS - \Id_d\|_{S_\gamma}$}{}}\label{subsec:close_identity}

\noindent
We now apply the zero-temperature analysis of Section~\ref{subsec:large_beta} to
the case $V(x) = |x-1|^\gamma$, for some $\gamma > 1$. This corresponds to the solutions to the convex optimization problem
\begin{equation*}
    \min_{\bS \in \mcC(\bx)} \|\bS - \Id_d\|_{S_\gamma},
\end{equation*}
with $\mcC(\bx) \coloneqq \{\bS \in \mcS_d \, : \, \Tr[\bW_\mu \bS] = (d-\Tr[\bS])/\sqrt{d}, \, \, \forall \mu \in [n]\}$.
Since $\gamma > 1$, $V(x)$ is not linear on any interval of $\bbR$: therefore we must be in the case~\ref{zero_entropy_gs} described in Section~\ref{subsec:large_beta}.
We now show the statements of Claim~\ref{claim:explicit_constructions} corresponding to this case, and detail the picture of Fig.~\ref{subfig:threshold_min_S_gamma_shifted}.
 
\begin{figure}[!t]
     \centering
     \begin{subfigure}[t]{0.45\textwidth}
         \centering
         \includegraphics[width=\textwidth]{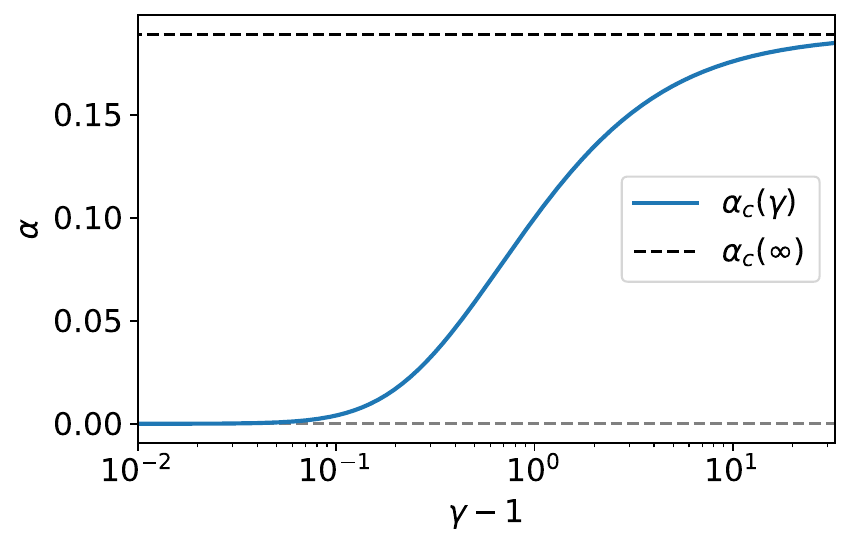}
         \caption{Evolution of $\alpha_c(\gamma)$ as a function of $\gamma$. 
         In dashed black we show the asymptotic computed in eq.~\eqref{eq:alphac_gamma_infty}.}
         \label{subfig:alphac_shift_id_gamma}
     \end{subfigure}
     \hfill
     \begin{subfigure}[t]{0.54\textwidth}
         \centering
         \includegraphics[width=\textwidth]{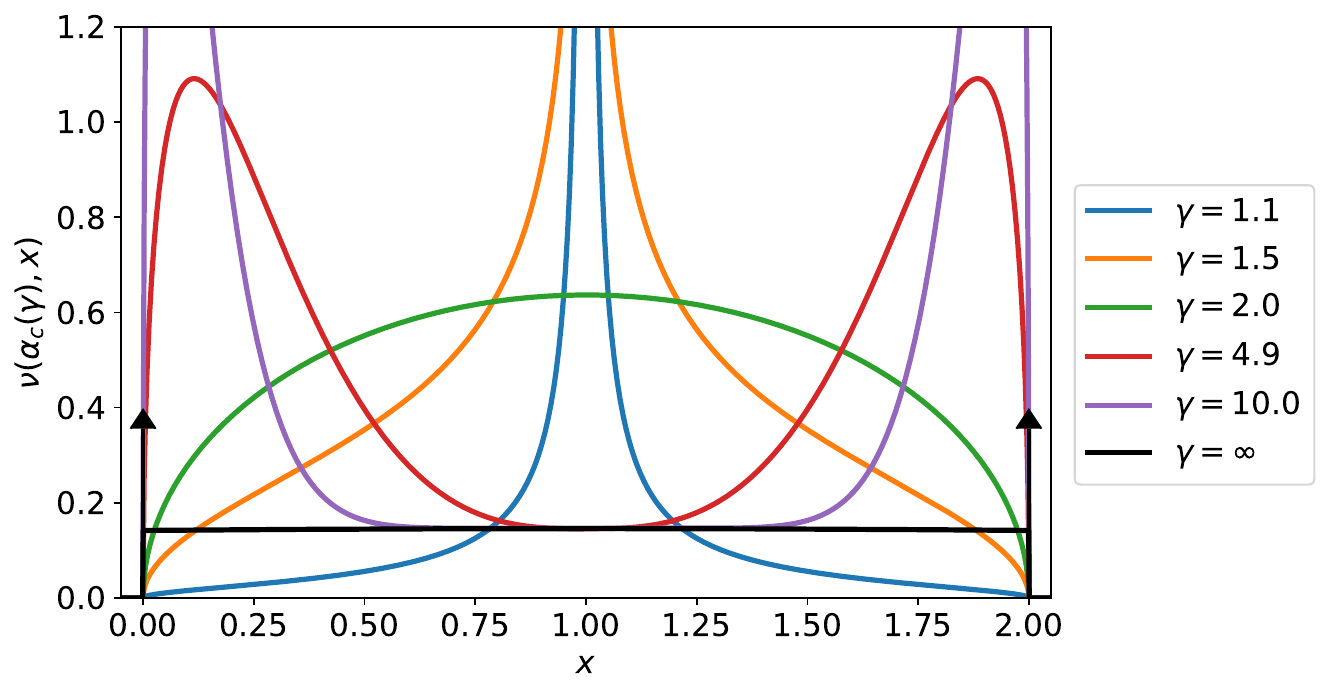}
         \caption{
            The density $\nu(x)$ for $\alpha = \alpha_c(\gamma)$, as a function of $\gamma$. 
            In the limit $\gamma \to \infty$ we represent as arrows the delta peaks in the density in $0$ and $1$.
         We notice that the black density is not uniform in $(0,2)$, but a truncation of a large semicircular law, see eq.~\eqref{eq:nu_limit_gamma_infty}.
         Note that for $\gamma < 2$ the density diverges in $x = 1$, but remains integrable.}
         \label{subfig:nu_shift_id_alphac_gamma}
     \end{subfigure}
        \caption{
        \label{fig:alphac_shift_id_gamma_complete}
        Illustration of our results for $\min \|\bS - \Id_d\|_{S_\gamma}$ at the transition point $\alpha = \alpha_c(\gamma)$.
        }
\end{figure}

\myskip
We have by eq.~\eqref{eq:se_beta_inf_spectrum}:
\begin{equation}\label{eq:F_nu_shift_identity}
    F_\nu(x) = F_{\sci}\left[ \frac{2\gamma}{\sqrt{\hq_0}} \frac{|x-1|^\gamma}{x-1} + h\sqrt{\hq_0} x + \frac{2 \lambda_1 }{\sqrt{\hq_0}}\right].
\end{equation}
Because of the form of $V(x)$, the equations~\eqref{eq:se_beta_inf} satisfied by $\nu$ are symmetric around $1$: 
if $\nu(x)$ is a solution, then $\tau(x) = \nu(2-x)$ is also a solution. 
Therefore, we will look for $\nu(x)$ as symmetric around $1$.
This implies that $\lambda_1 = -h \hq_0/2$, and eq.~\eqref{eq:se_beta_inf_4} is automatically satisfied.
In the end, we have as free parameters $h$ and $\hq_0$, and they are solutions to the equations (we define $\tnu(z) = \nu(z+1)$): 
\begin{equation}\label{eq:min_Sgamma_1}
    \begin{dcases}
    \frac{\hq_0}{2} \left(1 + \int \tnu(x) \, x^2 \, \rd x \right) &= \frac{\alpha}{h^2}, \\ 
    h \hq_0 &= 2 \gamma \int \tnu(x) |x|^\gamma \rd x, \\ 
    \tnu(x) &= \left(\frac{2 \gamma (\gamma-1)}{\sqrt{\hq_0}} |x|^{\gamma-2}+h\sqrt{\hq_0}\right)\rho_\sci\left[ \frac{2\gamma}{\sqrt{\hq_0}} \frac{|x|^\gamma}{x} + h\sqrt{\hq_0} x\right].
    \end{dcases}
\end{equation}
Eq.~\eqref{eq:min_Sgamma_1} may be rewritten as:
\begin{equation}\label{eq:min_Sgamma_2}
    \begin{dcases}
    \hq_0 &= \frac{2 \gamma^2}{\alpha} \left(1 + \int \tnu(x) x^2 \rd x\right) \left[\int \tnu(x) |x|^\gamma \rd x\right]^2, \\ 
    h &= \frac{\alpha}{\gamma} \frac{1}{\left(1 + \int \tnu(x) x^2 \rd x\right) \left[\int \tnu(x) |x|^\gamma \rd x\right]}, \\
    \tnu(x) &= \left(\frac{2 \gamma (\gamma-1)}{\sqrt{\hq_0}} |x|^{\gamma-2}+h\sqrt{\hq_0}\right)\rho_\sci\left[ \frac{2\gamma}{\sqrt{\hq_0}} \frac{|x|^\gamma}{x} + h\sqrt{\hq_0} x\right].
    \end{dcases}
\end{equation}
Eq.~\eqref{eq:min_Sgamma_2} can be solved efficiently numerically for any given $\alpha$, see Fig.~\ref{fig_app:nu_gamma_alpha_shift_identity} for examples of solutions (in appendix).
By eq.~\eqref{eq:F_nu_shift_identity}, the infimum $\widetilde{\lambda}_{\min}$ of the support of $\tnu$ is the solution to 
\begin{equation*}
    \frac{2\gamma}{\sqrt{\hq_0}} \frac{|\widetilde{\lambda}_{\min}|^\gamma}{\widetilde{\lambda}_{\min}} + h\sqrt{\hq_0} \, \widetilde{\lambda}_{\min} = - 2.
\end{equation*}
The support of $\nu$ ceases to be a subset of $\bbR_+$ when $\widetilde{\lambda}_{\min} = -1$, i.e.\ when 
\begin{equation}\label{eq:condition_threshold_Sgamma_shift_identity}
     \frac{2\gamma}{\sqrt{\hq_0}}  + h\sqrt{\hq_0} = 2. 
\end{equation}
Solving eq.~\eqref{eq:condition_threshold_Sgamma_shift_identity} on top of eq.~\eqref{eq:min_Sgamma_2} yields the predicted threshold $\alpha_c(\gamma)$ 
such that $\nu$ is not positively supported for $\alpha > \alpha_c(\gamma)$.
Using a numerical solver of these equations, we show in Fig.~\ref{fig:alphac_shift_id_gamma_complete} how $\alpha_c(\gamma)$ evolves with $\gamma$, as well as the typical spectral density of solutions 
at this threshold.

\myskip
We end our analysis of this case by describing two limits in which the above equations 
can be solved analytically, and one can get an explicit formula for $\alpha_c(\gamma)$.

\myskip 
\subsubsection{\texorpdfstring{$\gamma = 2$}{}}
In this case, one can easily show that $\nu(x)$ is a semicircular density, with mean $m = 1$, 
and variance $\sigma^2$ given by eq.~\eqref{eq:complete_sol_LS}. 
This means that the typical spectrum of this solution (which is a least-squares projection of the identity matrix) 
is exactly the spectrum of the actual least-squares solution described in Section~\ref{subsec:nn}. 
In particular, the critical value $\alpha_c(\gamma = 2)$ at which the solution ceases to be positive semidefinite is also $\alpha_c(\gamma = 2) = 1/10$.

\myskip 
\subsubsection{\texorpdfstring{The limit $\gamma \to \infty$}{}}
In this limit, the problem becomes $\min_{\bS \in \mcC(\bx)} \|\bS - \Id_d\|_\op$.
Eqs.~\eqref{eq:min_Sgamma_2} and \eqref{eq:condition_threshold_Sgamma_shift_identity} allow to fully characterize the typical asymptotic spectrum of the minimizer, as a function of 
$\alpha$, and to derive the limit $\alpha_c(\gamma = \infty)$.
Indeed, let us assume that $\alpha = \alpha_c(\gamma)$, for some large $\gamma \gg 1$, so that $\tnu$ is supported in $[-1,1]$.
By eq.~\eqref{eq:condition_threshold_Sgamma_shift_identity}, for any $x \in (-1, 1)$:
\begin{equation}\label{eq:nu_gamma}
    \tnu(x) = \left(\frac{2 \gamma (\gamma-1)}{\sqrt{\hq_0}} |x|^{\gamma-2}+2 \left[1 - \frac{\gamma}{\sqrt{\hq_0}}\right]\right)\rho_\sci\left[ \frac{2\gamma}{\sqrt{\hq_0}} \frac{|x|^\gamma}{x} + 2 \left(1 - \frac{\gamma}{\sqrt{\hq_0}}\right) x\right].
\end{equation}
Moreover, we have
$\gamma/\sqrt{\hq_0} \in (0,1)$, and
\begin{equation*}
    2 \left(1 - \frac{\gamma}{\sqrt{\hq_0}}\right) = h \sqrt{\hq_0} = \sqrt{2\alpha} \left[1 + \int \tnu(x) \, x^2 \, \rd x\right]^{-1/2} \geq \sqrt{\alpha},
\end{equation*}
since $\lambda_{\max}(\tnu) = 1$.
Thus, for any $x \in (-1, 1)$, we have as $\gamma \to \infty$ that: 
\begin{equation*}
    \frac{2\gamma}{\sqrt{\hq_0}} \frac{|x|^\gamma}{x} \ll 2 \left(1 - \frac{\gamma}{\sqrt{\hq_0}}\right) x.
\end{equation*}
Therefore, for any $x \in (-1,1)$, $\tnu(x) \sim_{\gamma \to \infty} 2 (1-\frac{\gamma}{\sqrt{\hq_0}}) \rho_\sci[2 (1-\frac{\gamma}{\sqrt{\hq_0}}) x]$.
In order for $\nu$ to have total mass equal to one while still having $\lambda_{\max}(\nu) = 1$, 
$\tnu(x)$ therefore has a delta peak in $\pm 1$, with weight compensating for the missing mass.
More precisely, let us denote $t_\gamma \coloneqq [2(1- \gamma/\sqrt{\hq_0})] =  \sqrt{2\alpha / [1 + \int \tnu(x) \, x^2 \, \rd x]}$ and $t = t_{\gamma \to \infty}$, 
then we have as $\gamma \to \infty$:
\begin{equation}\label{eq:nu_limit_gamma_infty}
   \nu(x) \to \omega(t) \{\delta(x-1) + \delta(x+1)\} + t \rho_\sci(t x) \indi\{|x| < 1\}.
\end{equation}
Here, $\omega(t) \coloneqq \int_t^2 \rho_\sci(u) \rd u$.
Note that $t \leq \sqrt{2 \alpha} \leq 1$ for $\alpha \leq 1/2$.
We can now close the equations using the first line of eq.~\eqref{eq:min_Sgamma_2}:
\begin{align}\label{eq:closure_gamma_inf_2}
    \nonumber
    t_\gamma &=  2 \left[1 - \sqrt{\frac{\alpha}{2 \left(1 + \int \tnu(x) \, x^2 \, \rd x\right)}} \frac{1}{\int \tnu(x) |x|^\gamma \rd x}\right], \\ 
    &=2 \left[1 - \frac{t_\gamma}{2 \int \tnu(x) |x|^\gamma \rd x}  \right].
\end{align}
By eq.~\eqref{eq:nu_gamma} we have (writing $\tnu_\gamma$ to explicit its dependency on $\gamma$):
\begin{equation*}
    \tnu_\gamma(x) = \left[(\gamma - 1)(2 - t_\gamma) |x|^{\gamma-2} + t_\gamma\right] \rho_\sci\left[(2-t_\gamma) \frac{|x|^\gamma}{x} + t_\gamma x\right].
\end{equation*}
Therefore, we get (recall $\alpha = \alpha_c(\gamma)$ so $\lambda_{\max}(\tnu) = 1$):
\begin{align*}
    \int \tnu_\gamma(x) |x|^\gamma \rd x &= 2 \int_0^1 \left[(\gamma - 1)(2 - t_\gamma) x^{\gamma-2} + t_\gamma\right] x^\gamma \rho_\sci\left[(2-t_\gamma) x^{\gamma-1} + t_\gamma x\right] \rd x, \\ 
    &= \frac{2}{\gamma} \int_0^1 u^{1/\gamma} [(\gamma-1)(2-t_\gamma) u^{1-2/\gamma} + t_\gamma] \rho_\sci[(2-t_\gamma)u^{1-1/\gamma} + t_\gamma u^{1/\gamma}] \rd u,
\end{align*}
using the change of variables $u = x^\gamma$. One can take the limit $\gamma \to \infty$ in this expression, and we get:
\begin{align*}
    \lim_{\gamma \to \infty} \int \tnu_\gamma(x) |x|^\gamma \rd x &= 2(2-t) \int_0^1 u \, \rho_\sci[(2-t)u + t]\, \rd u,\\
    &= \frac{1}{1-t/2} \int_t^2 (v-t) \, \rho_\sci(v)\, \rd v.
\end{align*}
Using this result in eq.~\eqref{eq:closure_gamma_inf_2} we finally reach: 
\begin{equation}\label{eq:closure_gamma_inf_3}
    t = 2 \left[1 - \frac{t}{2} \left(1 - \frac{t}{2}\right)\frac{1}{\int_t^2 (v-t) \, \rho_\sci(v)\, \rd v}\right].
\end{equation}
Eq.~\eqref{eq:closure_gamma_inf_3} can easily be solved numerically, and we obtain $t \simeq 0.4576$.
In turn, we get by eq.~\eqref{eq:nu_limit_gamma_infty}:
\begin{equation}\label{eq:alphac_gamma_infty}
    \alpha_c(\gamma = \infty) = \frac{t^2}{2} \left[1 + \int \tnu(x) x^2 \rd x\right] = \frac{t^2}{2} \left[1 + 2 \omega(t) + \frac{2}{t^2} \int_0^t \rho_\sci(u) u^2 \rd u\right]
    \simeq 0.1892,
\end{equation}
the value presented in Fig.~\ref{subfig:threshold_min_S_gamma_shifted}.
In Fig.~\ref{fig_app:shift_op_comparison_numerics} (in appendix) we compare the prediction for $\nu(x)$ as $\gamma \to \infty$ to numerical simulations of 
$\min_{\bS \in \mcC(\bx)} \|\bS - \Id_d\|_\op$, finding again good agreement with our analysis.

\subsection{Rotationally-invariant vectors with fluctuating norm}\label{subsec:rot_inv_fluctuating_norm}

\noindent
In this section we show Claim~\ref{claim:rot_inv}: we investigate the case in which $(\bx_\mu)_{\mu=1}^n$ are taken i.i.d.\ not from a standard Gaussian distribution, 
but from a more general rotationally-invariant distribution with a fluctuating norm.
We focus our discussion on the characterization of the SAT/UNSAT transition of the ellipsoid fitting problem, 
but one can also generalize all our analysis summarized in Claim~\ref{claim:explicit_constructions} to this setup.

\myskip
Recall that we consider the i.i.d.\ model for $(\bx_\mu)_{\mu=1}^n$ given by eq.~\eqref{eq:model_rot_inv}: 
\begin{equation*}
    \bx_\mu = \sqrt{\chi_\mu} \bomega_\mu.
\end{equation*}
We define $\Delta \coloneqq \sqrt{d}[\chi-1]$, so that $\EE[\Delta] = 0$ and $\tau = \EE[\Delta^2]$ (omitting the implicit large $d$ limit).
The fluctuations of the norm are of order $\Delta = \mcO(1)$:
for $\bx_\mu \sim \mcN(0, \Id_d)$ we have $\Delta \sim \mcN(0, 2)$ as $d \to \infty$.
Finally, we emphasize that we still have $\Tr[\bS] / d \simeq 1$ for all solutions to the linear constraints, by the same arguments used to derive eq.~\eqref{eq:trace_close_1}. 

\myskip
We denote $\alpha_c(\tau)$ the threshold such that ellipsoid fitting for $\bx_\mu$ given by eq.~\eqref{eq:model_rot_inv} 
is feasible for $\alpha < \alpha_c(\tau)$ and unfeasible for $\alpha > \alpha_c(\tau)$.

\subsubsection{Changes in the free entropy}

We now adapt the replica calculation of Section~\ref{sec:replica} to this more general setting, highlighting the main differences.
The partition function still has the form of eq.~\eqref{eq:partition_function}:
\begin{equation*}
    \mcZ(\beta, \bW) \coloneqq \int_{\mcS_d} \rd \bS \, e^{- \beta d \Tr[V(\bS)]} \, \prod_{\mu=1}^n \delta\left(\Tr[\bW_\mu \bS] - \frac{d - \Tr \, \bS}{\sqrt{d}}\right).
\end{equation*}
We still have $\bW_\mu = (\bx_\mu \bx_\mu^\T - \Id_d)/\sqrt{d}$, however notice that here the first two moments of $\bW_\mu$ 
do not match the one of a GOE matrix if $\tau \neq 2$! 
As before, we reach after some manipulation eq.~\eqref{eq:simplification_logZ}:
\begin{equation*}
    \Phi(\alpha, \beta) = \lim_{d \to \infty} \frac{1}{d^2} \EE \log \int \rd \bR \, \int \rd u \,\delta\left(\frac{1}{d} \Tr[\bR] - 1\right) \, e^{-\beta d \Tr[V(\bR)]} \prod_{\mu=1}^n\delta\left(\Tr[\bW_\mu \bR] + u\right).
\end{equation*}
The main difference with the Gaussian setting occurs when computing the replicated average of the constraint terms, i.e.\ the term in eq.~\eqref{eq:average_constraint}:
\begin{equation*}
    \EE_{\bW} \prod_{a=1}^r \delta\left(\Tr[\bW \bR^a] + u^a\right),
\end{equation*}
where $\{\bR^a,u^a\}_{a=1}^r$ are replicas introduced by the $r$-th power of the partition function. 
As in the Gaussian setting we introduce the overlap matrix $Q_{ab} \coloneqq (1/d) \Tr[\bR^a \bR^b]$.

\myskip
We wish to characterize the asymptotic joint distribution of $(z^a)_{a=1}^r$, with $z^a \coloneqq \Tr[\bW \bR^a]$.
We decompose (recall $\Delta = \sqrt{d}[\chi-1]$ and $\Tr[\bR^a] = d$):
\change{
\begin{equation*}
    z^a = \left[1 + \frac{\Delta}{\sqrt{d}}\right] \left[\frac{\bomega^\T \bR^a \bomega}{\sqrt{d}}\right] - \sqrt{d} = \left[1 + \frac{\Delta}{\sqrt{d}}\right] \left[\frac{\bomega^\T \bR^a \bomega - d}{\sqrt{d}}\right] + \Delta.
\end{equation*}
}
Let us denote $v^a \coloneqq (\bomega^\T \bR^a \bomega - d)/\sqrt{d} = [\bomega^\T (\bR^a - \Id_d) \bomega]/\sqrt{d}$. 
$(v^a)_{a=1}^r$ are independent of $\Delta$, and one can show (cf.\ Appendix~\ref{subsec_app:rot_inv}) that their joint distribution approaches a Gaussian as $d \to \infty$, 
with mean $0$ and variance $\EE[v^a v^b] \simeq 2 (Q_{ab} - 1)$.
In the end, as $d \to \infty$, we have 
\begin{equation*}
    z^a \simeq \Delta + y^a,
\end{equation*}
with $\by \sim \mcN(0, 2 [\bQ - \ones_r \ones_r^\T])$, independently\footnote{Notice that in the Gaussian case $\Delta \sim \mcN(0, 2)$, so we recover $\bz \sim \mcN(0, 2\bQ)$.} of $\Delta$.
We reach then the counterpart to eq.~\eqref{eq:phi_beta_r_after_average}:
\begin{align*}
    &\Phi(\alpha,\beta ; r) = \frac{1}{d^2} \log \int \prod_{a \leq b} \rd Q_{ab} \prod_{a=1}^r \rd \bR^a  \, \rd u^a \, \delta\left(\frac{1}{d} \Tr[\bR^a] - 1\right) e^{-\beta d \sum_{a=1}^r \Tr[V(\bR^a)]}
    \\ & 
    \times \prod_{a \leq b} \delta(d \Tr[\bR^a \bR^b] - d^2 Q_{ab}) 
    \times e^{-\frac{nr}{2} \log 4 \pi - \frac{n}{2} \log \det (\bQ - \ones_r \ones_r^\T)} 
    \times \left\{\EE_\Delta e^{-\frac{1}{4} (\bu + \Delta \ones_r)^\T (\bQ - \ones_r \ones_r^\T)^{-1}(\bu + \Delta \ones_r) } \right\}^n.
\end{align*}
We then perform Laplace's method as $d \to \infty$, and assume the replica symmetric ansatz, as in the Gaussian setting.
This time, we also apply Laplace's method to $(u^a)_{a=1}^r$, and assume that the supremum over $u^a$ is reached in a replica-symmetric point with $u^a = u$.
Under this assumption:
\begin{equation*}
    \begin{dcases} 
        \det(\bQ - \ones_r \ones_r^\T) &= (Q-q)^{r-1} [Q + (r-1)q - r], \\
        (\bu + \Delta\ones_r)^\T (\bQ - \ones_r \ones_r^\T)^{-1} (\bu + \Delta\ones_r) &= \frac{r(u+\Delta)^2}{Q-q} + \mcO(r^2).
    \end{dcases}
\end{equation*}
We finally get the counterpart to eq.~\eqref{eq:phi_r_final} (recall the definition of $J(\hQ, \hq)$ in eq.~\eqref{eq:J_hQ_analytic}), 
omitting additive constants:
\begin{align*}
    \Phi(\alpha,\beta;r) &= \extr_{Q, q,\hQ, \hq} \sup_{u} \Bigg[- \frac{\alpha (r-1)}{2} \log(Q-q) - \frac{\alpha}{2} \log(Q+(r-1)q - r\Big)+ \frac{r}{4} Q \hQ - \frac{r(r-1)}{4} q \hq \\ 
    \nonumber
    & + \alpha \log \EE_\Delta \left\{e^{-\frac{r(u+\Delta)^2}{4(Q-q)} + \mcO(r^2)}\right\} + J(\hQ, \hq) \Bigg].
\end{align*}
Taking now the derivative with respect to $r$ and the limit $r \to 0$ gives the free entropy as in eq.~\eqref{eq:phi_RS}:
\begin{align}\label{eq:phi_RS_fluctuating_norm}
    \nonumber
    &\Phi(\alpha,\beta) = \extr_{Q, q,\hQ, \hq} \sup_u \Bigg[- \frac{\alpha}{2} \log(Q-q) - \frac{\alpha(q-1)}{2(Q-q)} + \frac{1}{4} Q \hQ + \frac{1}{4} q \hq - \frac{\alpha (u^2 + \tau)}{4(Q-q)} + \frac{\partial}{\partial r}[J(\hQ, \hq)]_{r=0} \Bigg], \\ 
    &= \extr_{Q, q,\hQ, \hq} \Bigg[- \frac{\alpha}{2} \log (Q-q) - \frac{\alpha q}{2(Q-q)} -\frac{\alpha(\tau - 2)}{4(Q-q)} + \frac{1}{4} Q \hQ + \frac{1}{4} q \hq  + \lim_{d \to \infty} \frac{1}{d^2} \EE_{\bY} \log I(\bY) \Bigg],
\end{align}
since $\EE[\Delta] = 0$, $\EE[\Delta^2] = \tau$, and the supremum over $u$ is then clearly reached in $u = 0$.
And we recall the definition of $I(\bY)$ in eq.~\eqref{eq:phi_RS}.

\myskip 
\textbf{Final change in the free entropy --}
As one sees clearly in eq.~\eqref{eq:phi_RS_fluctuating_norm},
the final conclusion is that for rotationally-invariant vectors with fluctuating norms, the free entropy is identical to one given by eq.~\eqref{eq:phi_RS} in the Gaussian setting, 
with the addition of an extra term $-\alpha(\tau - 2) / [4(Q-q)]$.
Using the expression of $I(\bY)$ given in eq.~\eqref{eq:E_logI}, the replica equations of eq.~\eqref{eq:se_general} 
become (with $t \coloneqq q/Q$): 
\begin{subnumcases}{\label{eq:se_general_fnorm}}
    \label{eq:se_general_fnorm_1}
    \frac{Q \hq}{2} = \frac{\alpha t}{(1-t)^2} + \frac{\alpha(\tau - 2)}{2 Q(1-t)^2}, & \\ 
    \label{eq:se_general_fnorm_3}
    (1-t) Q = \frac{1}{\sqrt{\hq}} \partial_\theta I_\HCIZ(\theta = \sqrt{\hq}, \mu, \sigma_{\sci}), &\\
    \label{eq:se_general_fnorm_2}
    \hQ + t \hq = \frac{2\alpha}{Q} - \frac{\alpha(\tau-2)}{Q^2 (1-t)}, & \\
    \label{eq:se_general_fnorm_4}
    \int \mu(x) \rd x = 1, & \\ 
    \label{eq:se_general_fnorm_5}
    \int \mu(x) \, x  \, \rd x = 1, & \\
    \label{eq:se_general_fnorm_6}
    Q = \int \mu(x) \, x^2  \, \rd x, & \\
    \label{eq:se_general_fnorm_7}
    \int \mu(y) \log | x-y |\, \rd y - \beta V(x) - \frac{\hQ + \hq}{4} x^2 - \lambda_1 x - \lambda_0 + \frac{1}{2} \frac{\partial}{\partial \mu(x)}  I_\HCIZ(\sqrt{\hq}, \mu, \sigma_{\sci}) = 0.&
\end{subnumcases}
Eq.~\eqref{eq:se_general_fnorm_7} is valid for all $x \in \supp(\mu)$. 
In particular if $V(x) = +\infty$ then $x \notin \supp(\mu)$.

\myskip
\subsubsection{The SAT/UNSAT transition point}
Let us now derive the transition point $\alpha_c(\tau)$ from eq.~\eqref{eq:se_general_fnorm}, similarly to Section~\ref{subsec:sat_unsat}.
We assume that $V(x) = +\infty \times \indi\{x < 0\}$, and we pick $\beta = 1$.
As in the Gaussian setting, when $\alpha \uparrow \alpha_c(\tau)$, the convex solution space shrinks to a point: $t = q/Q \to 1$. 
Moreover, we assume again that the spectral density of solutions has a finite limit for $\alpha \uparrow \alpha_c(\tau)$.
Using the dilute limit of the HCIZ integral of eq.~\eqref{eq:dilute_expansion_hciz} in eq.~\eqref{eq:se_general_fnorm}, 
we reach that $\hq \sim \hq_0(1-t)^{-2}$, $\hQ + \hq \sim R_0 (1-t)^{-1}$, and we have in the limit $\alpha \uparrow \alpha_c(\tau)$:
\begin{subnumcases}{\label{eq:se_transition_fnorm}}
    \label{eq:se_transition_fnorm_1}
    \hq_0 = \frac{2 \alpha}{Q} + \frac{\alpha(\tau - 2)}{Q^2}, & \\ 
    \label{eq:se_transition_fnorm_3}
    Q \sqrt{\hq_0} = \int_0^1 X_\mu(p) X_\sci(p) \rd p, &\\
    \label{eq:se_transition_fnorm_2}
    R_0 = \frac{2\alpha}{Q}, & \\
    \label{eq:se_transition_fnorm_4}
    \int \mu(x) \rd x = 1, & \\ 
    \label{eq:se_transition_fnorm_5}
    \int \mu(x) \, x  \, \rd x = 1, & \\
    \label{eq:se_transition_fnorm_6}
    Q = \int \mu(x) \, x^2  \, \rd x, & \\
    \label{eq:se_transition_fnorm_7}
    F_\mu(x) = F_\sci\left[\frac{R_0}{\sqrt{\hq_0}} x + \frac{2\lambda_1}{\sqrt{\hq_0}}\right]. &
\end{subnumcases}
As in the Gaussian case, we rescaled the Lagrange multipliers $\lambda_1, \lambda_0$ as $t \to 1$.
Since eq.~\eqref{eq:se_transition_fnorm_7} is derived exactly in the same way as eq.~\eqref{eq:F_mu_transition}, we leave its derivation to the reader.
Letting $m \coloneqq - 2 \lambda_1 / R_0$ and $\sigma \coloneqq \sqrt{\hq}_0 / R_0$, we have
$F_\mu(x) = F_\sci[(x-m)/\sigma]$ with $F_\mu$ the CDF of $\mu$, i.e.\ $\mu(x)$ is a truncated semicircle density (positively supported), 
the original semicircle having variance $\sigma$ and mean $m$, and the missing mass being put as a delta peak in $x = 0$.
Moreover, by eq.~\eqref{eq:se_transition_fnorm_3} we have 
$Q \sqrt{\hq_0} = \sigma^{-1} \int \mu(x) x^2 \rd x - (m/\sigma) \int \mu(x) x \rd x = (Q - m)/\sigma$.
Therefore, $m = Q [R_0 - \hq_0] / R_0$, and thus 
\begin{equation}\label{eq:m_fluctuating_norm}
    m = 1 - \frac{\tau}{2}.
\end{equation}
Furthermore, combining eqs.~\eqref{eq:se_transition_fnorm_1} and \eqref{eq:se_transition_fnorm_2} we obtain 
\begin{equation}\label{eq:Q_fluctuating_norm}
    Q = 2 \alpha \sigma^2 + 1 - \frac{\tau}{2}.
\end{equation}
Using eq.~\eqref{eq:se_transition_fnorm_5} and eq.~\eqref{eq:m_fluctuating_norm} we reach that 
$X(\tau) \coloneqq m /\sigma$ is given by the solution to the equation: 
\begin{equation}\label{eq:X_fluctuating_norm}
    X(\tau) = \left(1 - \frac{\tau}{2}\right)  \int_{-X(\tau)}^2 \rho_\sci(v) [X(\tau)+v] \rd v.
\end{equation}
And finally by eq.~\eqref{eq:se_transition_fnorm_6} and eq.~\eqref{eq:Q_fluctuating_norm}, 
$\alpha_c(\tau)$ is given by 
\begin{equation}\label{eq:alpha_SAT_tau}
    \alpha_c(\tau) = -\frac{1}{2}\left(1 - \frac{\tau}{2}\right) \left(\int_{-X(\tau)}^2 \rho_\sci(v) [X(\tau)+v] \rd v\right)^2 + \frac{1}{2} \int_{-X(\tau)}^2 \rho_\sci(v) [X(\tau) + v]^2 \rd v.
\end{equation}
\textbf{A sanity check: $\tau = 2$ -- } When $\tau = 2$, the variance of the norm fluctuations matches the one of the Gaussian case.
In this case, the trivial solution to eq.~\eqref{eq:X_fluctuating_norm} is given by $X(2) = 0$. Plugging it into eq.~\eqref{eq:alpha_SAT_tau} 
we get $\alpha_c(2) = (1/2) \int_0^2 \rho_\sci(v) v^2 \rd v = 1/4$, consistently with our previous derivation.

\section{Concluding remarks: towards a rigorous treatment}\label{sec:conclusion}
\noindent
We conclude by discussing two approaches to leverage our non-rigorous replica analysis
as a guide towards a rigorous proof of the SAT/UNSAT transition of the ellipsoid fitting of Gaussian random vectors at $\alpha = 1/4$.

\subsection{Free entropy universality}

\noindent
A careful reading of the replica computation of Section~\ref{sec:replica} reveals
that the free entropy does not depend on the specific distribution of the matrices $\bW_\mu$: the randomness of a sample $\bW = \bW_1$ only entered 
when we considered the variables $z^a \coloneqq \Tr[\bW \bS^a]$ (for $a \in \{1,\cdots,r\}$ the replica index), and argued that by the central limit theorem, 
these variables became as $d \to \infty$ approximately Gaussian with mean $0$ and covariance $\EE[z^a z^b] = 2 Q^{ab} = 2 \Tr[\bS^a \bS^b]/d$, see eq.~\eqref{eq:average_constraint_after_clt}.

\myskip
More generally, the condition needed on the matrix $\bW$ is a \emph{uniform CLT of its low-dimensional projections}, of the type:
\begin{equation}\label{eq:1d_clt}
    (\Tr[\bW \bS^a])_{a=1}^r \sim_{d\to \infty} \mcN(0, 2 \bQ),
\end{equation}
with $Q_{ab} = 2 \Tr[\bS^a \bS^b] / d$, and which must be valid \emph{for all positive definite matrices} $\bS^1, \cdots, \bS^r$.
The condition of eq.~\eqref{eq:1d_clt} can be made mathematically precise, and similar uniform CLTs of low-dimensional projections have
been shown in previous problems to imply universality of the free entropy, see e.g.\ \cite{hu2022universality,montanari2022universality,gerace2024gaussian,dandi2024universality} and references therein. 

\myskip 
We have justified that eq.~\eqref{eq:1d_clt} holds when $\bW = (\bx \bx^\T - \Id_d)/\sqrt{d}$ (i.e.\ for ellipsoid fitting). 
Moreover, it is immediate to see that eq.~\eqref{eq:1d_clt} also holds when $\bW \sim \GOE(d)$.
However, in this latter setting, the original problem is much simpler, as the affine subspace of solutions to the linear constraints
$\mcC(\bx) \coloneqq \{\bS \in \mcS_d \, : \, \Tr[\bW_\mu \bS] = (d-\Tr[\bS])/\sqrt{d}, \, \, \forall \mu \in [n]\}$
is randomly oriented \emph{uniformly over all directions}. 
One can then use Gordon's theorem and an important existing literature on phase transitions in random convex programs \cite{gordon1988milman,amelunxen2014living} (see Section~\ref{subsec:literature})
to analyze this simpler setting, and try to leverage the free entropy universality to transpose the conclusions to the original ellipsoid fitting model.
In a companion work \cite{maillard2023fitting}, the first author and A.\ Bandeira use this idea to mathematically prove that a slightly modified version of the ellipsoid fitting problem has indeed a SAT/UNSAT transition at $\alpha = 1/4$.

\subsection{Convexity and replica symmetry}

\noindent
We have emphasized that in all the problems we considered, the solution space is convex, i.e.\ the Gibbs measure is log-concave. 
While the fact that such measures satisfy replica symmetry is folklore in the statistical physics literature,  
it is also often possible to show this implication rigorously \cite{talagrand2010mean,barbier2022strong}. 

\myskip
This might provide a more direct route to prove our replica-symmetric equations~\eqref{eq:se_general}, which could then be analyzed close to the SAT/UNSAT transition through a rigorous treatment 
of the extensive-rank HCIZ integrals, which are complicated but well-known objects in the random matrix theory literature \cite{guionnet2022rare}.

\section*{Acknowledgements}

\noindent
The authors are grateful to Afonso S.\ Bandeira, Bruno Loureiro and Florent Krzakala for insightful discussions and suggestions.

\bibliography{refs}

\begin{thebibliography}{10}

\bibitem{saunderson2011subspace}
James Saunderson.
\newblock {\em Subspace identification via convex optimization}.
\newblock PhD thesis, Massachusetts Institute of Technology, 2011.

\bibitem{saunderson2012diagonal}
James Saunderson, Venkat Chandrasekaran, Pablo~A Parrilo, and Alan~S Willsky.
\newblock Diagonal and low-rank matrix decompositions, correlation matrices,
  and ellipsoid fitting.
\newblock {\em SIAM Journal on Matrix Analysis and Applications},
  33(4):1395--1416, 2012.

\bibitem{saunderson2013diagonal}
James Saunderson, Pablo~A Parrilo, and Alan~S Willsky.
\newblock Diagonal and low-rank decompositions and fitting ellipsoids to random
  points.
\newblock In {\em 52nd IEEE Conference on Decision and Control}, pages
  6031--6036. IEEE, 2013.

\bibitem{potechin2023near}
Aaron Potechin, Paxton~M Turner, Prayaag Venkat, and Alexander~S Wein.
\newblock Near-optimal fitting of ellipsoids to random points.
\newblock In {\em The Thirty Sixth Annual Conference on Learning Theory}, pages
  4235--4295. PMLR, 2023.

\bibitem{kane2023nearly}
Daniel Kane and Ilias Diakonikolas.
\newblock A nearly tight bound for fitting an ellipsoid to gaussian random
  points.
\newblock In {\em The Thirty Sixth Annual Conference on Learning Theory}, pages
  3014--3028. PMLR, 2023.

\bibitem{hsieh2023ellipsoid}
Jun-Ting Hsieh, Pravesh~K Kothari, Aaron Potechin, and Jeff Xu.
\newblock Ellipsoid fitting up to a constant.
\newblock In {\em 50th International Colloquium on Automata, Languages, and
  Programming (ICALP 2023)}. Schloss Dagstuhl--Leibniz-Zentrum f{\"u}r
  Informatik, 2023.

\bibitem{tulsiani2023ellipsoid}
Madhur Tulsiani and June Wu.
\newblock Ellipsoid fitting up to constant via empirical covariance estimation.
\newblock {\em arXiv preprint arXiv:2307.10941}, 2023.

\bibitem{bandeira2023fitting}
Afonso~S Bandeira, Antoine Maillard, Shahar Mendelson, and Elliot Paquette.
\newblock Fitting an ellipsoid to a quadratic number of random points.
\newblock {\em arXiv preprint arXiv:2307.01181}, 2023.

\bibitem{gordon1988milman}
Yehoram Gordon.
\newblock On {M}ilman's inequality and random subspaces which escape through a
  mesh in $\mathbb{R}^n$.
\newblock In {\em Geometric Aspects of Functional Analysis: Israel Seminar
  (GAFA) 1986--87}, pages 84--106. Springer, 1988.

\bibitem{amelunxen2014living}
Dennis Amelunxen, Martin Lotz, Michael~B McCoy, and Joel~A Tropp.
\newblock Living on the edge: Phase transitions in convex programs with random
  data.
\newblock {\em Information and Inference: A Journal of the IMA}, 3(3):224--294,
  2014.

\bibitem{harish1957differential}
Harish-Chandra.
\newblock Differential operators on a semisimple {L}ie algebra.
\newblock {\em American Journal of Mathematics}, 79:87--120, 1957.

\bibitem{itzykson1980planar}
Claude Itzykson and J-B Zuber.
\newblock The planar approximation. {II}.
\newblock {\em Journal of Mathematical Physics}, 21(3):411--421, 1980.

\bibitem{matytsin1994large}
A~Matytsin.
\newblock On the large-n limit of the itzykson-zuber integral.
\newblock {\em Nuclear Physics B}, 411(2-3):805--820, 1994.

\bibitem{guionnet2002large}
Alice Guionnet and Ofer Zeitouni.
\newblock Large deviations asymptotics for spherical integrals.
\newblock {\em Journal of functional analysis}, 188(2):461--515, 2002.

\bibitem{bun2014instanton}
Joel Bun, Jean-Philippe Bouchaud, Satya~N Majumdar, and Marc Potters.
\newblock Instanton approach to large-$n$ harish-chandra-itzykson-zuber
  integrals.
\newblock {\em Physical review letters}, 113(7):070201, 2014.

\bibitem{maillard2023fitting}
Antoine Maillard and Afonso~S Bandeira.
\newblock Exact threshold for approximate ellipsoid fitting of random points.
\newblock {\em arXiv preprint arXiv:2310.05787}, 2023.

\bibitem{vershynin2018high}
Roman Vershynin.
\newblock {\em High-dimensional probability: An introduction with applications
  in data science}, volume~47.
\newblock Cambridge university press, 2018.

\bibitem{podosinnikova2019overcomplete}
Anastasia Podosinnikova, Amelia Perry, Alexander~S Wein, Francis Bach,
  Alexandre d’Aspremont, and David Sontag.
\newblock Overcomplete independent component analysis via {SDP}.
\newblock In {\em The 22nd International Conference on Artificial Intelligence
  and Statistics}, pages 2583--2592. PMLR, 2019.

\bibitem{ghosh2020sum}
Mrinalkanti Ghosh, Fernando~Granha Jeronimo, Chris Jones, Aaron Potechin, and
  Goutham Rajendran.
\newblock Sum-of-squares lower bounds for {S}herrington-{K}irkpatrick via
  planted affine planes.
\newblock In {\em 2020 IEEE 61st Annual Symposium on Foundations of Computer
  Science (FOCS)}, pages 954--965. IEEE, 2020.

\bibitem{mezard1987spin}
Marc M{\'e}zard, Giorgio Parisi, and Miguel~Angel Virasoro.
\newblock {\em Spin glass theory and beyond: An Introduction to the Replica
  Method and Its Applications}, volume~9.
\newblock World Scientific Publishing Company, 1987.

\bibitem{charbonneau2023spin}
Patrick Charbonneau, Enzo Marinari, Giorgio Parisi, Federico Ricci-tersenghi,
  Gabriele Sicuro, Francesco Zamponi, and Marc Mezard.
\newblock {\em Spin Glass Theory and Far Beyond: Replica Symmetry Breaking
  after 40 Years}.
\newblock World Scientific, 2023.

\bibitem{sourlas1989spin}
Nicolas Sourlas.
\newblock Spin-glass models as error-correcting codes.
\newblock {\em Nature}, 339(6227):693--695, 1989.

\bibitem{nishimori2001statistical}
Hidetoshi Nishimori.
\newblock {\em Statistical physics of spin glasses and information processing:
  an introduction}.
\newblock Clarendon Press, 2001.

\bibitem{montanari2007modern}
Andrea Montanari and Rudiger Urbanke.
\newblock Modern coding theory: The statistical mechanics and computer science
  point of view.
\newblock {\em Complex Systems, Les Houches lecture notes}, page~69, 2007.

\bibitem{mezard2009information}
Marc Mezard and Andrea Montanari.
\newblock {\em Information, physics, and computation}.
\newblock Oxford University Press, 2009.

\bibitem{krzakala2007landscape}
Florent Krzakala and Jorge Kurchan.
\newblock Landscape analysis of constraint satisfaction problems.
\newblock {\em Physical Review E}, 76(2):021122, 2007.

\bibitem{mezard2009constraint}
Marc M{\'e}zard and Thierry Mora.
\newblock Constraint satisfaction problems and neural networks: A statistical
  physics perspective.
\newblock {\em Journal of Physiology-Paris}, 103(1-2):107--113, 2009.

\bibitem{zdeborova2016statistical}
Lenka Zdeborov{\'a} and Florent Krzakala.
\newblock Statistical physics of inference: Thresholds and algorithms.
\newblock {\em Advances in Physics}, 65(5):453--552, 2016.

\bibitem{fyodorov2004complexity}
Yan~V Fyodorov.
\newblock Complexity of random energy landscapes, glass transition, and
  absolute value of the spectral determinant of random matrices.
\newblock {\em Physical review letters}, 92(24):240601, 2004.

\bibitem{maillard2020landscape}
Antoine Maillard, G{\'e}rard~Ben Arous, and Giulio Biroli.
\newblock Landscape complexity for the empirical risk of generalized linear
  models.
\newblock In {\em Mathematical and Scientific Machine Learning}, pages
  287--327. PMLR, 2020.

\bibitem{aubin2019committee}
Benjamin Aubin, Antoine Maillard, Jean Barbier, Florent Krzakala, Nicolas
  Macris, and Lenka Zdeborov{\'a}.
\newblock The committee machine: computational to statistical gaps in learning
  a two-layers neural network.
\newblock {\em Journal of Statistical Mechanics: Theory and Experiment},
  2019(12):124023, 2019.

\bibitem{gabrie2020mean}
Marylou Gabri{\'e}.
\newblock Mean-field inference methods for neural networks.
\newblock {\em Journal of Physics A: Mathematical and Theoretical},
  53(22):223002, 2020.

\bibitem{mannelli2020marvels}
Stefano~Sarao Mannelli, Giulio Biroli, Chiara Cammarota, Florent Krzakala,
  Pierfrancesco Urbani, and Lenka Zdeborov{\'a}.
\newblock Marvels and pitfalls of the langevin algorithm in noisy
  high-dimensional inference.
\newblock {\em Physical Review X}, 10(1):011057, 2020.

\bibitem{guerra2002thermodynamic}
Francesco Guerra and Fabio~Lucio Toninelli.
\newblock The thermodynamic limit in mean field spin glass models.
\newblock {\em Communications in Mathematical Physics}, 230:71--79, 2002.

\bibitem{talagrand2006parisi}
Michel Talagrand.
\newblock The parisi formula.
\newblock {\em Annals of mathematics}, pages 221--263, 2006.

\bibitem{panchenko2013parisi}
Dmitry Panchenko.
\newblock The parisi ultrametricity conjecture.
\newblock {\em Annals of Mathematics}, pages 383--393, 2013.

\bibitem{auffinger2013random}
Antonio Auffinger, Gerard Ben~Arous, and Jiri Cerny.
\newblock Random matrices and complexity of spin glasses.
\newblock {\em Communications on Pure and Applied Mathematics}, 66(2):165--201,
  2013.

\bibitem{subag2017complexity}
Eliran Subag.
\newblock The complexity of spherical p-spin models-a second moment approach.
\newblock {\em Annals of Probability}, 45(5):3385--3450, 2017.

\bibitem{barbier2019optimal}
Jean Barbier, Florent Krzakala, Nicolas Macris, L{\'e}o Miolane, and Lenka
  Zdeborov{\'a}.
\newblock Optimal errors and phase transitions in high-dimensional generalized
  linear models.
\newblock {\em Proceedings of the National Academy of Sciences},
  116(12):5451--5460, 2019.

\bibitem{montanari2016semidefinite}
Andrea Montanari and Subhabrata Sen.
\newblock Semidefinite programs on sparse random graphs and their application
  to community detection.
\newblock In {\em Proceedings of the forty-eighth annual ACM symposium on
  Theory of Computing}, pages 814--827, 2016.

\bibitem{javanmard2016phase}
Adel Javanmard, Andrea Montanari, and Federico Ricci-Tersenghi.
\newblock Phase transitions in semidefinite relaxations.
\newblock {\em Proceedings of the National Academy of Sciences},
  113(16):E2218--E2223, 2016.

\bibitem{khot2012grothendieck}
Subhash Khot and Assaf Naor.
\newblock Grothendieck-type inequalities in combinatorial optimization.
\newblock {\em Communications on Pure and Applied Mathematics}, 65:0992--1035,
  2012.

\bibitem{talagrand2010mean}
Michel Talagrand.
\newblock {\em Mean field models for spin glasses: Volume I: Basic examples},
  volume~54.
\newblock Springer Science \& Business Media, 2010.

\bibitem{boucheron2013concentration}
St{\'e}phane Boucheron, G{\'a}bor Lugosi, and Pascal Massart.
\newblock {\em Concentration inequalities: A nonasymptotic theory of
  independence}.
\newblock Oxford university press, 2013.

\bibitem{guionnet2022rare}
Alice Guionnet.
\newblock Rare events in random matrix theory.
\newblock In {\em Proceedings of ICM}, 2022.

\bibitem{github_repo}
Antoine Maillard and Dmitriy Kunisky.
\newblock Numerical code used to produce the figures.
\newblock \url{https://github.com/AnMaillard/fitting_ellipsoid_replicas}, 2023.

\bibitem{anderson2010introduction}
Greg~W Anderson, Alice Guionnet, and Ofer Zeitouni.
\newblock {\em An introduction to random matrices}.
\newblock Cambridge university press, 2010.

\bibitem{livan2018introduction}
Giacomo Livan, Marcel Novaes, and Pierpaolo Vivo.
\newblock {\em Introduction to Random Matrices: Theory and Practice},
  volume~26.
\newblock Springer, 2018.

\bibitem{arous1997large}
G\'erard Ben~Arous and Alice Guionnet.
\newblock Large deviations for wigner's law and voiculescu's non-commutative
  entropy.
\newblock {\em Probability theory and related fields}, 108:517--542, 1997.

\bibitem{voiculescu1993analogues}
Dan Voiculescu.
\newblock The analogues of entropy and of fisher's information measure in free
  probability theory, i.
\newblock {\em Communications in mathematical physics}, 155(1):71--92, 1993.

\bibitem{guionnet2005fourier}
Alice Guionnet, M~Ma{\i}, et~al.
\newblock A fourier view on the r-transform and related asymptotics of
  spherical integrals.
\newblock {\em Journal of functional analysis}, 222(2):435--490, 2005.

\bibitem{collins2007new}
Benoit Collins and Piotr {\'S}niady.
\newblock New scaling of itzykson--zuber integrals.
\newblock In {\em Annales de l'Institut Henri Poincare (B) Probability and
  Statistics}, volume~43, pages 139--146. Elsevier, 2007.

\bibitem{maillard2022perturbative}
Antoine Maillard, Florent Krzakala, Marc M{\'e}zard, and Lenka Zdeborov{\'a}.
\newblock Perturbative construction of mean-field equations in extensive-rank
  matrix factorization and denoising.
\newblock {\em Journal of Statistical Mechanics: Theory and Experiment},
  2022(8):083301, 2022.

\bibitem{troiani2022optimal}
Emanuele Troiani, Vittorio Erba, Florent Krzakala, Antoine Maillard, and Lenka
  Zdeborov{\'a}.
\newblock Optimal denoising of rotationally invariant rectangular matrices.
\newblock In {\em Mathematical and Scientific Machine Learning}, pages 97--112.
  PMLR, 2022.

\bibitem{pourkamali2023matrix}
Farzad Pourkamali, Jean Barbier, and Nicolas Macris.
\newblock Matrix inference in growing rank regimes.
\newblock {\em arXiv preprint arXiv:2306.01412}, 2023.

\bibitem{hu2022universality}
Hong Hu and Yue~M Lu.
\newblock Universality laws for high-dimensional learning with random features.
\newblock {\em IEEE Transactions on Information Theory}, 69(3):1932--1964,
  2022.

\bibitem{montanari2022universality}
Andrea Montanari and Basil~N Saeed.
\newblock Universality of empirical risk minimization.
\newblock In {\em Conference on Learning Theory}, pages 4310--4312. PMLR, 2022.

\bibitem{gerace2024gaussian}
Federica Gerace, Florent Krzakala, Bruno Loureiro, Ludovic Stephan, and Lenka
  Zdeborov{\'a}.
\newblock Gaussian universality of perceptrons with random labels.
\newblock {\em Physical Review E}, 109(3):034305, 2024.

\bibitem{dandi2024universality}
Yatin Dandi, Ludovic Stephan, Florent Krzakala, Bruno Loureiro, and Lenka
  Zdeborov{\'a}.
\newblock Universality laws for gaussian mixtures in generalized linear models.
\newblock {\em Advances in Neural Information Processing Systems}, 36, 2024.

\bibitem{barbier2022strong}
Jean Barbier, Dmitry Panchenko, and Manuel S{\'a}enz.
\newblock Strong replica symmetry for high-dimensional disordered log-concave
  gibbs measures.
\newblock {\em Information and Inference: A Journal of the IMA},
  11(3):1079--1108, 2022.

\end{thebibliography}

\newpage
\appendix 
\addtocontents{toc}{\protect\setcounter{tocdepth}{1}} %Only sections titles in the TOC for the appendix
\section{Derivation of \texorpdfstring{eq.~\eqref{eq:almost_exact_fit}}{}}\label{subsec_app:almost_exact_fit}

\change{
    Let $\Delta_\mu \coloneqq \sqrt{d}(\|\bx_\mu\|^2/d - 1)$. 
    Then we have $\EE[\Delta_\mu] = 0$, and by a direct computation we get 
    \begin{align*}
        \begin{dcases}
            \EE[\Delta_\mu^2] &= 2, \\
            \mathrm{Var}[\Delta_\mu^2] &= 8 + \frac{48}{d} \leq 9.
        \end{dcases}
    \end{align*}
    Therefore, by Chebyshev's inequality and the independence of the $\Delta_\mu$'s, we have for any $t > 0$: 
    \begin{align*}
        \bbP\left[\left|\frac{d}{n} \sum_{\mu=1}^n \left(\frac{\|\bx_\mu\|^2}{d}-1\right)^2 - 2 \right| \geq t \right] 
        = 
        \bbP\left[\left|\frac{1}{n} \sum_{\mu=1}^n \Delta_\mu^2 - 2 \right| \geq t \right] 
        &\leq \frac{81}{n^2 t^2}.
    \end{align*}
    Taking e.g.\ $t = 0.5$ ends the proof of eq.~\eqref{eq:almost_exact_fit}.
}

\section{Derivation of \texorpdfstring{eq.~\eqref{eq:interpretation_Q_q}}{}}\label{subsec_app:interpretation_q_Q}

\noindent
We compute the stationary equations on $\hQ,\hq$ from eq.~\eqref{eq:phi_RS}. Recall the definition of the effective Gibbs measure of eq.~\eqref{eq:def_effective_gibbs}. 
We get from the equation on $\hQ$: 
\begin{equation*}
    Q = - \frac{4}{d^2} \EE_\bY \left[\frac{\partial_{\hQ} I(\bY)}{I(\bY)}\right] = \frac{1}{d} \EE \langle \Tr[\bR^2] \rangle_\eff.
\end{equation*}
From the equation on $\hq$ we have, using integration by parts on the Gaussian variables $Y_{ij}$
\begin{align*}
    q &= - \frac{4}{d^2} \EE_\bY \left[\frac{\partial_{\hq} I(\bY)}{I(\bY)}\right], \\ 
    &= \frac{1}{d} \EE \langle \Tr[\bR^2] \rangle_\eff - \frac{1}{d \sqrt{\hq}}  \EE \langle \Tr[\bR \bY] \rangle_\eff, \\
    &= \frac{1}{d} \EE \langle \Tr[\bR^2] \rangle_\eff - \frac{2}{d \sqrt{\hq}} \Bigg(\sum_{i < j} \EE[Y_{ij} \langle R_{ij} \rangle_\eff] + \frac{1}{2}\sum_{i} \EE[Y_{ii} \langle R_{ii} \rangle_\eff] \Bigg), \\
    &= \frac{1}{d} \EE \langle \Tr[\bR^2] \rangle_\eff - \frac{2}{d} \Bigg(\sum_{i < j} \EE[\langle R_{ij}^2 \rangle_\eff - \langle R_{ij} \rangle_\eff^2] + \frac{1}{2}\sum_{i} \EE[\langle R_{ii}^2 \rangle_\eff - \langle R_{ii} \rangle_\eff^2] \Bigg), \\
    &= \frac{1}{d} \EE \Tr[\langle \bR \rangle_\eff^2].
\end{align*}

\section{Asymptotics of the HCIZ integral}\label{subsec_app:hciz}

\noindent
We provide here the exact asymptotic expression of $I_\HCIZ(\theta, \rho_A, \rho_B)$.
It was first studied by Matytsin \cite{matytsin1994large} and later proven by Guionnet and Zeitouni \cite{guionnet2002large}.
Note that by rescaling $\rho_A$, one can assume without loss of generality that $\theta = 1$.
We state the theorem informally, and we refer e.g.\ to \cite{guionnet2022rare} (Theorem~3.1) for a mathematically complete account of the theorem.

\begin{theorem}[Extensive-rank HCIZ integral -- informal \protect\cite{matytsin1994large,guionnet2002large}]\label{thm:hciz_extensive_rank}
    \noindent
   Let $d \geq 1$, and $\bA,\bB \in \mcS_d$.
   We assume that the empirical spectral distributions of $\bA$ and $\bB$ both converge as $d \to \infty$
   to probability measures $\rho_A, \rho_B \in \mathcal{M}_1^+(\bbR)$. Then:
   \begin{align*}
      I_\HCIZ(1, \rho_A, \rho_B) &\coloneqq \lim_{d \to \infty} \frac{2}{d^2} \log \int_{\mcO(d)} \mcD\bO \, \exp \Big\{\frac{d}{2} \mathrm{Tr}[\bA \bO \bB \bO^\dagger]\Big\} \\ 
      &= - \frac{3}{4} + J(\rho_A) + J(\rho_B) - \frac{1}{2} \int_0^1  {\rm d}t \int {\rm d}x \, \rho_t(x) \Big[v_t(x)^2 + \frac{\pi^2}{3} \rho_t(x)^2\Big].
   \end{align*}
   Here we have defined:
   \begin{equation*}
      J(\rho) \coloneqq \frac{1}{2} \int \rho(x)\, x^2 \, \rd x - \frac{1}{2} \iint \rho(x) \rho(y) \log |x-y| \, \rd x \, \rd y.
   \end{equation*}
   Moreover, $f(t, x) \coloneqq v_t(x) + i \pi \rho_t(x)$ satisfies a complex Burgers' equation with prescribed boundary conditions: 
   \begin{equation*}
    \begin{dcases}
        \partial_t f + f \partial_x f &= 0, \\
        f(t = 0, x) &= \rho_A(x), \\
        f(t = 1, x) &= \rho_B(x).
    \end{dcases}
   \end{equation*}
\end{theorem}

\section{On the concentration of the spectral density of \texorpdfstring{$\bS$}{}}\label{subsec_app:concentration_spectrum}

\change{
We give here a more precise argument on how the replica computation in the main text also allows to deduce the statements of Claims~\ref{claim:solution_space} and \ref{claim:explicit_constructions} 
concerning the asymptotic spectral density of ellipsoid fits. 
Recall that our main computation is based on a non-rigorous analytical method,
so we give this argument at a mathematically informal level, and do not claim that it forms a formal proof.
Throughout this appendix, we use the notation $\tr(\bS) \coloneqq (1/d) \Tr[\bS]$ for the normalized trace.
}

\myskip
\change{
We fix a bounded Lipschitz function $f : \bbR \to \bbR$.
First, let us notice that
one can immediately repeat the replica computation that gave rise to eq.~\eqref{eq:phi_RS} by adding a term $\exp\{\lambda d^2 \, \tr f(\bS)\}$ in the Gibbs measure of eq.~\eqref{eq:def_Gibbs}, 
which will in turn add a term $e^{\lambda d^2 \tr f(\bR)}$ in the definition of $I(\bY)$.
Indeed, recall that we had defined $\bS = (1+d^{-1/2} u) \bR$, with $u = \mcO(1)$, so the spectrums of $\bS$ and $\bR$ coincide at leading order.
Moreover, the asymptotic expression of $(1/d^2) \EE_\bY \log I(\bY)$ in eq.~\eqref{eq:E_logI} can also be 
straightforwardly generalized to this modification by adding an extra term $+\lambda \int f(x) \mu(\rd x)$ to the variational problem.
Comparing the first and second derivatives at $\lambda = 0$ of the free entropy and of the modification of eq.~\eqref{eq:E_logI} shows that:
\begin{align}\label{eq:limit_spectrum_S}
        \begin{dcases}
        \lim_{d\to \infty}\EE_\bW [\EE_{\bS \sim \bbP_{\beta,\bW}}(\tr f(\bS))] = \int f(x) \mu(x) \rd x, \\ 
        \EE_\bW [\EE_{\bS \sim \bbP_{\beta,\bW}}((\tr f(\bS))^2)] - \EE_\bW [\EE_{\bS \sim \bbP_{\beta,\bW}}(\tr f(\bS))]^2 
        &= \mcO\left(\frac{1}{d^2}\right).
    \end{dcases}
\end{align}
By Chebyshev's inequality, for $\bS \sim \EE_{\bW}[\bbP_{\beta,\bW}(\cdot)]$, eq.~\eqref{eq:limit_spectrum_S} implies 
that $\tr f(\bS) \to_{d \to \infty} \int f(x) \mu(\rd x)$, in probability. Since this is true for all test functions $f$, 
the empirical spectral density of $\bS$ thus converges weakly (in probability) to $\mu$.
}

\section{Non-commutative entropy of the eigenvalue density near the SAT/UNSAT threshold}\label{subsec_app:nc_entropy_near_sat_unsat}

\noindent
We assume that the delta peak in $0$ develops polynomially in $(1-t)^{-1}$ -- a natural assumption as the terms in eq.~\eqref{eq:se_general_7} grow polynomially in $(1-t)^{-1}$ --, 
i.e.\ that we have 
\begin{equation*}
    \mu(z) \simeq \mu_{\textrm{cont.}}(z) + (1-t)^{-\tau} \mu_{\textrm{sing.}}[(1-t)^{-\tau} z], 
\end{equation*}
with $\tau > 0$, so that the second part goes to \change{$\mu_{\textrm{sing.}}(0) \delta(z)$}.
Then we have as $t \to 1$: 
\begin{equation*}
    \Sigma(\mu) \simeq \Sigma(\mu_\textrm{cont.}) + \Sigma(\mu_\textrm{sing.}) + 2 \int \mu_\textrm{cont.}(z) \log |z| \rd z + \tau \log(1-t).
\end{equation*}

\section{\texorpdfstring{Derivation of eq.~\eqref{eq:se_NN_mds}}{}}\label{subsec_app:derivation_se_NN_mds}

\noindent
We define, for $k \geq 0$ and $u \in \bbR$: 
\begin{equation*}
    H_k(u) \coloneqq \int_{-2}^{\min(2, u)} \rho_\sci(y) \, (u - y)^k \, \rd y .
\end{equation*}
Notice that $H_0 = F_{\sci}$.
The functions $(H_k)$ have the following expression for $k \in \{1, 2\}$:
\begin{align}
    \label{eq_app:H1}
    H_1(u) &= 
    \begin{dcases} 
        0 & \textrm{ if } u \leq -2, \\ 
        u & \textrm{ if } u \geq 2, \\
        \frac{(8+u^2)\sqrt{4 - u^2} + 24 u \arctan \left[\frac{2 + u}{\sqrt{4-u^2}}\right]}{12 \pi}
    &\textrm{ otherwise}.
    \end{dcases} \\
    \label{eq_app:H2}
    \change{H_2(u)} &= 
    \begin{dcases} 
        0 & \textrm{ if } u \leq -2, \\ 
        1 + u^2 & \textrm{ if } u \geq 2, \\
        \frac{u(26+u^2)\sqrt{4 - u^2} + 48 (1+u^2) \arctan \left[\frac{2 + u}{\sqrt{4-u^2}}\right]}{24 \pi}
         & \textrm{ otherwise}.
    \end{dcases}.
\end{align}
Let us now show how eq.~\eqref{eq:se_NN_unsat} leads to eq.~\eqref{eq:se_NN_mds}.
\begin{itemize}
    \item \textbf{Eq.~\eqref{eq:se_NN_unsat_1} --}
    We have 
    \begin{align*}
       &\frac{\indi\{m-\delta+2\sigma \geq 0\}}{2 \pi \sigma^2} \int_{0}^{m - \delta + 2 \sigma} \!\! x^2 \, \sqrt{4 \sigma^2 - (x - m + \delta)^2} \, \rd x \\
       &= \indi\{m-\delta+2\sigma \geq 0\} \sigma^2 H_2\left(\frac{m-\delta}{\sigma}\right) = \sigma^2 H_2\left(\frac{m-\delta}{\sigma}\right),
    \end{align*}
    since $H_2(u) = 0$ for $u \leq -2$.
    And similarly:
    \begin{align*}
       &\frac{\indi\{m+\delta-2\sigma \leq 0\}}{2 \pi \sigma^2} \int_{m + \delta - 2 \sigma}^0 \, x^2 \, \sqrt{4 \sigma^2 - (x - m - \delta)^2} \, \rd x \\
       &= \indi\{m+\delta-2\sigma \leq 0\} \sigma^2 \left[1 + \left(\frac{m+\delta}{\sigma}\right)^2 - H_2\left(\frac{m+\delta}{\sigma}\right) \right], \\ 
       &= \sigma^2 \left[1 + \left(\frac{m+\delta}{\sigma}\right)^2 - H_2\left(\frac{m+\delta}{\sigma}\right) \right],
    \end{align*}
    since $H_2(u) = 1 + u^2$ for $u \geq 2$.
    Thus, we get 
    \begin{equation*}
        \frac{1}{\sigma^2}\int \nu(x) \, x^2 \, \rd x = H_2\left(\frac{m-\delta}{\sigma}\right) + 1 + \left(\frac{m+\delta}{\sigma}\right)^2 - H_2\left(\frac{m+\delta}{\sigma}\right).
    \end{equation*}
    As $\sigma^{-1} = h \sqrt{\hq_0}$, we have by eq.~\eqref{eq:se_NN_unsat_1} that $\int \nu(x) x^2 \rd x = 2 \alpha \sigma^2$, giving the first equation of eq.~\eqref{eq:se_NN_mds}.
    \item \textbf{Eq.~\eqref{eq:se_NN_unsat_2} --}  
    Let us denote $Q \coloneqq \int \nu(x) \, x^2 \, \rd x = 2\alpha \sigma^2$.
    By changing variables to $z = X_\nu(p)$:
    \begin{align*}
        Q h \sqrt{\hq_0} &= \int X_\sci[F_\nu(z)] \, z \, \nu(z)  \rd z, \\ 
        &= \frac{1}{\sigma} \int [z - m + \delta \sign(z)] z \nu(z)  \rd z, \\ 
        &= \frac{Q - m + \delta \int \nu(z) |z| \rd z}{\sigma},
    \end{align*}
    using eq.~\eqref{eq:se_NN_unsat_3} and the definition of $Q$.
    Notice that 
    \begin{align*}
       \frac{\indi\{m-\delta+2\sigma \geq 0\}}{2 \pi \sigma^2} \int_{0}^{m - \delta + 2 \sigma} \! \! x \, \sqrt{4 \sigma^2 - (x - m + \delta)^2} \, \rd x 
       &= \indi\{m-\delta+2\sigma \geq 0\} \sigma H_1\left(\frac{m-\delta}{\sigma}\right), \\ 
       &= \sigma H_1\left(\frac{m-\delta}{\sigma}\right),
    \end{align*}
    since $H_1(u) = 0$ for $u \leq -2$.
    In the same way, we can compute
    \begin{align*}
       &\frac{\indi\{m+\delta-2\sigma \leq 0\}}{2 \pi \sigma^2} \int_{m + \delta - 2 \sigma}^0 \!\! x \, \sqrt{4 \sigma^2 - (x - m - \delta)^2} \, \rd x \\
       &= \indi\{m+\delta-2\sigma \leq 0\} \sigma \left[\frac{m+\delta}{\sigma} - H_1\left(\frac{m+\delta}{\sigma}\right) \right], \\ 
       &= \sigma \left[\frac{m+\delta}{\sigma} - H_1\left(\frac{m+\delta}{\sigma}\right) \right],
    \end{align*}
    since $H_1(u) = u$ for $u \geq 2$.
    Thus, we get:
    \begin{equation*}
        \int \nu(z) |z| \rd z = \sigma \left[-\frac{m+\delta}{\sigma} + H_1\left(\frac{m+\delta}{\sigma}\right) + H_1\left(\frac{m-\delta}{\sigma}\right) \right].
    \end{equation*}
    Recall that we defined $\sigma^{-1} = h \sqrt{\hq_0}$, so that we reach the second equation of eq.~\eqref{eq:se_NN_mds}.
    \item \textbf{Eq.~\eqref{eq:se_NN_unsat_3} --} 
    By the same arguments as above, we easily get from eq.~\eqref{eq:se_NN_unsat_3} the last equation of eq.~\eqref{eq:se_NN_mds}.
\end{itemize}

\section{Technicalities of Section~\ref{subsec:rot_inv_fluctuating_norm}}
\label{subsec_app:rot_inv}

\noindent
Let us justify that if $v^a \coloneqq \bomega^\T \bM^a \bomega / \sqrt{d}$, with $\Tr[\bM^a] = 0$ and $\Tr[\bM^a \bM^b]/d = \sigma_{ab}$, 
then $\bv \deq \mcN(0, 2 \bsigma)$ as $d \to \infty$.
Applying it to $\bM^a = \bR^a - \Id_d$ will then yield our claim, since $\Tr[\bR^a] = d$.
We simply notice that $\bomega \deq  \sqrt{d} \, \bg / \|\bg\|$, with $\bg \sim \mcN(0, \Id_d)$, so that 
\begin{equation*}
    v^a \deq \sqrt{d} \, \frac{\bg^\T \bM^a \bg}{\|\bg\|^2}.
\end{equation*}
It is well-known \change{that} the squared norm of $\bg$ concentrates as $\|\bg\|^2 = d + \mcO(\sqrt{d})$, where the leading order term in $\mcO(\sqrt{d})$ are given by Gaussian fluctuations.
Moreover, by the same computation made in Section~\ref{sec:replica}, and since $\Tr[\bM^a] = 0$, 
we know that as $d \to \infty$: $(\bg^\T \bM^a \bg / \sqrt{d})_{a=1}^r \sim \mcN(0, 2 \bsigma)$.
Thus, at leading order we also reach that $v^a \sim \mcN(0, 2\bsigma)$ as $d \to \infty$.

\section{\texorpdfstring{Derivation of eq.~\eqref{eq:alphac_lstar_kappa}}{}}\label{subsec_app:kappa}

\noindent
Let us first rewrite eqs.~\eqref{eq:F_mu_transition},\eqref{eq:constraints_spectrum_transition}:
\begin{equation}\label{eq:constraints_spectrum_transition_kappa}
    \begin{dcases}
        F_\mu(x) &= F_{\sci}\left(\frac{x-m}{\sigma}\right),\\
        \int \mu(x) \, x \, \rd x &= 1, \\
        \int \mu(x) \, x^2 \, \rd x &= 2 \alpha \sigma^2, \\
        2 \alpha \int \mu(x) \, x^2 \, \rd x &= \left(\int_0^1 X_\mu(p) X_\sci(p) \rd p\right)^2.
    \end{dcases}
\end{equation}
In other words, $\mu(x)$ is a semicircular density with mean $m$ and variance $\sigma^2$, 
truncated so that its support is included in $[\kappa, \infty)$, and the missing mass put as a delta peak in $x = \kappa$.
Defining the functions
\begin{equation*}
    G_k(r) \coloneqq \int_{-2}^r \rho_\sci(u) \, u^k \, \rd u,
\end{equation*}
we then have $\mu(x) = G_0[(\kappa-m)/\sigma] \delta(x-\kappa) + \sigma^{-1} \rho_\sci[(x-m)/\sigma] \indi\{x \geq \kappa\}$, 
and one computes
\begin{align*}
    &\int \mu(x) \, x \, \rd x = m + (\kappa-m) G_0\left[\frac{\kappa-m}{\sigma}\right] - \sigma G_1\left[\frac{\kappa-m}{\sigma}\right],  \\ 
    &\int \mu(x) \, x^2 \, \rd x = m^2 + \sigma^2 + (\kappa^2 - m^2) G_0\left[\frac{\kappa-m}{\sigma}\right] - 2 m \sigma G_1\left[\frac{\kappa-m}{\sigma}\right] - \sigma^2 G_2\left[\frac{\kappa-m}{\sigma}\right], \\
    &\int_0^1 \! X_\mu(p) X_\sci(p) \rd p = \kappa G_1\left[\frac{\kappa-m}{\sigma}\right] + \frac{1}{\sigma} \left(\int \! \mu(x) x^2 \rd x - m \int \! \mu(x) x \rd x\right) - \frac{\kappa(\kappa - m)}{\sigma} G_0\left[\frac{\kappa-m}{\sigma}\right].
\end{align*}
The first term of the third equation is obtained by recalling that $X_\mu(p) = \kappa$ for any $p \leq  G_0[(\kappa-m)/\sigma]$.
Plugging these equations back into eq.~\eqref{eq:constraints_spectrum_transition_kappa}, 
we get easily that 
\begin{equation*}
    m = -\frac{\kappa}{1 - \kappa},
\end{equation*}
\change{so} that $\sigma$ is the solution to: 
\begin{equation*}
    1 = m + (\kappa - m) G_0\left[\frac{\kappa-m}{\sigma}\right] - \sigma G_1\left[\frac{\kappa-m}{\sigma}\right],
\end{equation*}
and finally $\alpha = \alpha_c(\kappa)$ is given by 
\begin{equation*}
    \alpha_c(\kappa) = \frac{m^2 + \sigma^2}{2\sigma^2} + \frac{\kappa^2 - m^2}{2\sigma^2} G_0\left[\frac{\kappa-m}{\sigma}\right] - \frac{m}{\sigma} G_1\left[\frac{\kappa-m}{\sigma}\right] - \frac{1}{2} G_2\left[\frac{\kappa-m}{\sigma}\right].
\end{equation*}
This ends the derivation of eq.~\eqref{eq:alphac_lstar_kappa}.
Using analytical expressions for $G_k(r)$ -- which are readily obtained similarly as for $H_k(r)$ in eqs.~\eqref{eq_app:H1},\eqref{eq_app:H2} -- 
eq.~\eqref{eq:alphac_lstar_kappa} can easily be solved numerically for $(m, \sigma, \alpha_c)$ as a function of $\kappa$.

\section{Additional figures}\label{subsec_app:additional_figures}

\begin{figure}
    \centering
     \begin{subfigure}[t]{\textwidth}
    \includegraphics[width=\textwidth]{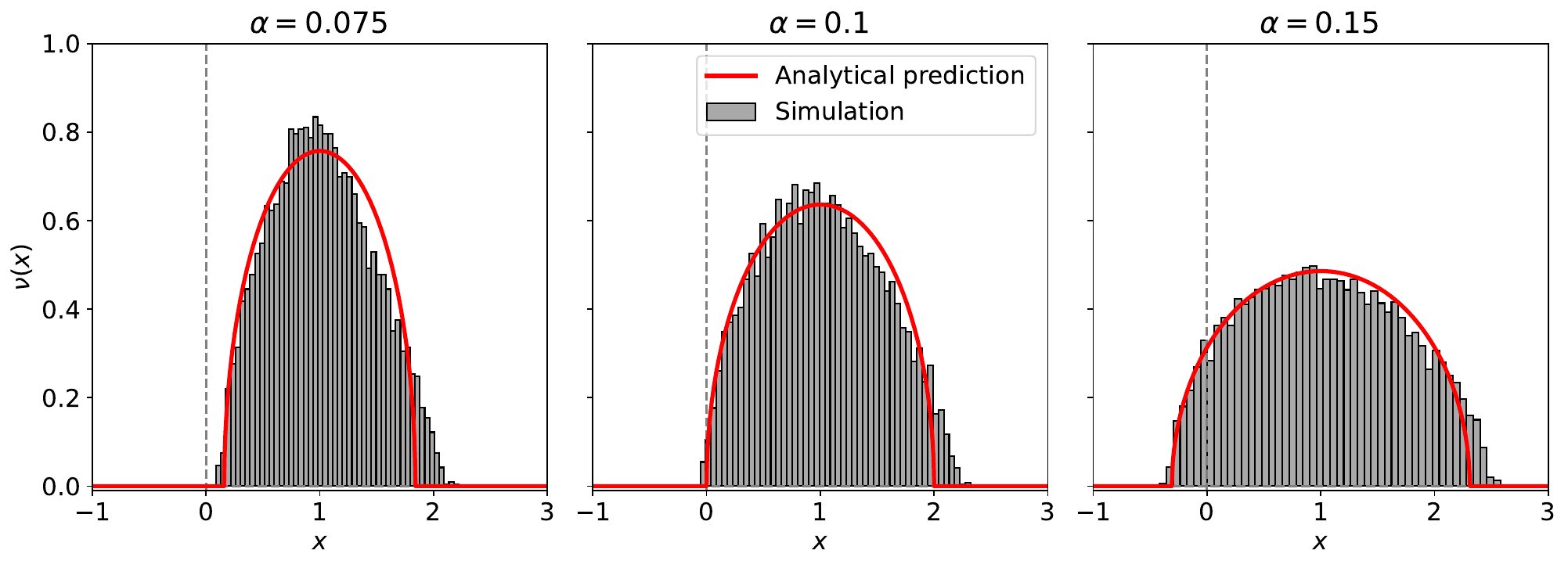}
    \caption{
        The typical spectral density of the least-squares solution for $\alpha \in \{0.075, 0.1, 0.15\}$
         For each $\alpha$, we compare with finite-size simulations using $d = 100$, combining $50$ realizations.
        The predicted threshold is $\alpha_c(\gamma = 2) = 0.1$, in good agreement with our numerical simulations.
    \label{fig_app:ls_comparison_numerics}
    }
    \end{subfigure}
     \vfill
     \begin{subfigure}[t]{\textwidth}
    \centering
    \includegraphics[width=0.5\textwidth]{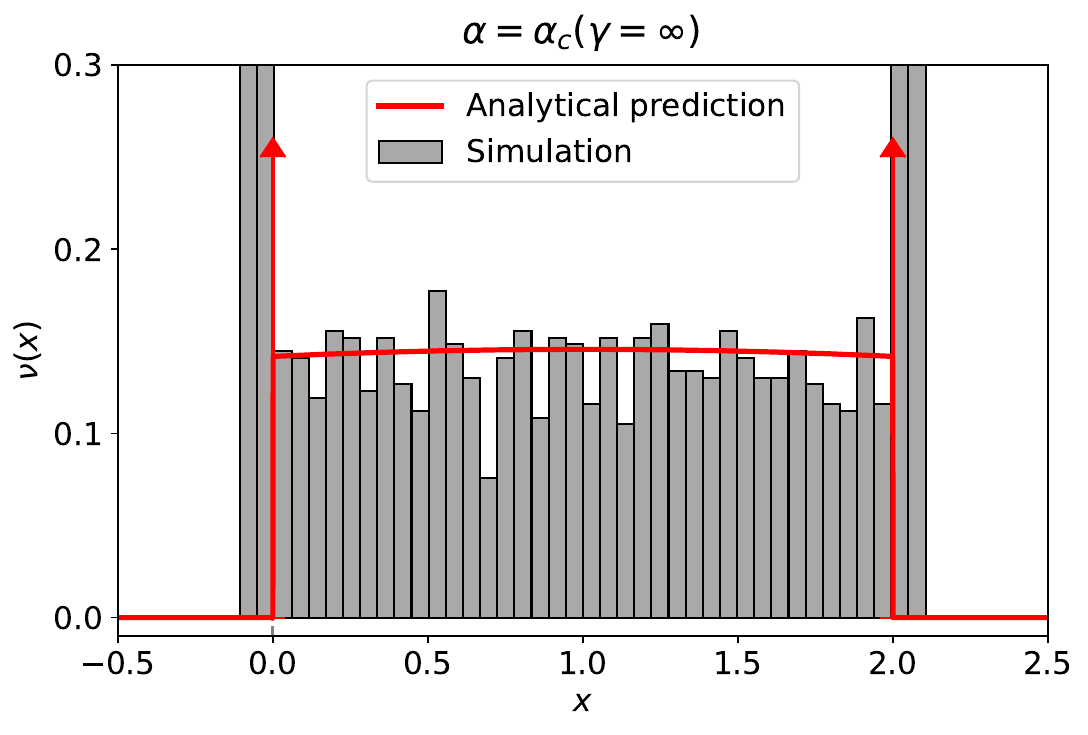}
    \caption{
    The typical spectral density of the solution to $\min_{\bS \in \mcC(\bx)} \|\bS - \Id_d\|_\op$ for $\alpha = \alpha_c(\gamma = \infty) \simeq 0.1892$.
    We compare with finite-size simulations using $d = 100$, $n/d^2 = 0.1892$, and combining $50$ realizations.
   \label{fig_app:shift_op_comparison_numerics}
    }
     \end{subfigure}
     \caption{Comparison of the analytical predictions of the typical spectral density with the numerical simulations, for the least-squares estimator $\min_{\bS \in \mcC(\bx)} \|\bS\|_F$ (top), 
     and the estimator $\min_{\bS \in \mcC(\bx)} \|\bS - \Id_d\|_\op$ (bottom).}
\end{figure}

\begin{figure}
     \centering
     \begin{subfigure}[t]{0.49\textwidth}
         \centering
         \includegraphics[width=\textwidth]{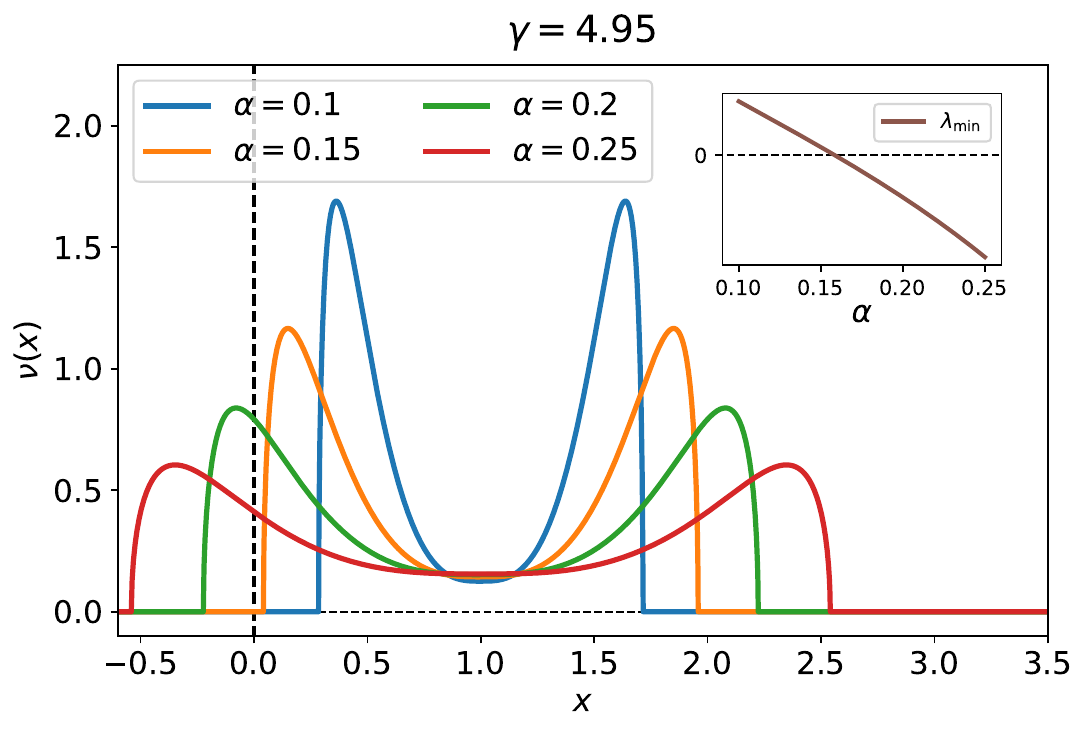}
         \caption{Evolution of $\nu(x)$ as a function of $\alpha$ for a fixed value of $\gamma = 4.95$.
         In inset, we show the evolution of $\lambda_{\min}$ as $\alpha$ grows, deducing the threshold $\alpha_c(\gamma)$ 
         at which the smallest eigenvalue crosses zero.}
     \end{subfigure}
     \hfill
     \begin{subfigure}[t]{0.49\textwidth}
         \centering
         \includegraphics[width=\textwidth]{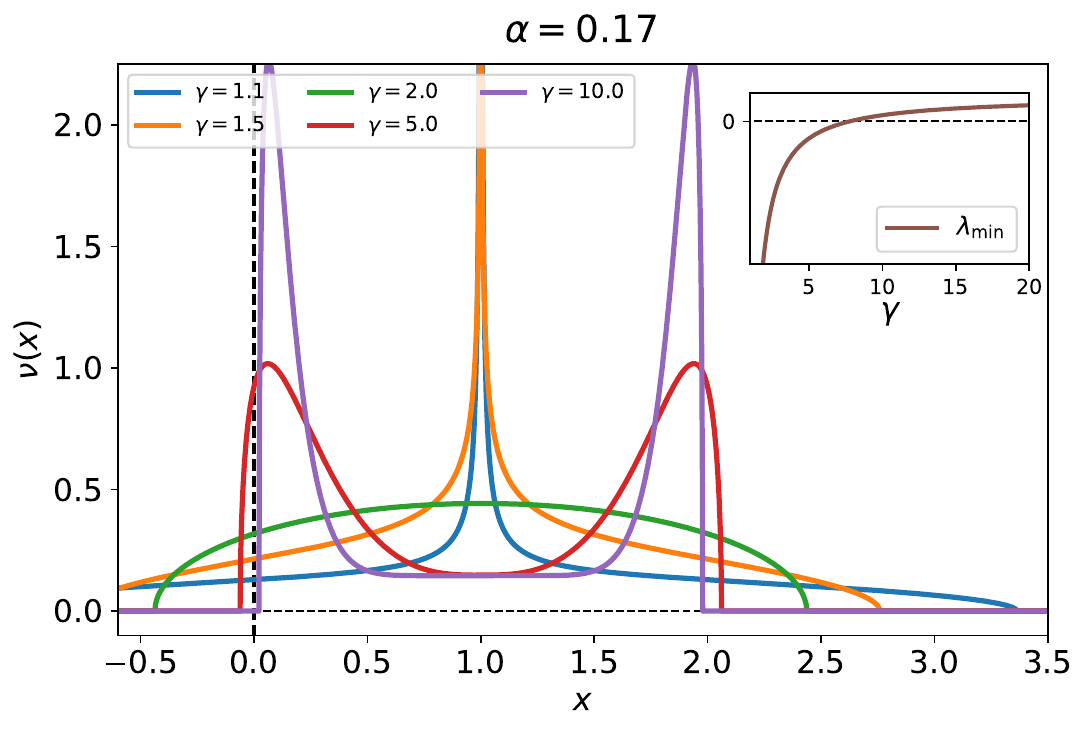}
         \caption{Evolution of $\nu(x)$ as a function of $\gamma$ for a fixed value of $\alpha = 0.17$.
         In inset, we show the evolution of $\lambda_{\min}$ as $\gamma$ grows.}
     \end{subfigure}
        \caption{
         \label{fig_app:nu_gamma_alpha_shift_identity}
        For the estimator $\argmin_{\bS \in \mcC(\bx)} \|\bS - \Id_d\|_{S_\gamma}$ (Section~\ref{subsec:close_identity}), we show the evolution of the asymptotic spectral density $\nu(x)$ as a function of $\alpha$ for fixed $\gamma$ (left) and as a function of $\gamma$ for fixed $\alpha$ (right).
        $\lambda_{\min}$ is the infimum of the support of $\nu$.
        }
\end{figure}

\end{document}